\documentclass[prd,preprint,showpacs,preprintnumbers,nofootinbib,eqsecnum,superscriptaddress]{revtex4}

 \usepackage[dvips,final]{graphicx}
  \usepackage{amssymb}
   \usepackage{amsmath}
    \usepackage{amsfonts}
     \usepackage{epsfig}
      \usepackage{bm}% bold math

\usepackage{mathpazo}

\usepackage[section]{placeins}

\usepackage{multirow}
\usepackage{ctable}
\usepackage{booktabs}
\usepackage{array}
\usepackage{tabularx}
\usepackage{xcolor}
\usepackage{pstricks}
\begin{document}

%\vfill
\title{Charm quark and meson production in association\\ with single-jet
at the LHC
}

\author{Rafa{\l} Maciu{\l}a}
\email{rafal.maciula@ifj.edu.pl} \affiliation{Institute of Nuclear
Physics, Polish Academy of Sciences, Radzikowskiego 152, PL-31-342 Krak{\'o}w, Poland}

\author{Antoni Szczurek\footnote{also at University of Rzesz\'ow, PL-35-959 Rzesz\'ow, Poland}}
\email{antoni.szczurek@ifj.edu.pl} \affiliation{Institute of Nuclear
Physics, Polish Academy of Sciences, Radzikowskiego 152, PL-31-342 Krak{\'o}w, Poland}

\date{\today}

\begin{abstract}
We discuss charm quark/antiquark and charmed meson production
in association with one extra jet (gluon, quark, antiquark) at the LHC.
The calculations are performed both in collinear and $k_T$-factorization
approaches. Different unintegrated gluon distribution functions
are used in the $k_T$-factorization approach.
Several predictions for the LHC are presented. We show distributions in rapidity
and transverse momenta of $c$/$\bar c$ (or charmed mesons) and the
associated jet as well as some two-dimensional observables.
Interesting correlation effects are predicted, \textit{e.g.} in azimuthal angles $\varphi_{c\bar{c}}$ and $\varphi_{c\mathrm{\textit{-jet}}}$.
We have also discussed a relation of the $2 \to 2$ and $2 \to 3$
partonic calculations in the region of large transverse momenta of 
charm quarks/antiquarks as well as the similarity of the next-to-leading order collinear
approach and the $k_T$-factorization approach with the KMR unintegrated parton distribution functions.
Integrated cross sections for $D^{0}$ + jet production for ATLAS detector acceptance and for different cuts on jet transverse
momenta are also presented.
\end{abstract}


\maketitle

%-----------------------------------
\section{Introduction}
%-----------------------------------

Production of charm quarks/antiquarks or charmed mesons is interesting
topic from the point of view of applications of quantum chromodynamics.
At large energies and in particular at the LHC the gluon initiated
processes dominate. Therefore the charm production processes can be
very useful in testing and in extracting (hopefully in a future)
gluon distribution in proton. Since charm quarks are relatively light
one can get access to the region of rather small gluon momentum
fractions where the QCD dynamics is not fully understood.

Usually inclusive distributions of single charmed mesons are presented by
experimental collaborations \cite{ALICE:2011aa,Abelev:2012tca,Aaij:2013mga,Aad:2015zix,Aaij:2015bpa}.
Production of the $D$ meson pairs and correlations between them were discussed so far only
by the LHCb collaboration \cite{Aaij:2012dz}.
On the theoretical side the inclusive production is described by the collinear
next-to-leading order (NLO) approach \cite{Cacciari:2012ny,Kniehl:2012ti}. An interesting
alternative is the $k_T$-factorization approach. The latter approach is the
only one which was successfully used to describe the correlation observables
\cite{Maciula:2013wg,Karpishkov:2016gda}.

The high luminosity already achieved at the LHC potentially allows 
to study more complicated final states such as $D$ mesons and associated
jets. Such final states are also accessible at present on theoretical
side. The automatized methods (calculations) of multi-leg amplitudes
are in this context very important.
The main effort in this field was concentrated so far on multijet production \cite{Bern:2011ep,Badger:2012pf,Kutak:2016mik}
or production of Higgs boson or gauge bosons in association with
a few jets \cite{Greiner:2015jha,Berger:2010zx,Berger:2010vm}. Recently, also production of two charm quark-antiquark pairs has been carefully studied \cite{vanHameren:2015wva}.
So far not much attention was devoted to similar case when
jets are produced in association with charm. Even without any
calculations one can expect large cross sections for associated
production of charm and jets.
The new situation (the new multi-leg methods, high-luminosity) opens new possibilities in testing dynamics of the
pQCD processes. In our opinion it is a good time to explore the new
possibility.

In the present paper we start the new investigation program limiting to production
of charm quarks and/or $D$ mesons associated with single-jet.
We shall use the leading-order (LO) collinear approach as well as 
the $k_T$-factorization approach. The latter approach was successfully
used both for $c \bar c$ production \cite{Maciula:2013wg} and for
inclusive jet \cite{Kniehl:2011hc}, dijet \cite{Nefedov:2013ywa}
and even four-jet production very recently \cite{Kutak:2016mik}.
In our practical exploration we shall use an unique tool - A Very Handy LIBrary (AVHLIB)\footnote{available for download at https://bitbucket.org/hameren/avhlib} \cite{Bury:2015dla} 
prepared for calculating multi-leg processes, up to four-particle final states within both collinear and $k_T$-factorization frameworks.
In the following paper we wish to compare results obtained with
the two different approaches.

%-----------------------------------------------------------------
\section{A sketch of the theoretical formalism}
%-----------------------------------------------------------------

\begin{figure}[!h]
\begin{minipage}{0.32\textwidth}
 \centerline{\includegraphics[width=1.0\textwidth]{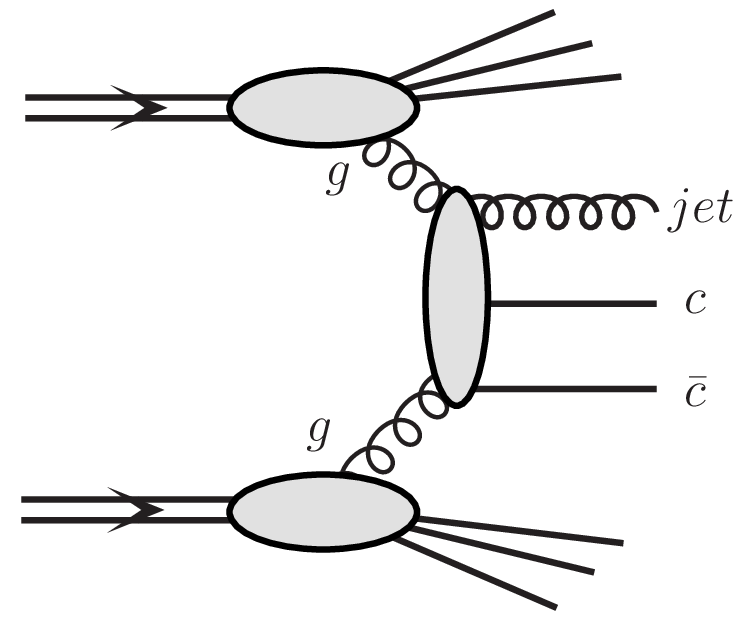}}
\end{minipage}
\hspace{0.1cm}
\begin{minipage}{0.32\textwidth}
 \centerline{\includegraphics[width=1.0\textwidth]{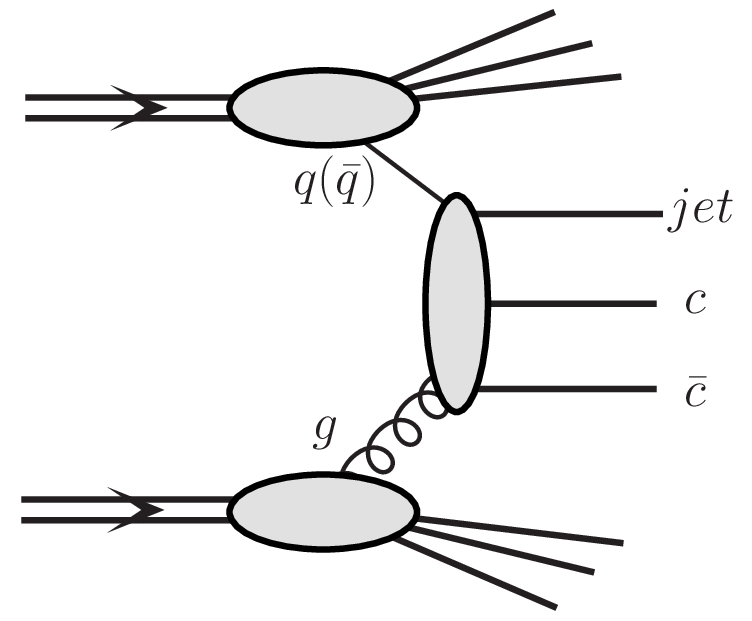}}
\end{minipage}
\hspace{0.1cm}
\begin{minipage}{0.32\textwidth}
 \centerline{\includegraphics[width=1.0\textwidth]{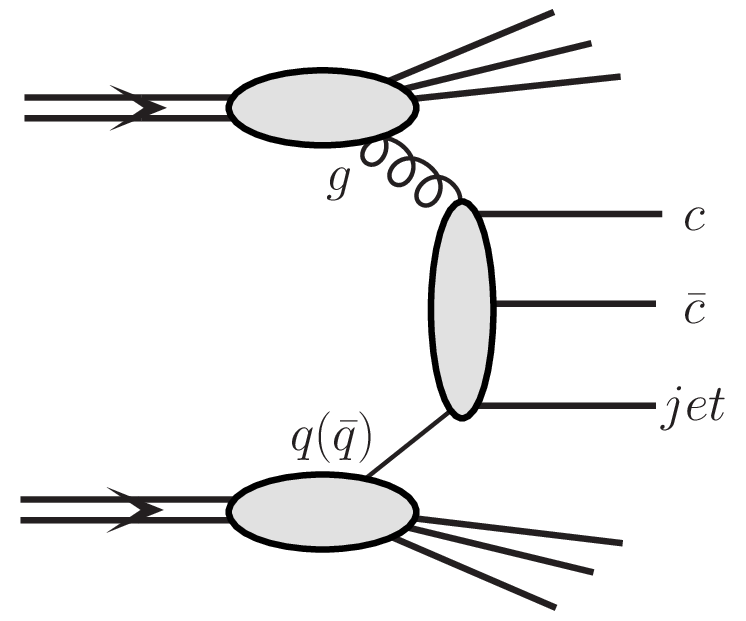}}
\end{minipage}
   \caption{A diagrammatic representation of the mechanisms considered
            for the $pp \to c\bar c + jet$ reaction.
\small 
 }
 \label{fig:diagrams}
\end{figure}

The diagrams under consideration relevant for the production of the $c \bar c$ pair in association with single-jet
are shown schematically in Fig.~\ref{fig:diagrams}. We include three clasess of the QCD $2\to 3$ partonic subprocesses:
$gg \to c\bar c g$, $g q(\bar q) \to c\bar c q(\bar q)$ and $q(\bar q) g \to c\bar c q(\bar q)$, working in the $n_{F}=3$
flavour scheme, where $q = u, d, s$ and $\bar q = \bar u, \bar d, \bar s$.

%-----------------------------------------------------------------------------
%------------------------------------------------------------------------------

The hadronic cross section for inclusive $pp \to c \bar c + jet$ reaction in the LO collinear approach can be written as: 
\begin{equation}
    \begin{split}
d \sigma (pp\to c \bar c + jet) & = \int d x_1 d x_2 \;
          \Big[ g(x_1,\mu_F^2) g(x_2,\mu_F^2) 
           \; d \hat{\sigma}_{gg \to c \bar c g}  \\
         &+  \sum_f \; q_f(x_1,\mu_F^2) g(x_2,\mu_F^2) 
           \; d \hat{\sigma}_{q g \to c \bar c q} 
         +  g(x_1,\mu_F^2) \sum_f \; q_f(x_2,\mu_F^2)
           \; d \hat{\sigma}_{ g q \to c \bar c q} \\
         &+  \sum_f \; \bar {q}_f(x_1,\mu_F^2) g(x_2,\mu_F^2) 
           \; d \hat{\sigma}_{\bar q g \to c \bar c \bar q} 
         +  g(x_1,\mu_F^2) \sum_f \; \bar{q}_f(x_2,\mu_F^2)
           \; d \hat{\sigma}_{g \bar q \to c \bar c \bar q}\Big] \; , 
    \end{split}
    \label{hadronic_cs_coll}
\end{equation}
where $g(x_{1,2},\mu_F^2)$, $q_{f}(x_{1,2},\mu_F^2)$ and $\bar{q}_{f}(x_{1,2},\mu_F^2)$ are the standard collinear parton distribution functions (PDFs) for gluons,
quarks and antiquarks, respectively, carrying $x_{1,2}$ momentum fractions of the proton and evaluated at the factorization scale $\mu_{F}$. Here, $d\hat{\sigma}$ are the elementary partonic cross sections for a given $2 \to 3$ subprocess.  

The elementary cross section, \textit{e.g} for the $gg \to c \bar c g$ mechanism has the following generic form:
\begin{equation}
d\hat{\sigma} = \frac{1}{2\hat{s}} \; \overline{|{\cal M}_{g g \rightarrow c \bar{c} g}|^2} \; \frac{d^3 p_1}{2 E_1 (2 \pi)^3} \frac{d^3 p_2}{2 E_2 (2 \pi)^3}
           \frac{d^3 p_3}{2 E_3 (2 \pi)^3} (2\pi)^{4}
           \delta^3 \left( p_1 + p_2 + p_3 - k_{1} - k_{2} \right) \; ,
\end{equation}
where ${\cal M}_{g g \rightarrow c \bar{c} g}$ is the partonic on-shell matrix element, $\hat{s}$ is the partonic center-of-mass energy squared, 
$p_1, p_2, p_3$ are four-momenta of final 
$c$ quark, $\bar c$ antiquark and gluon, respectively, and $k_{1}$ and $k_{2}$ are
four-momenta of incoming gluons.

Switching to the $k_{T}$-factorization approach, the analogous formula to (\ref{hadronic_cs_coll}) takes the following form:
\begin{equation}
    \begin{split}
d \sigma &(pp\to c \bar c + jet) = \int d x_1 \frac{d^2 k_{1t}}{\pi} d x_2 \frac{d^2 k_{2t}}{\pi} \; \Big[ 
           {\cal F}_{g}(x_1,k_{1t}^{2},\mu_F^2) {\cal F}_{g}(x_2,k_{2t}^{2},\mu_F^2) 
           \; d \hat{\sigma}_{g^*g^* \to c \bar c g} \\
         &+ {\cal F}_{q}(x_1,k_{1t}^{2},\mu_F^2) {\cal F}_{g}(x_2,k_{2t}^{2},\mu_F^2) 
           \; d \hat{\sigma}_{q^* g^* \to c \bar c q} 
         +  {\cal F}_{g}(x_1,k_{1t}^{2},\mu_F^2) {\cal F}_{q}(x_2,k_{2t}^{2},\mu_F^2)
           \; d \hat{\sigma}_{ g^* q^* \to c \bar c q} \\
         &+ {\cal F}_{\bar q}(x_1,k_{1t}^{2},\mu_F^2) {\cal F}_{g}(x_2,k_{2t}^{2},\mu_F^2) 
           \; d \hat{\sigma}_{\bar{q}^* g^* \to c \bar c \bar{q}} 
         +  {\cal F}_{g}(x_1,k_{1t}^{2},\mu_F^2) {\cal F}_{\bar q}(x_2,k_{2t}^{2},\mu_F^2)
           \; d \hat{\sigma}_{ g^* \bar{q}^* \to c \bar c \bar{q}} \Big] \; .
    \end{split}
        \label{hadronic_cs_kT}
\end{equation}
Here, $k_{1,2t}$ are transverse momenta of incident partons (new degrees of freedom compared to collinear approach) and ${\cal F}(x,k_{t}^{2},\mu_F^2)$'s are transverse momentum dependent, so-called, unintegrated parton distribution functions (uPDFs). Within this framework the elementary partonic cross sections are defined in terms of off-shell matrix elements, that take into account that both partons entering the hard process are off-shell with virtualities $k_{1}^{2} = - k_{1t}^2$ and $k_{2}^{2} = - k_{2t}^2$.

The off-shell matrix elements are known only in the LO and only for limited types of QCD $2 \to 2$ processes (see e.g. heavy quarks \cite{Catani:1990eg}, dijet \cite{Nefedov:2013ywa}, Drell-Yan \cite{Nefedov:2012cq}). Some first steps to calculate NLO corrections in the $k_{T}$-factorization framework have been done only very recently for diphoton production \cite{Nefedov:2015ara,Nefedov:2016clr}.
Here, we extend the standard scope of mechanisms usually studied in the $k_{T}$-factorization approach by analysing the three-particle $c \bar c + jet$ final state. 
Moving on to higher final state parton multiplicities, it is possible to generate relevant amplitudes analytically applying suitably defined Feynman rules \cite{vanHameren:2012if} or recursive methods, like  generalised  BCFW recursion \cite{vanHameren:2014iua}, or numerically with the help of methods of numerical BCFW recursion \cite{Bury:2015dla}
as implemented in AVHLIB. The latter method was already successfully applied even for $2 \to 4$ production mechanisms in the case of $c\bar c c\bar c$ \cite{vanHameren:2015wva} and four-jet \cite{Kutak:2016mik} final states.

In this Fortran library, scattering amplitudes are calculated numerically as a function of the
external four-momenta via Dyson-Schwinger recursion generalized to tree-level amplitudes with off-shell initial-state particles. This
recursive method exists in several explicit implementations with
on-shell initial-state particles (see e.g. Ref.~\cite{Kanaki:2000ey}). AVHLIB allows for various
choices of the representation of the external helicities and colors. The library includes a full
Monte Carlo program with an adaptive phase-space generator \cite{vanHameren:2007pt,vanHameren:2010gg} that deals with the integration variables related to both the
initial-state momenta and the final-state momenta. The program can also conveniently generate a file of unweighted events, which approach was used in the analysis presented in this paper.

In the numerical calculations below, we set charm quark mass $m_{c} = 1.5$ GeV and renormalization/factorization scales $\mu = \mu_{R}=\mu_{F} = \sqrt{\frac{m_{t,c}^2 + m_{t,\bar{c}}^2 + p_{t,jet}^2}{3}}$, where $m_{t} = \sqrt{p_{t}^2 + m_{c}^2}$ is the transverse mass of charm quark or antiquark. We use running strong-coupling $\alpha_{S}$ at next-to-leading order as implemented in the MMHT2014 set of PDFs \cite{Harland-Lang:2014zoa}.

%--------------------------------------------------
\section{Numerical results}
%--------------------------------------------------

%----------------------------------------
\subsection{Collinear approach}
%----------------------------------------

\begin{figure}[!h]
\begin{minipage}{0.47\textwidth}
 \centerline{\includegraphics[width=1.0\textwidth]{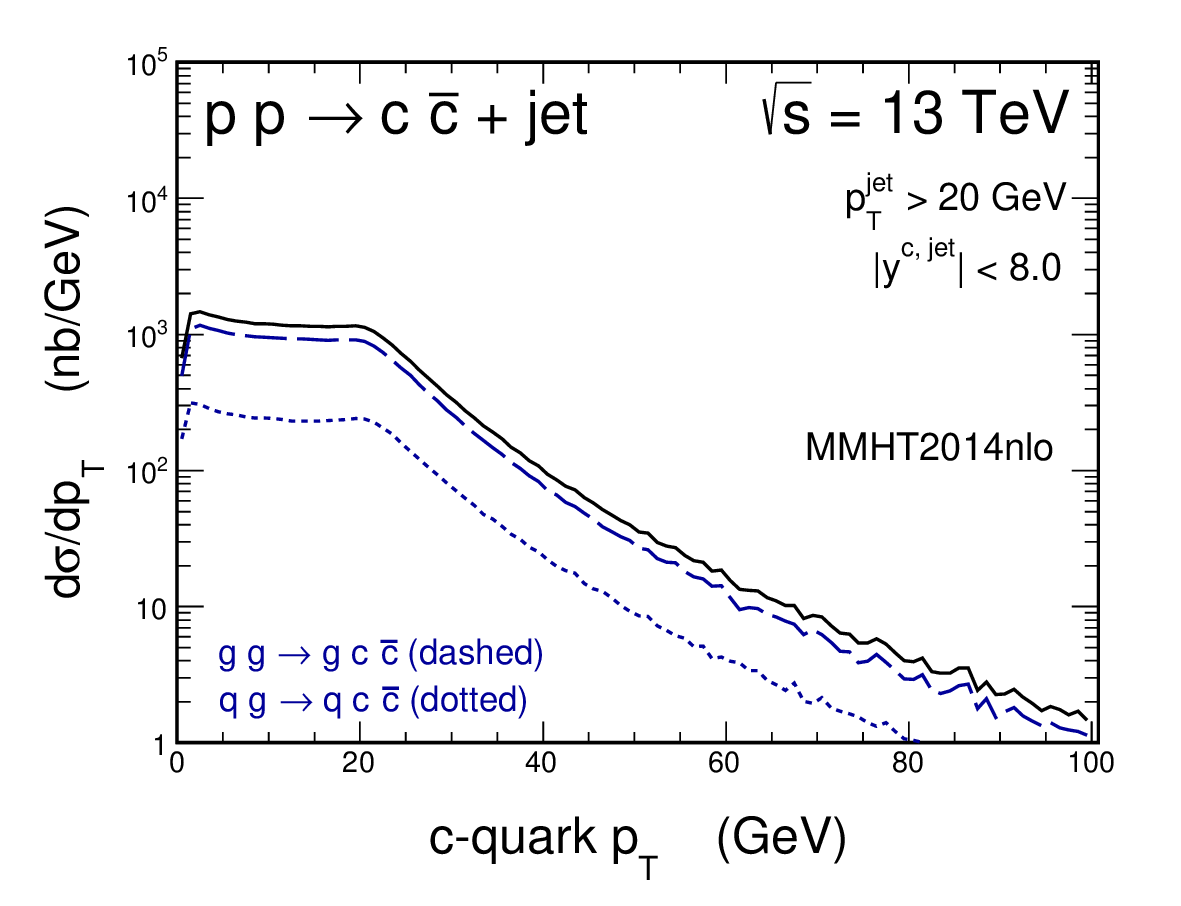}}
\end{minipage}
\hspace{0.5cm}
\begin{minipage}{0.47\textwidth}
 \centerline{\includegraphics[width=1.0\textwidth]{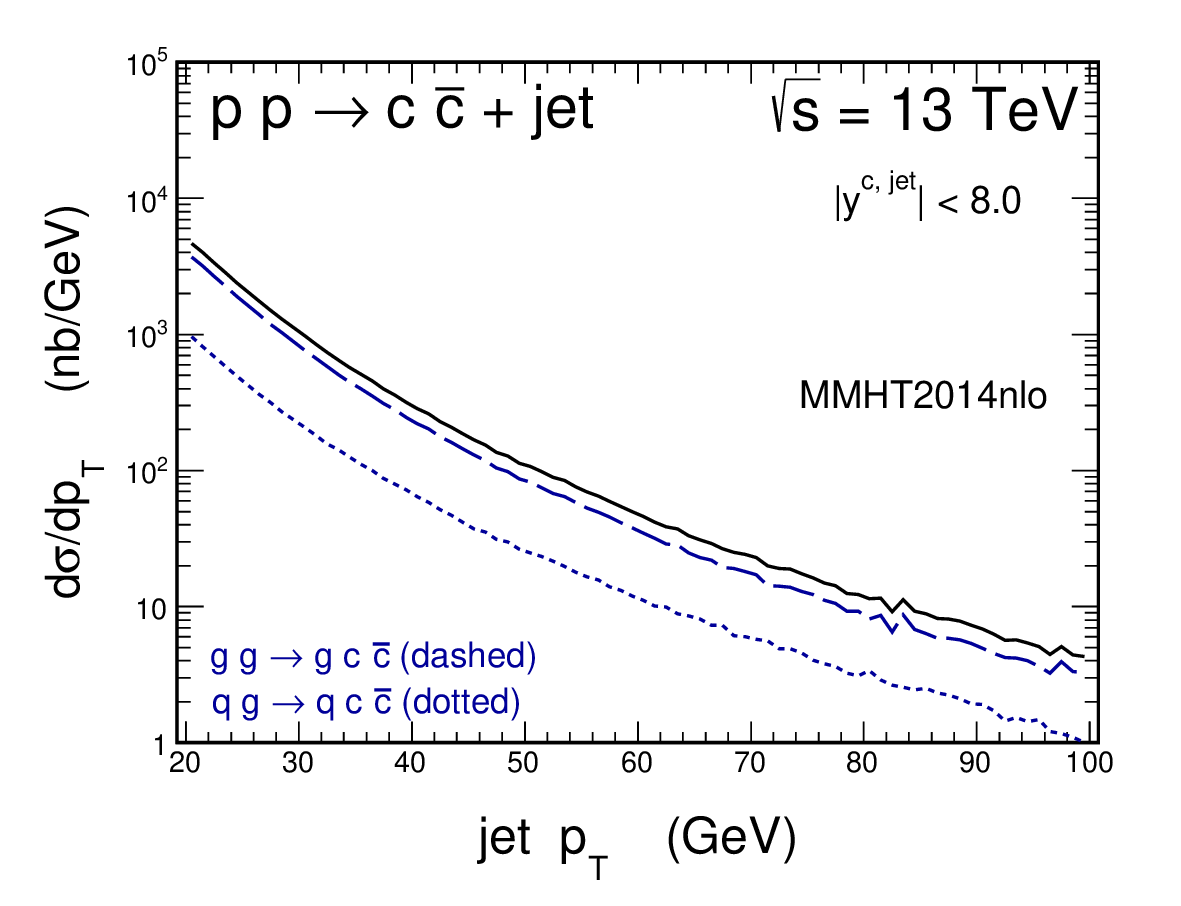}}
\end{minipage}
   \caption{
\small Transverse momentum distribution of c-quark (let panel) and
associated jet (right panel) in the collinear approach.
 }
 \label{fig:coll_ccbar_pt}
\end{figure}

\begin{figure}[!h]
\begin{minipage}{0.47\textwidth}
 \centerline{\includegraphics[width=1.0\textwidth]{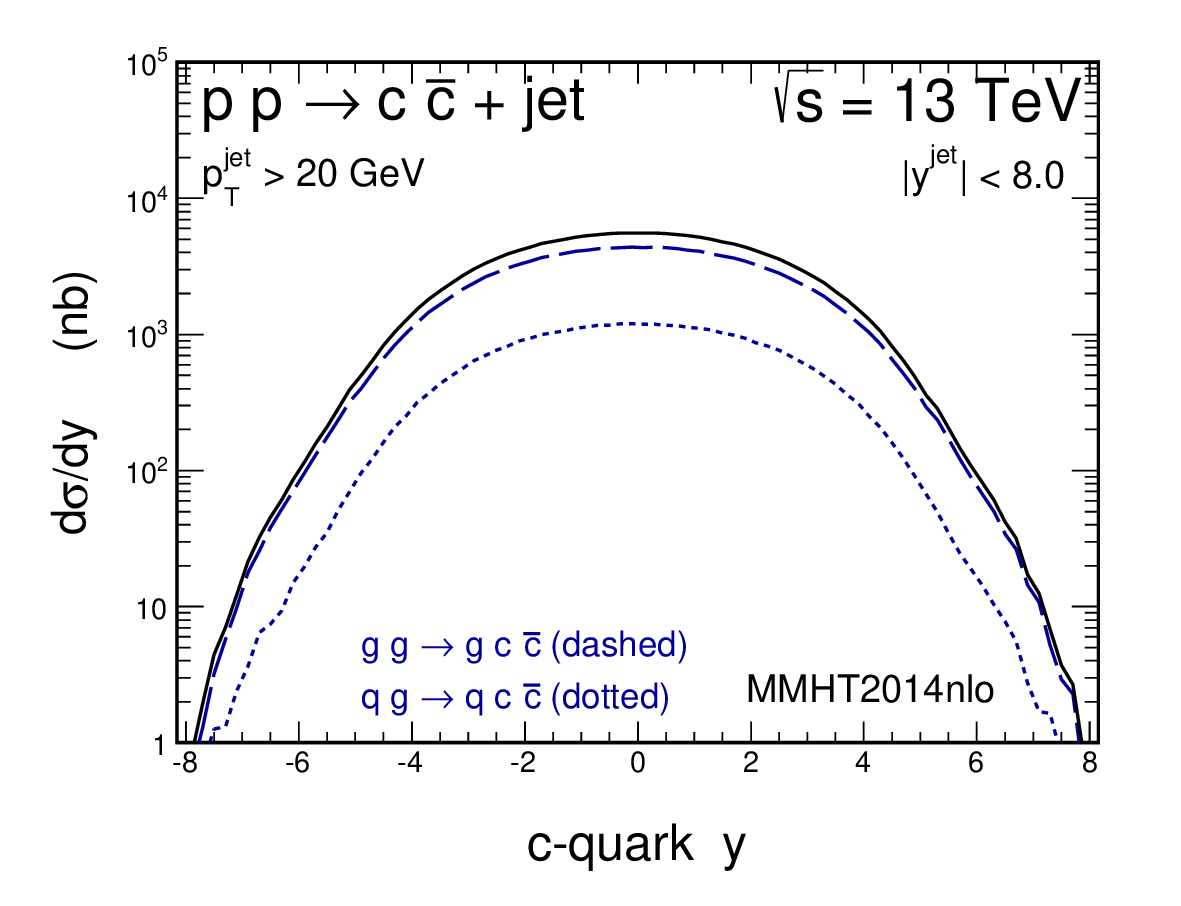}}
\end{minipage}
\hspace{0.5cm}
\begin{minipage}{0.47\textwidth}
 \centerline{\includegraphics[width=1.0\textwidth]{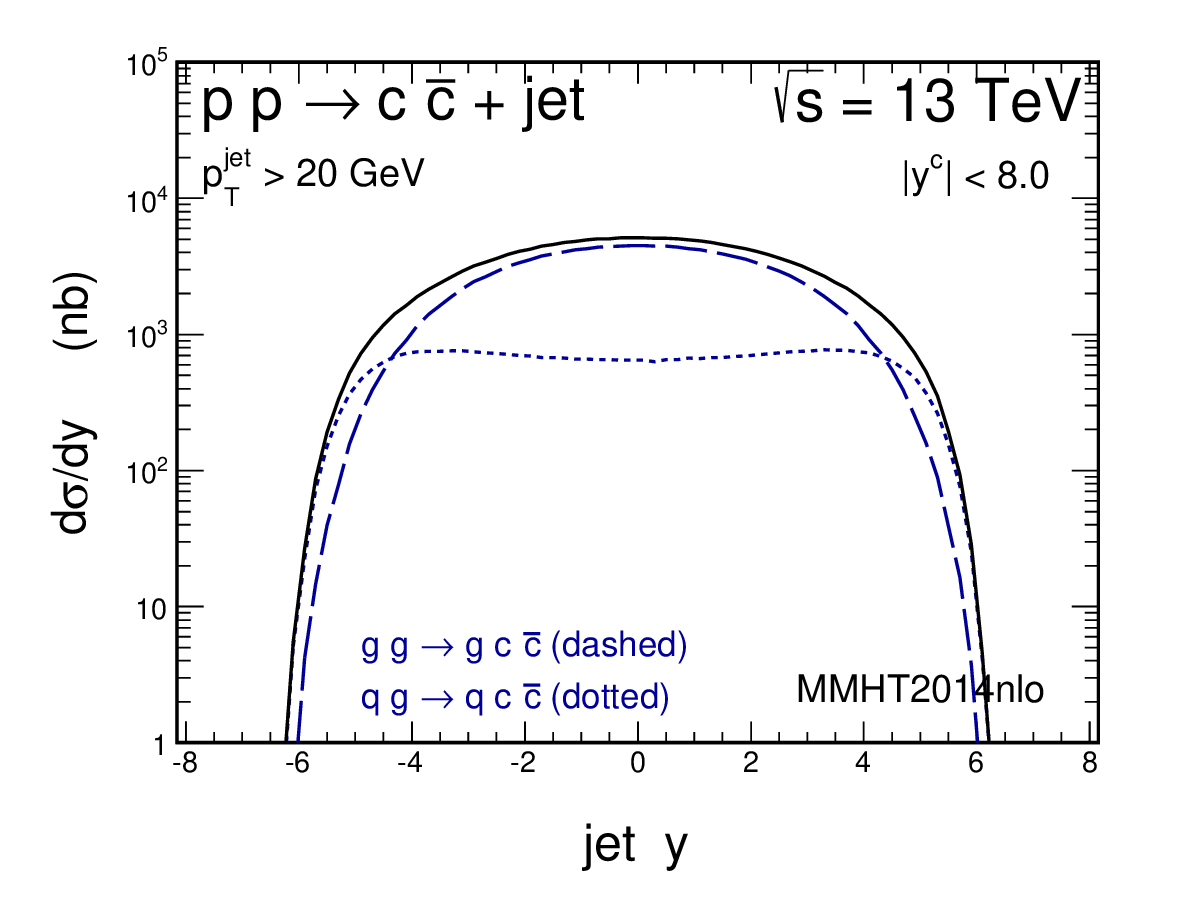}}
\end{minipage}
   \caption{
\small Rapidity distribution of c-quark (left panel) and
associated jet (right panel) in the collinear approach.
 }
 \label{fig:coll_ccbar_y}
\end{figure}

\begin{figure}[!h]
\begin{minipage}{0.47\textwidth}
 \centerline{\includegraphics[width=1.0\textwidth]{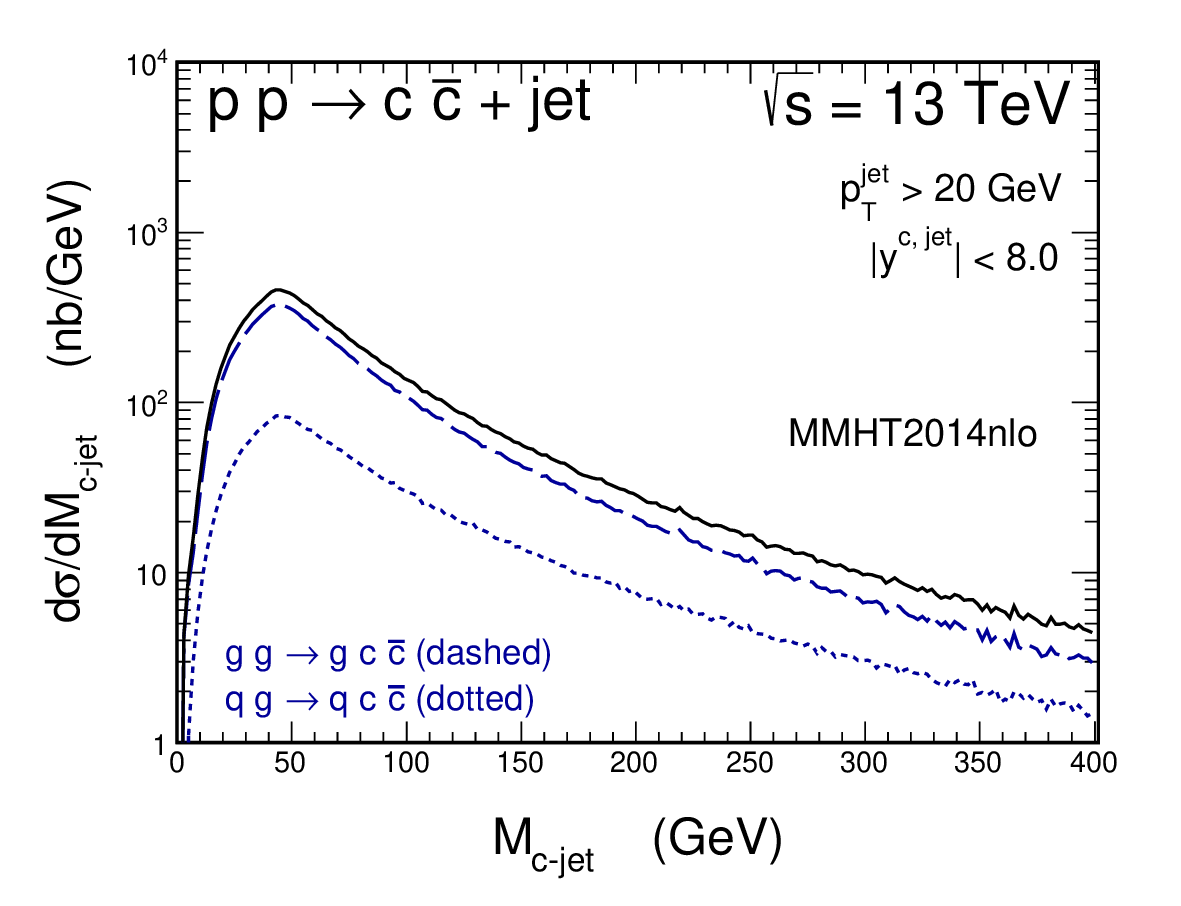}}
\end{minipage}
\hspace{0.5cm}
\begin{minipage}{0.47\textwidth}
 \centerline{\includegraphics[width=1.0\textwidth]{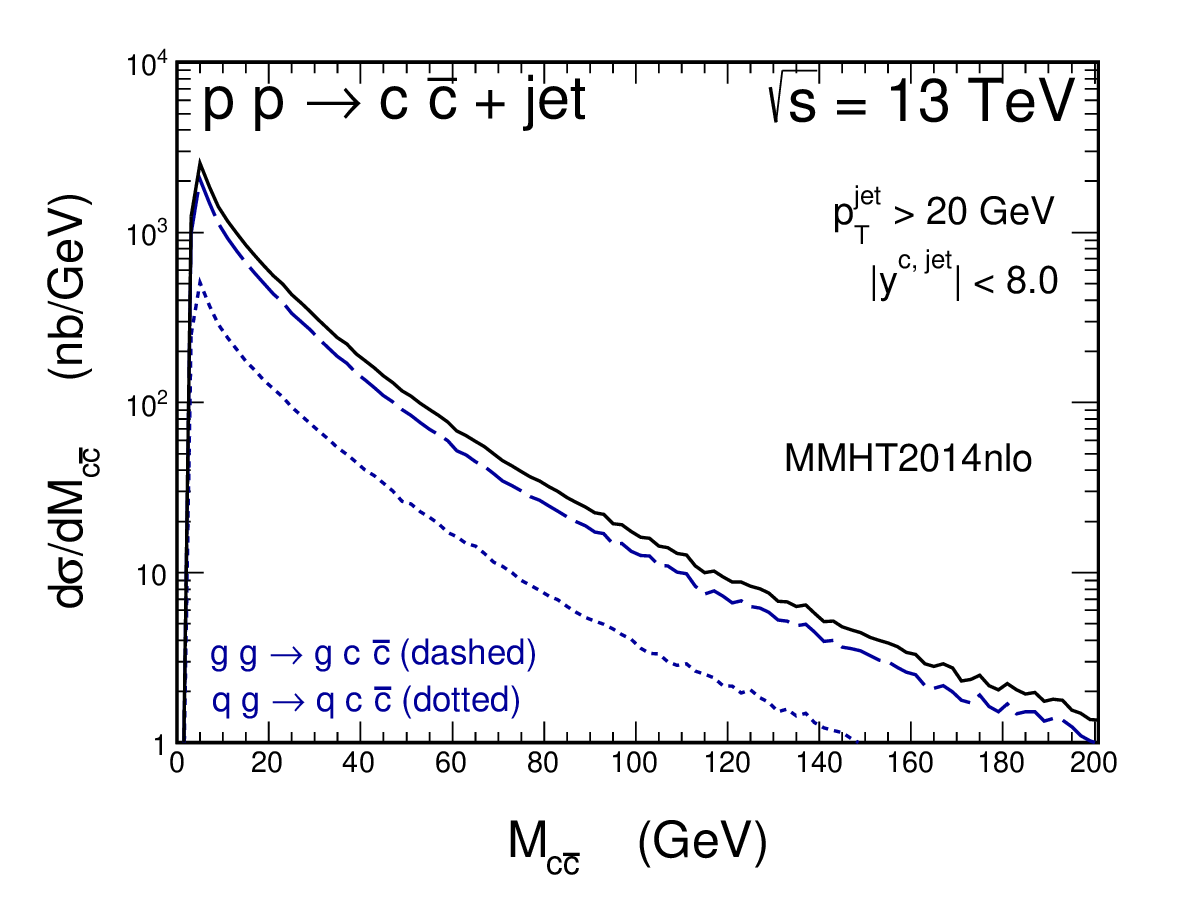}}
\end{minipage}
   \caption{
\small Distribution in invariant mass of the c-quark-jet (left panel)
       and $c \bar c$ system (right panel) in the collinear approach.
 }
 \label{fig:coll_ccbar_M}
\end{figure}

\begin{figure}[!h]
\begin{minipage}{0.47\textwidth}
 \centerline{\includegraphics[width=1.0\textwidth]{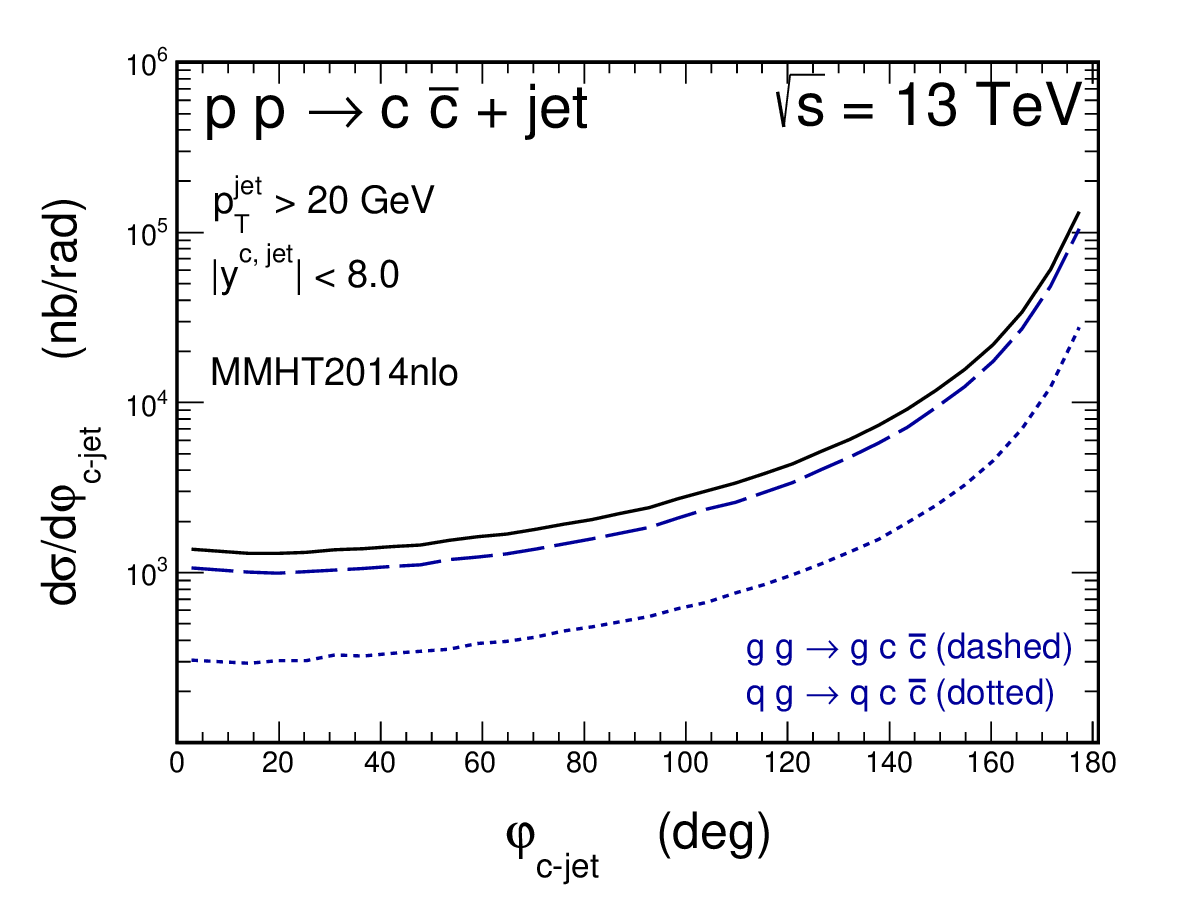}}
\end{minipage}
\hspace{0.5cm}
\begin{minipage}{0.47\textwidth}
 \centerline{\includegraphics[width=1.0\textwidth]{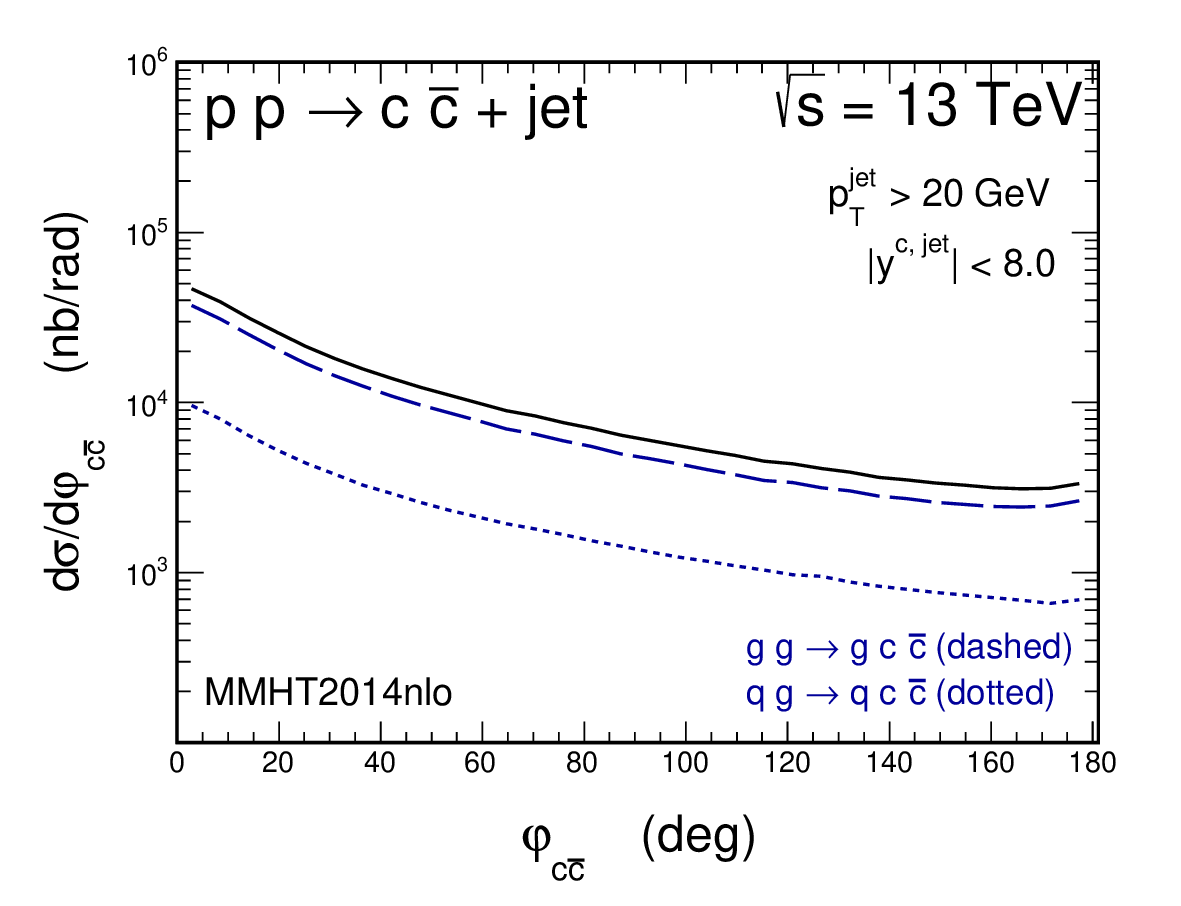}}
\end{minipage}
   \caption{
\small Distribution in azimuthal angle between c-quark and jet (left
panel) and between $c$-quark and ${\bar c}$-antiquark (right panel)
in the collinear approach.
 }
 \label{fig:coll_ccbar_phi}
\end{figure}

Here we start presentation of our results within the collinear approach
which is a good reference point for further calculations in 
the $k_T$-factorization. In this subsection we shall show results
in the full phase space at the parton level.
In Figs.~\ref{fig:coll_ccbar_pt}, \ref{fig:coll_ccbar_y}, \ref{fig:coll_ccbar_M} and \ref{fig:coll_ccbar_phi}
we show distributions in $c$/$\bar c$ transverse momentum, rapidity,
diparton invariant masses as well as in relative azimuthal angle
between different outgoing partons.
In these calculation we used MMHT2014nlo set of PDFs \cite{Harland-Lang:2014zoa} as an example.
Here by jets we understand partons (gluons, quarks, antiquarks) with
transverse momenta $p_{T}^{jet} > 20$ GeV.
We find that for the full phase space the contribution of processes
$g g \to c \bar c g$ is 3-4 times larger than that for
combined $q(\bar q) g \to c \bar c q(\bar q)$, $g q(\bar q) \to c \bar c
q(\bar q)$ partonic processes.
However, the processes with light quarks/antiquarks start to dominate
at large jet (pseudo)rapidities $|\eta| >$ 4.5.
We observe a broad plateau in transverse momentum distribution of $c$/$\bar c$ quarks/antiquarks
for $p_{T}^{c} < 20$ GeV. The effect is a consequence of the cut on jet transverse momentum $p_{T}^{jet} > 20$ GeV.
For larger cuts the plateau would be even broader. 
The distributions in $\varphi_{c\mathrm{\textit{-jet}}}$ and $\varphi_ {c \bar c}$ are
particularly interesting. While the first one has rather typical
dependence with the maximum at the back-to-back configuration,
the second one has maximum at $\varphi_{c \bar c} \sim$ 0, rather different
than in the case of inclusive $c\bar c$ production \cite{Maciula:2013wg}.

In the next subsection we show similar results obtained in the
$k_T$-factorization approach.

%-----------------------------------------------------------------------------
%------------------------------------------------------------------------------

%-----------------------------------------------------------------------------
%------------------------------------------------------------------------------

%-----------------------------------------------------------------------------
%------------------------------------------------------------------------------

%-----------------------------------------------------------------------------
%------------------------------------------------------------------------------

%------------------------------------------------------
\subsection{\boldmath{$k_{T}$}-factorizaton approach}
%------------------------------------------------------

\begin{figure}[!h]
\begin{minipage}{0.47\textwidth}
 \centerline{\includegraphics[width=1.0\textwidth]{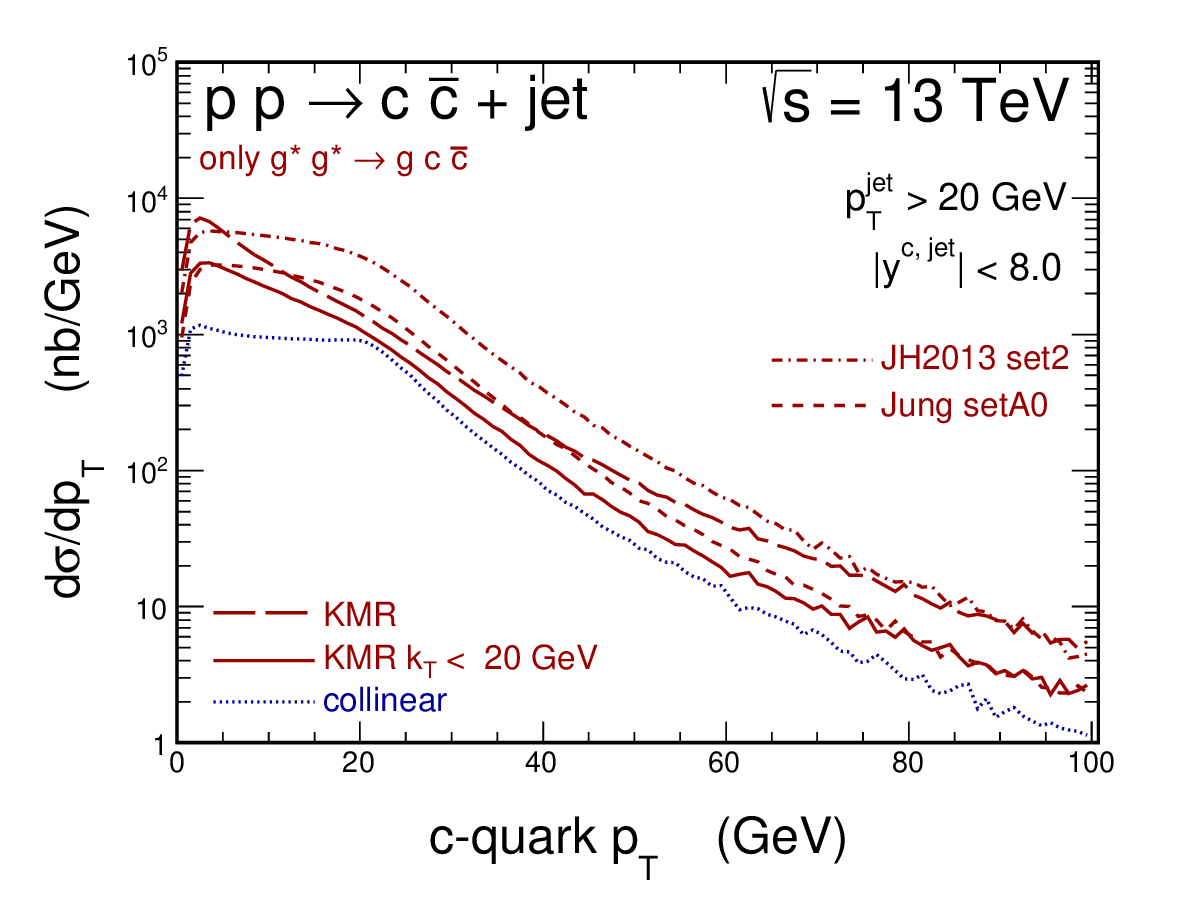}}
\end{minipage}
\hspace{0.5cm}
\begin{minipage}{0.47\textwidth}
 \centerline{\includegraphics[width=1.0\textwidth]{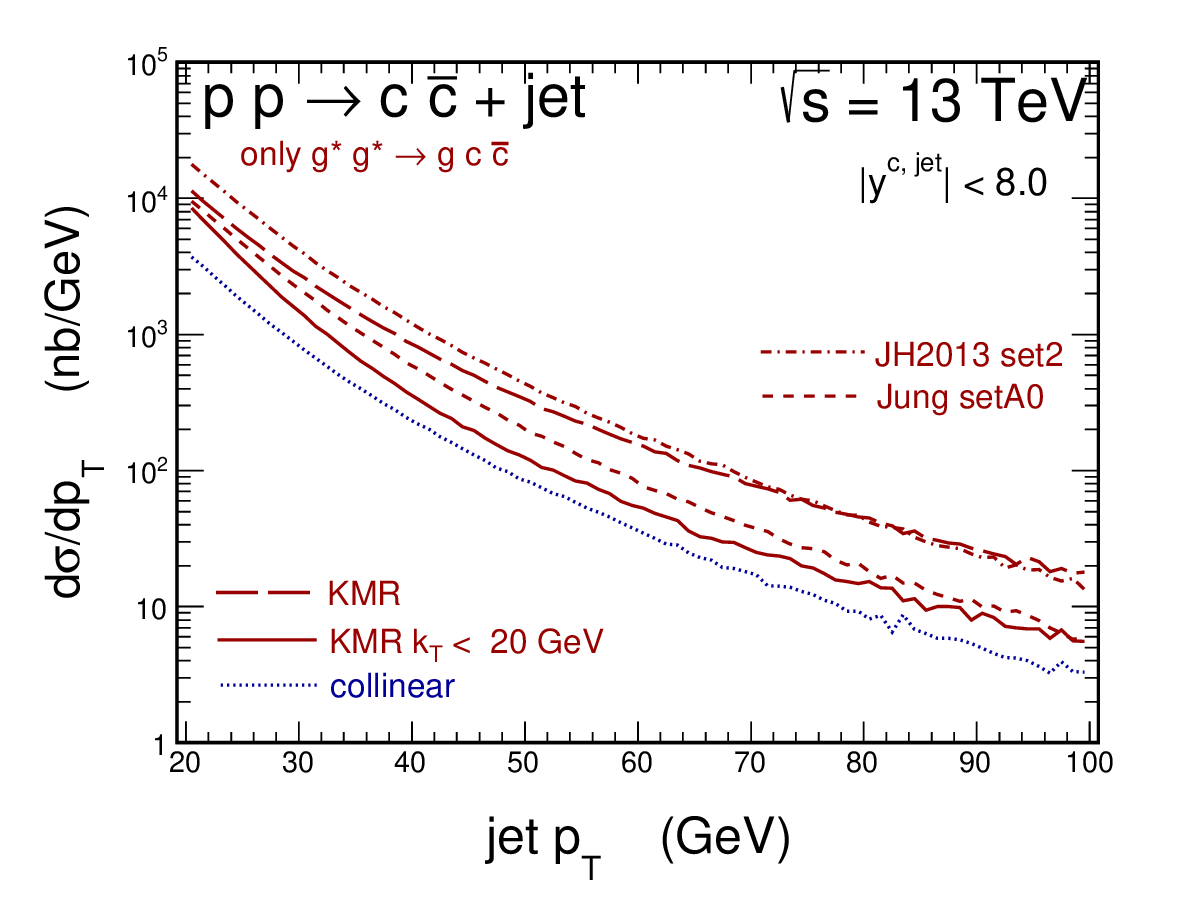}}
\end{minipage}
   \caption{
\small Transverse momentum distribution of c-quark (let panel) and
associated jet (right panel) in the $k_T$-factorization approach
with different uGDFs.}
 \label{fig:kt_ccbar_pt}
\end{figure}

\begin{figure}[!h]
\begin{minipage}{0.47\textwidth}
 \centerline{\includegraphics[width=1.0\textwidth]{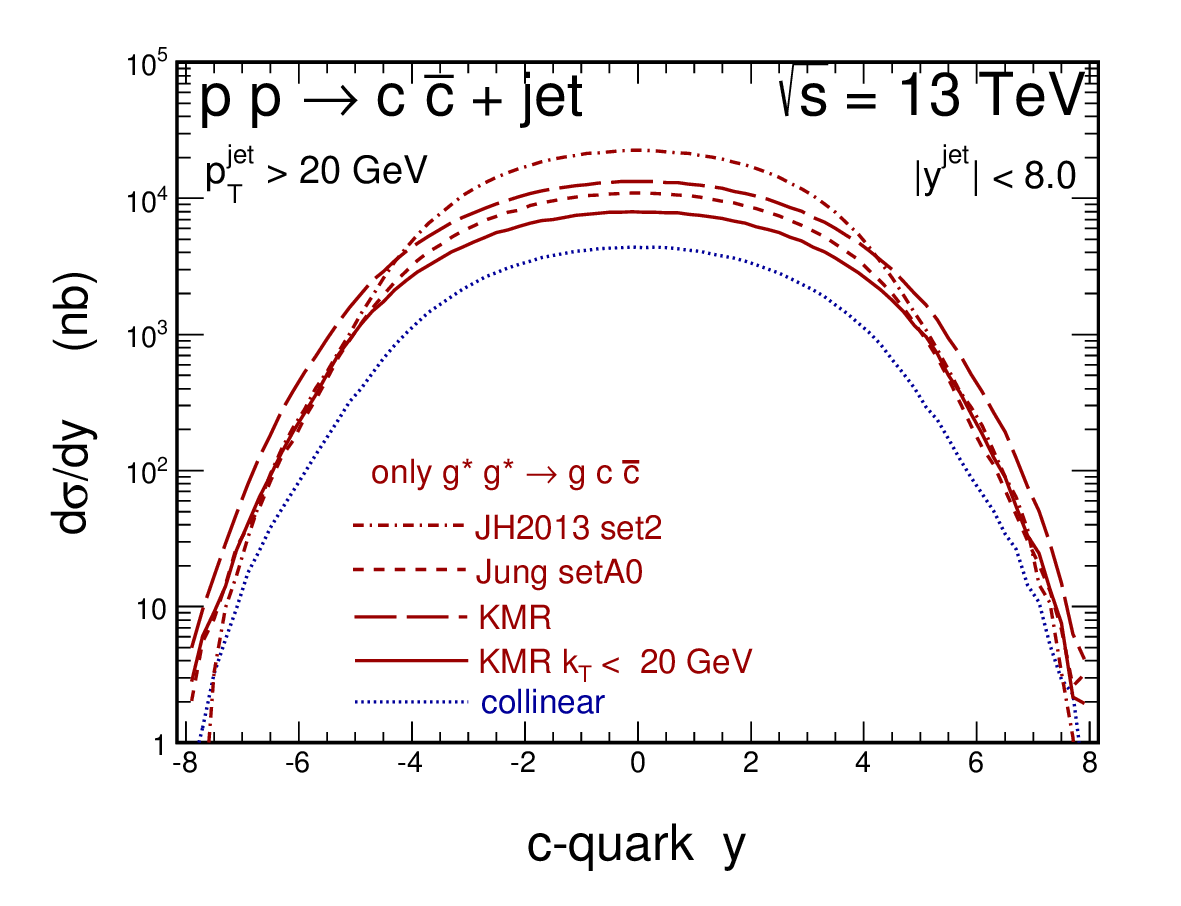}}
\end{minipage}
\hspace{0.5cm}
\begin{minipage}{0.47\textwidth}
 \centerline{\includegraphics[width=1.0\textwidth]{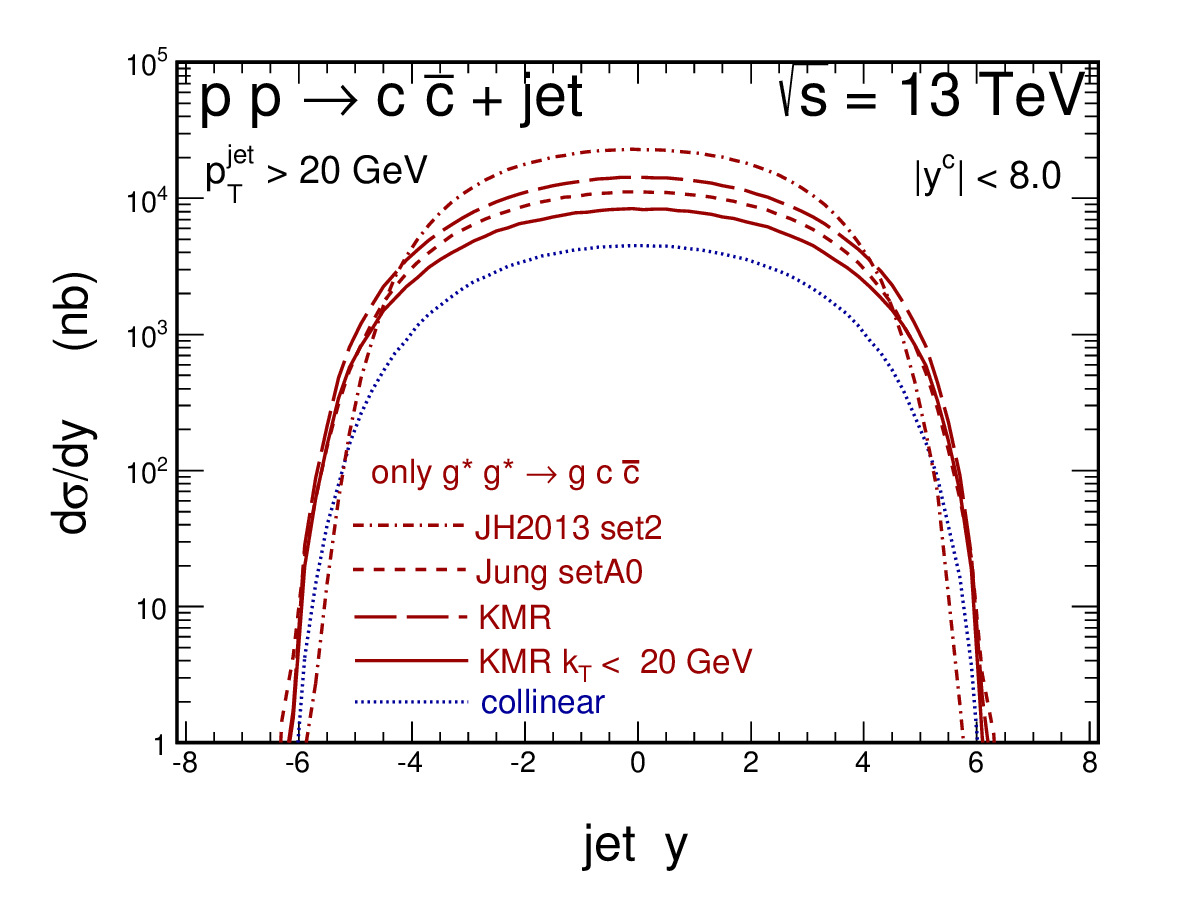}}
\end{minipage}
   \caption{
\small Rapidity distribution of c-quark (left panel) and
associated jet (right panel) in the $k_T$-factorization approach
with different uGDFs.
 }
 \label{fig:kt_ccbar_y}
\end{figure}

\begin{figure}[!h]
\begin{minipage}{0.47\textwidth}
 \centerline{\includegraphics[width=1.0\textwidth]{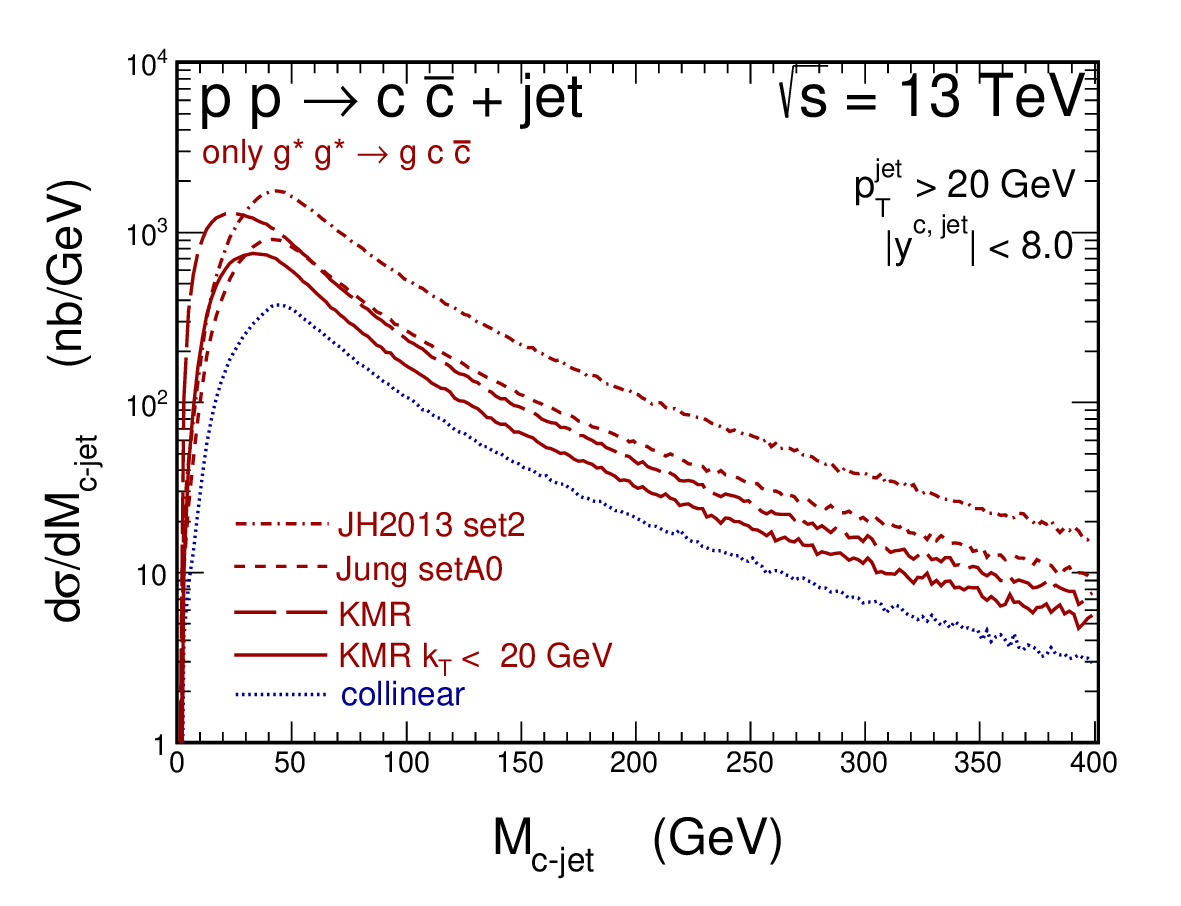}}
\end{minipage}
\hspace{0.5cm}
\begin{minipage}{0.47\textwidth}
 \centerline{\includegraphics[width=1.0\textwidth]{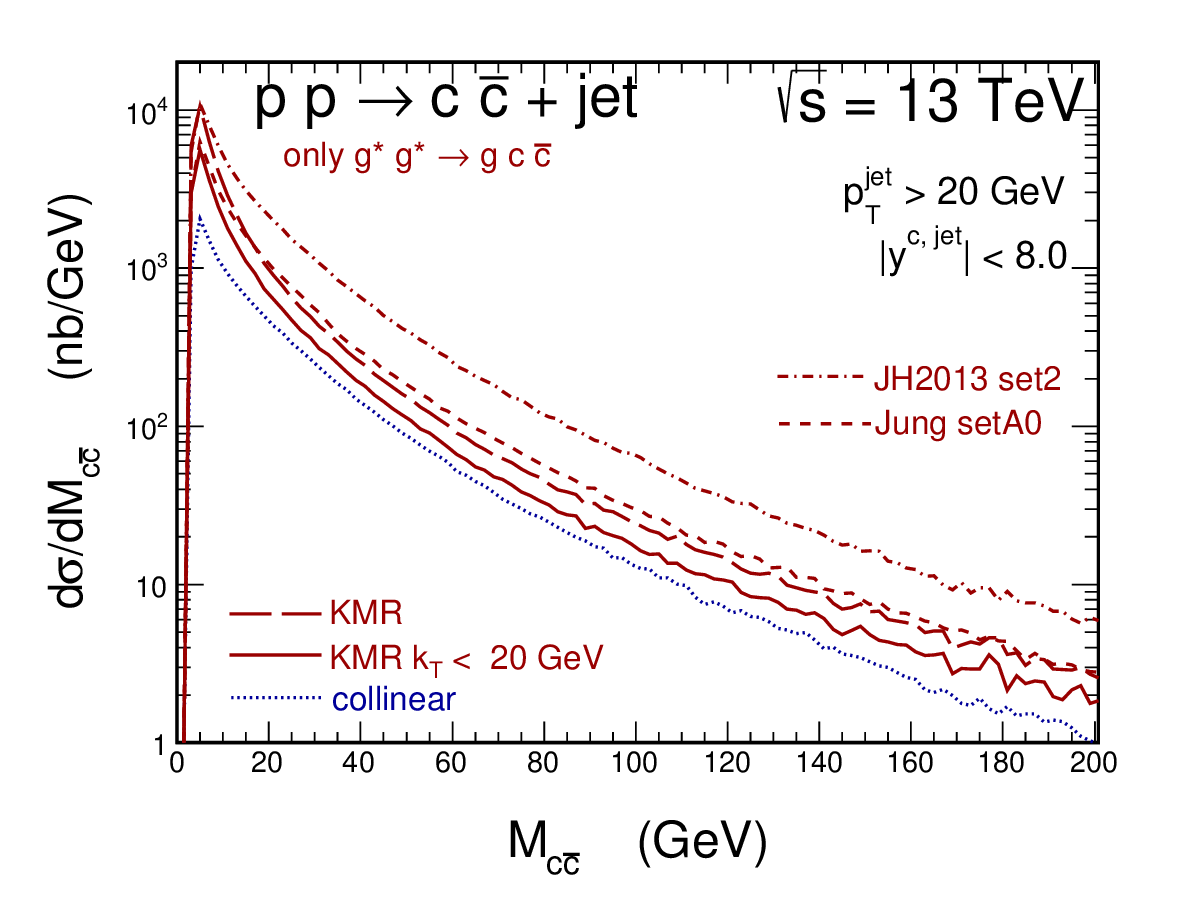}}
\end{minipage}
   \caption{
\small Distribution in invariant mass of the c-quark-jet (left panel)
       and $c \bar c$ system (right panel) in the $k_T$-factorization
       approach with different uGDFs.
 }
 \label{fig:kt_ccbar_M}
\end{figure}

\begin{figure}[!h]
\begin{minipage}{0.47\textwidth}
 \centerline{\includegraphics[width=1.0\textwidth]{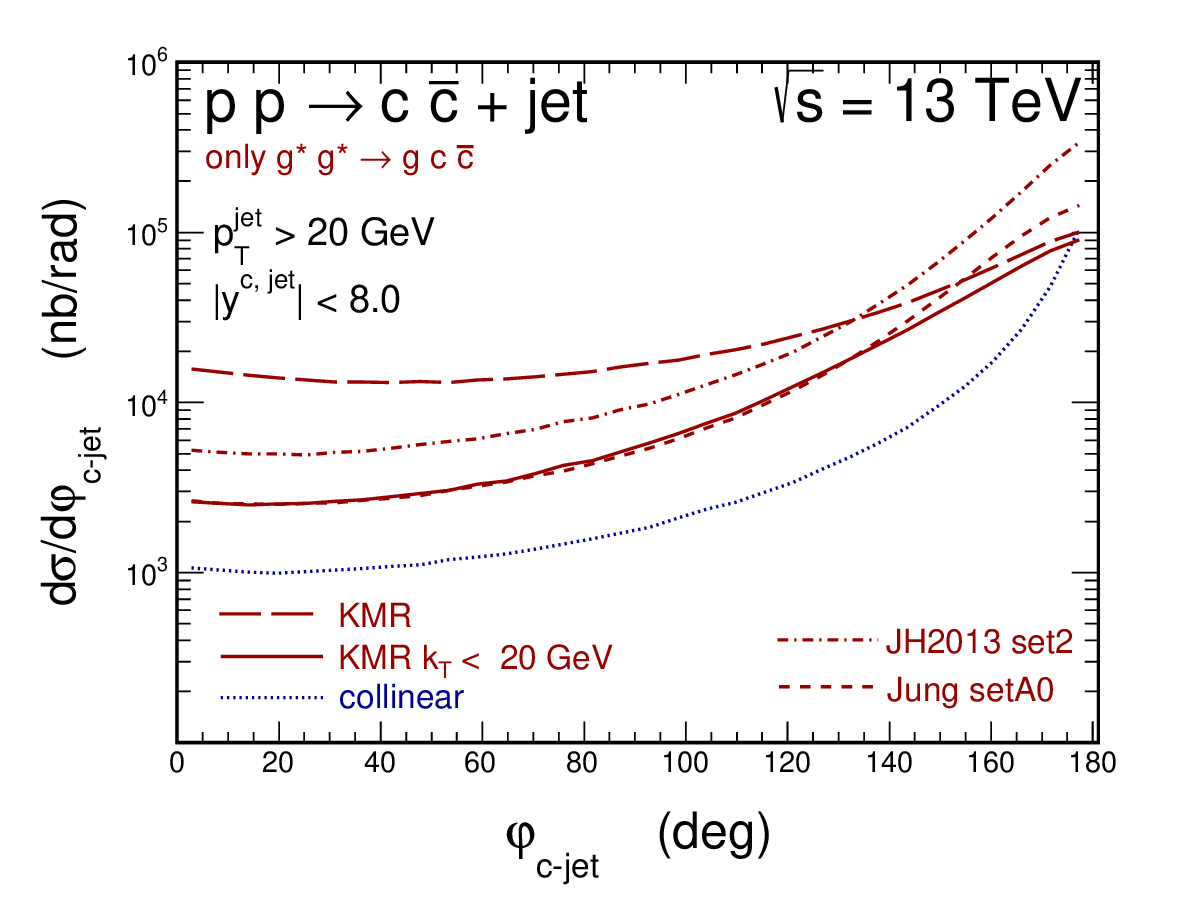}}
\end{minipage}
\hspace{0.5cm}
\begin{minipage}{0.47\textwidth}
 \centerline{\includegraphics[width=1.0\textwidth]{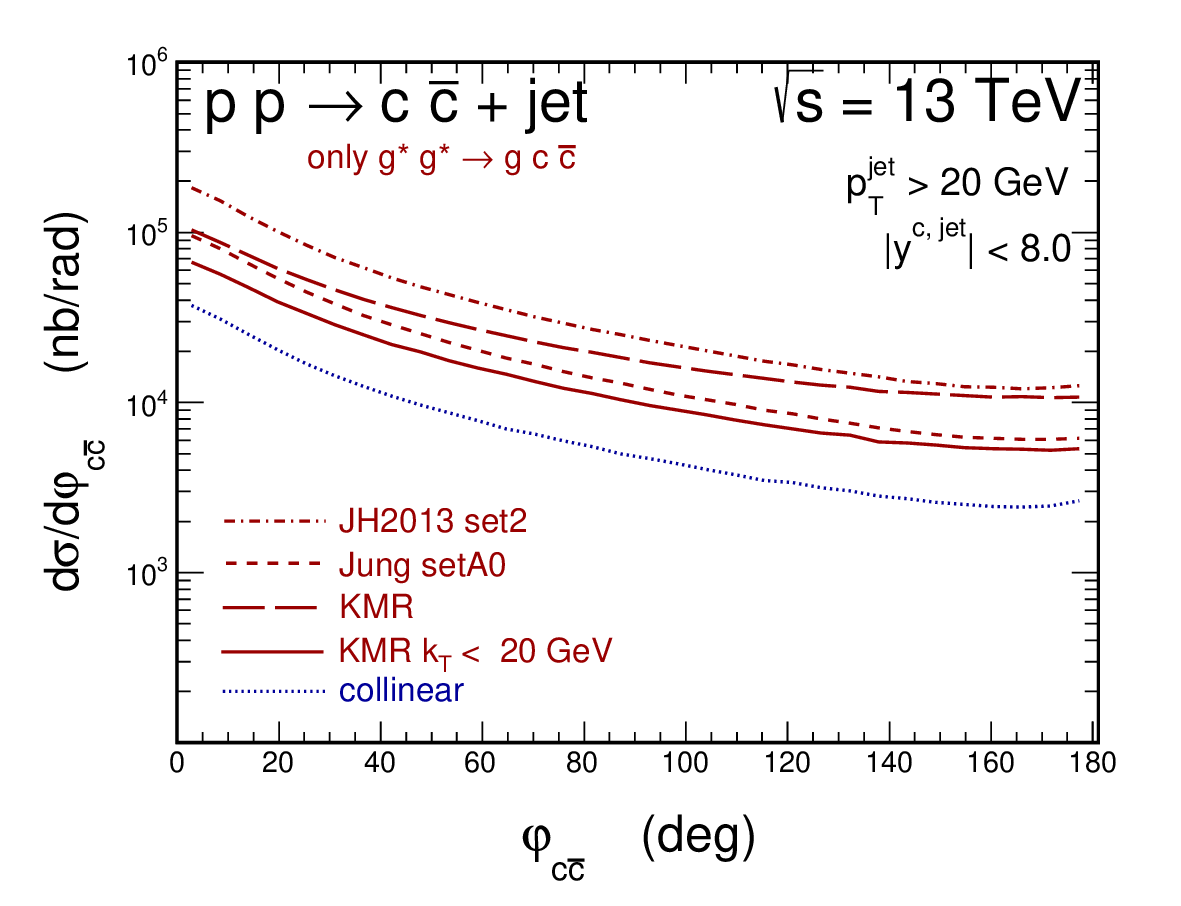}}
\end{minipage}
   \caption{
\small  Distribution in azimuthal angle between c-quark and jet (left
panel) and between $c$-quark and ${\bar c}$-antiquark (right panel)
in the $k_T$-factorization approach with different uGDFs.
 }
 \label{fig:kt_ccbar_phi}
\end{figure}

In this subsection we show similar results but in the
$k_T$-factorization approach (see Figs.~\ref{fig:kt_ccbar_pt}, \ref{fig:kt_ccbar_y}, \ref{fig:kt_ccbar_M} and \ref{fig:kt_ccbar_phi}).
Here we include only the dominant $g g$-initiated processes.
We show results for a few different unintegrated gluon distribution
functions (uGDFs) from the literature.
In the case of inclusive production of $c \bar c$ usually the 
Kimber-Martin-Ryskin (KMR) \cite{Kimber:2001sc,Watt:2003vf} distribution gives the best description of
the experimental data \cite{Maciula:2013wg}. For consistency, we use MMHT2014 collinear PDFs in calculation of the KMR transverse momentum dependent distributions. In the case of the KMR uGDF we show also
result when limiting transverse momenta of initial gluons (this will be discussed below).
The comparison of corresponding results show that the contribution
of large initial gluon transverse momenta (virtualities) is rather
large. For the KMR distribution it is very probable that
more than one jet (two or three) are produced.
A rigorous eliminating of such cases in the $k_T$-factorization
approach is not an easy and obvious task. We think that the cut on initial gluon
transverse momenta $k_{T} < p_{T,min}^{jet}$ takes this into account in an approximate way.
This procedure is reliable only for the KMR uGDFs when the range of (pseudo)rapidity coverage is large (as for ATLAS or CMS).
For other uGDFs it strongly depends on their construction.
For comparison, we also use here the JH2013 set2 \cite{Hautmann:2013tba} and the Jung setA0 \cite{Jung:2004gs} CCFM-based uGDFs. 

Compared to the LO collinear factorization approach no plateau at $p_{T}^{c} < 20$ GeV can be observed for the $k_{T}$-factorization approach (see left panel in Fig.~\ref{fig:kt_ccbar_pt}). 

%-----------------------------------------------------------------------------
%------------------------------------------------------------------------------

%-----------------------------------------------------------------------------
%------------------------------------------------------------------------------

%-----------------------------------------------------------------------------
%------------------------------------------------------------------------------

%-----------------------------------------------------------------------------
%------------------------------------------------------------------------------

\begin{figure}[!h]
\begin{minipage}{0.3\textwidth}
 \centerline{\includegraphics[width=1.0\textwidth]{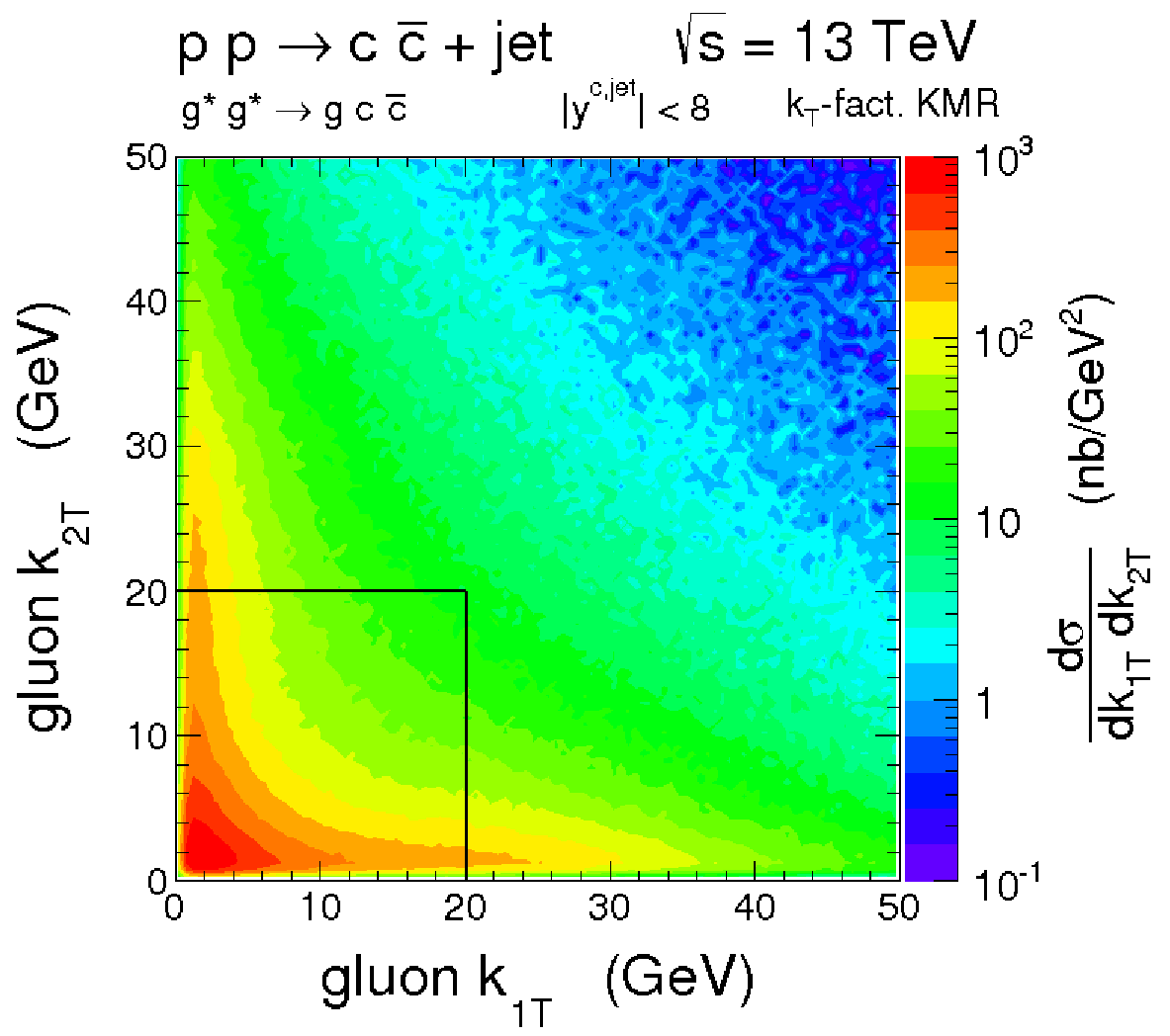}}
\end{minipage}
\hspace{0.5cm}
\begin{minipage}{0.3\textwidth}
 \centerline{\includegraphics[width=1.0\textwidth]{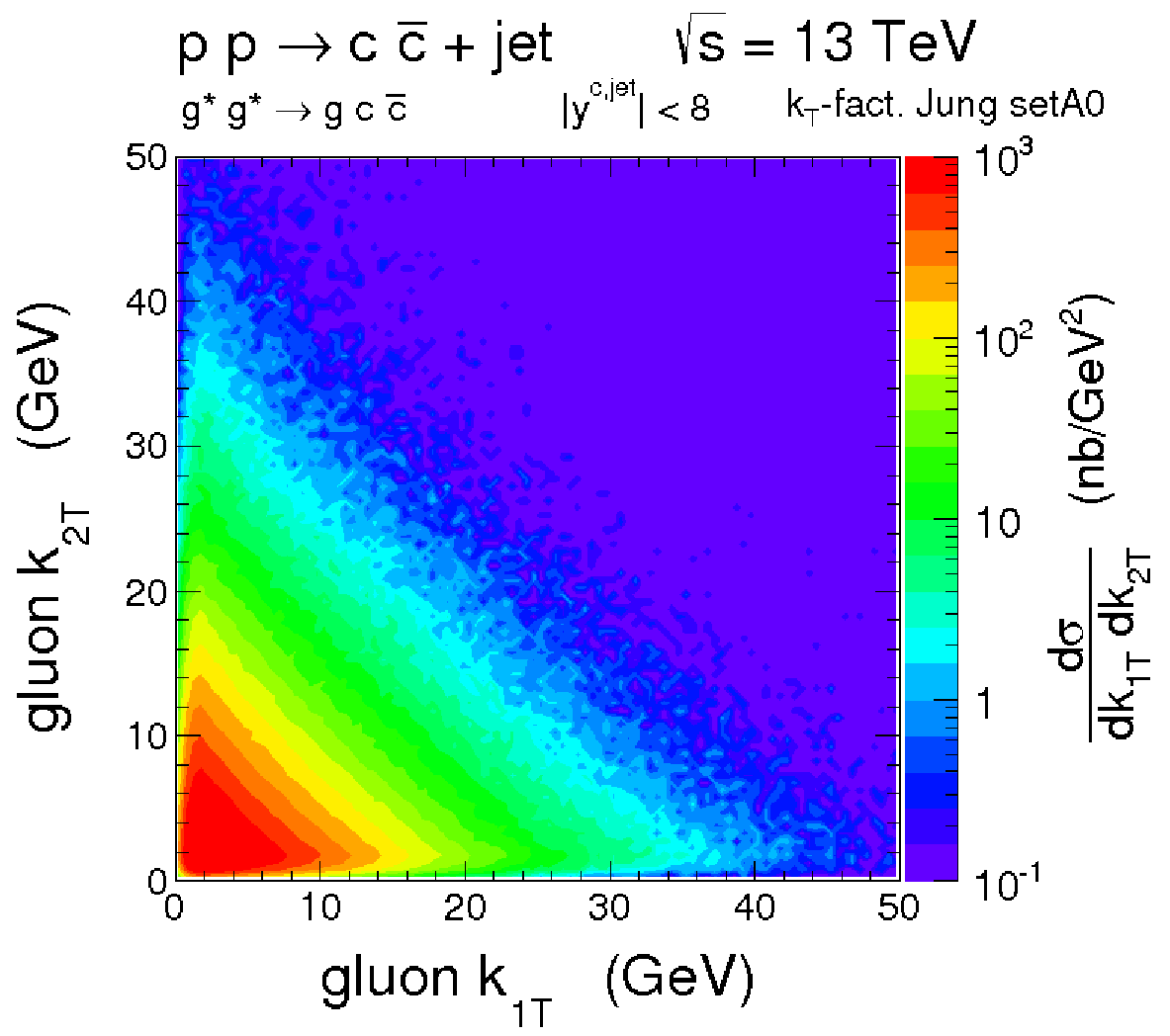}}
\end{minipage}
\begin{minipage}{0.3\textwidth}
 \centerline{\includegraphics[width=1.0\textwidth]{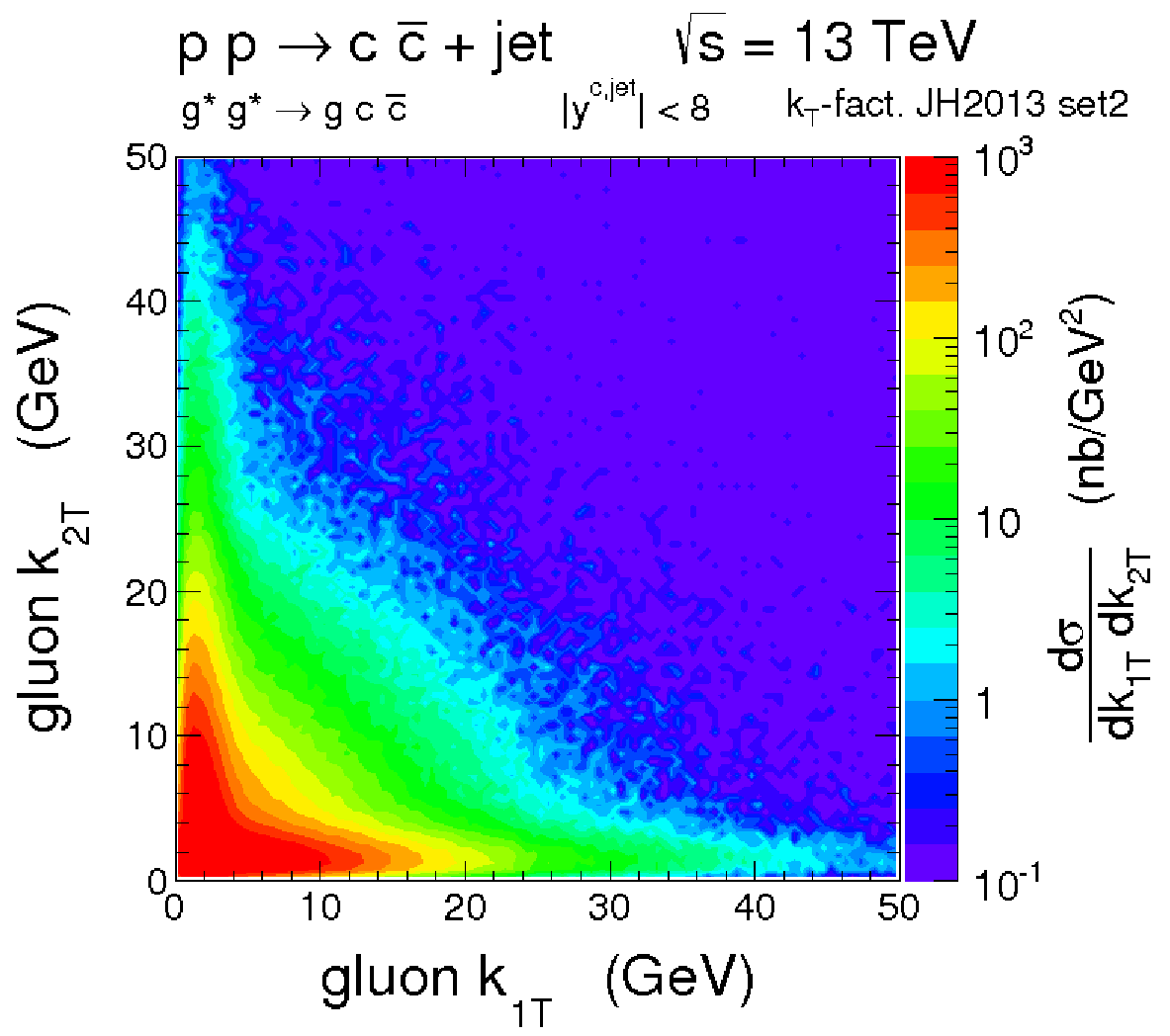}}
\end{minipage}
   \caption{
\small Two-dimensional distributions in transverse momenta of initial
gluons for three different uGDFs specified in the figure caption.
 }
 \label{fig:q1tq2t}
\end{figure}

In Fig.~\ref{fig:q1tq2t} we show two-dimensional distributions
in transverse momenta of both initial gluons.
Once again we observe that large transverse momenta (virtualities) of the initial
gluons are involved into production of our final state.

%-----------------------------------------------------------------------------
%------------------------------------------------------------------------------

\begin{figure}[!h]
\begin{minipage}{0.47\textwidth}
 \centerline{\includegraphics[width=1.0\textwidth]{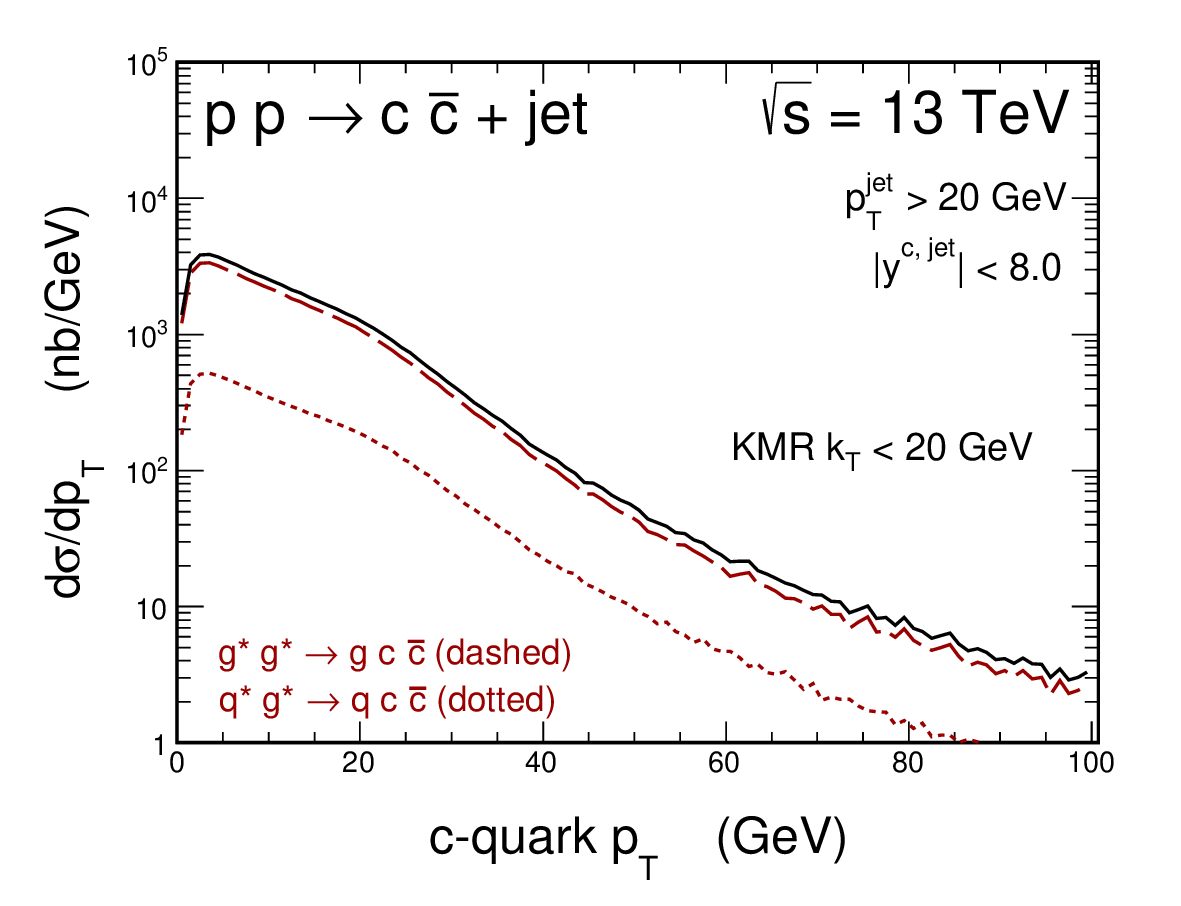}}
\end{minipage}
\hspace{0.5cm}
\begin{minipage}{0.47\textwidth}
 \centerline{\includegraphics[width=1.0\textwidth]{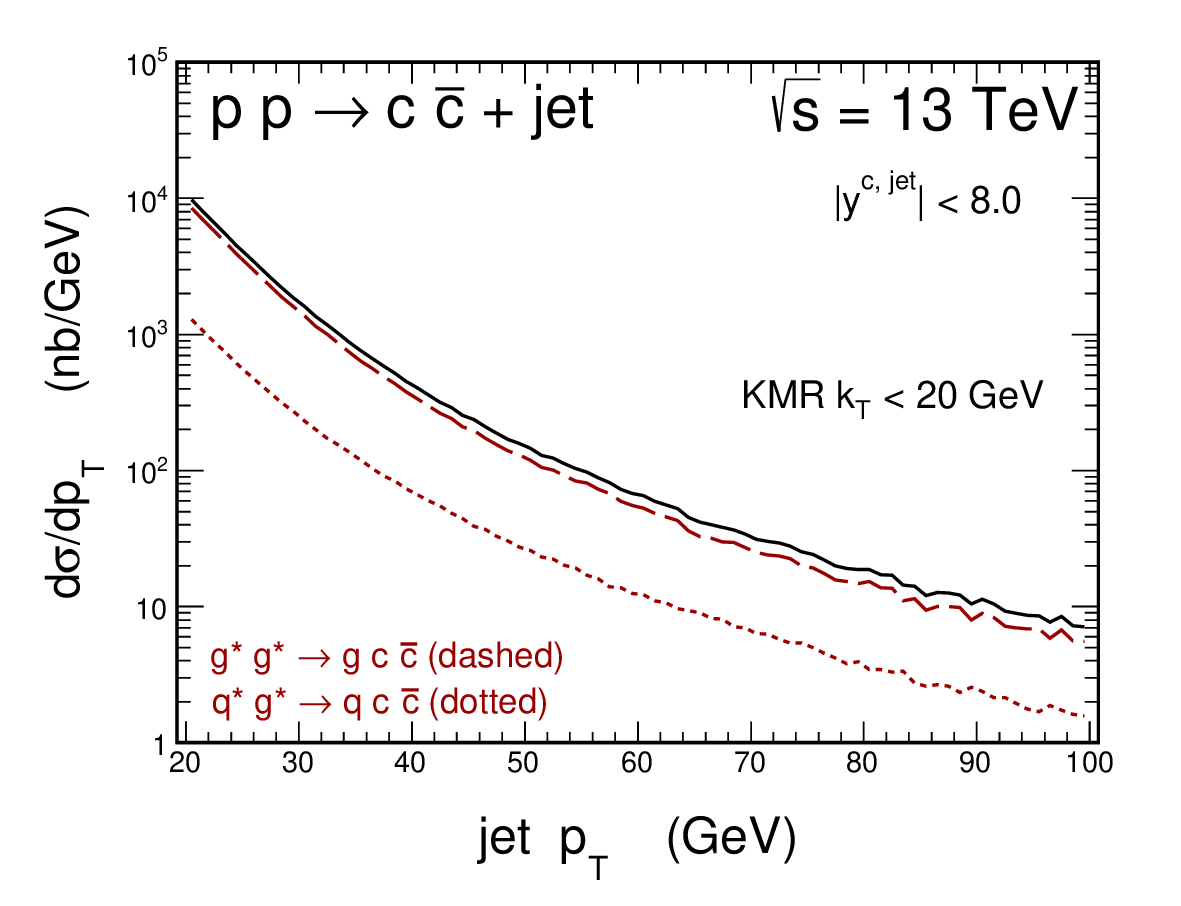}}
\end{minipage}
   \caption{
\small Transverse momentum distribution of c-quark (let panel) and
associated jet (right panel) in the $k_T$-factorization approach
with the KMR uGDF and the extra cut on initial gluon transverse momenta.
 }
 \label{fig:ktKMR_ccbar_pt}
\end{figure}

\begin{figure}[!h]
\begin{minipage}{0.47\textwidth}
 \centerline{\includegraphics[width=1.0\textwidth]{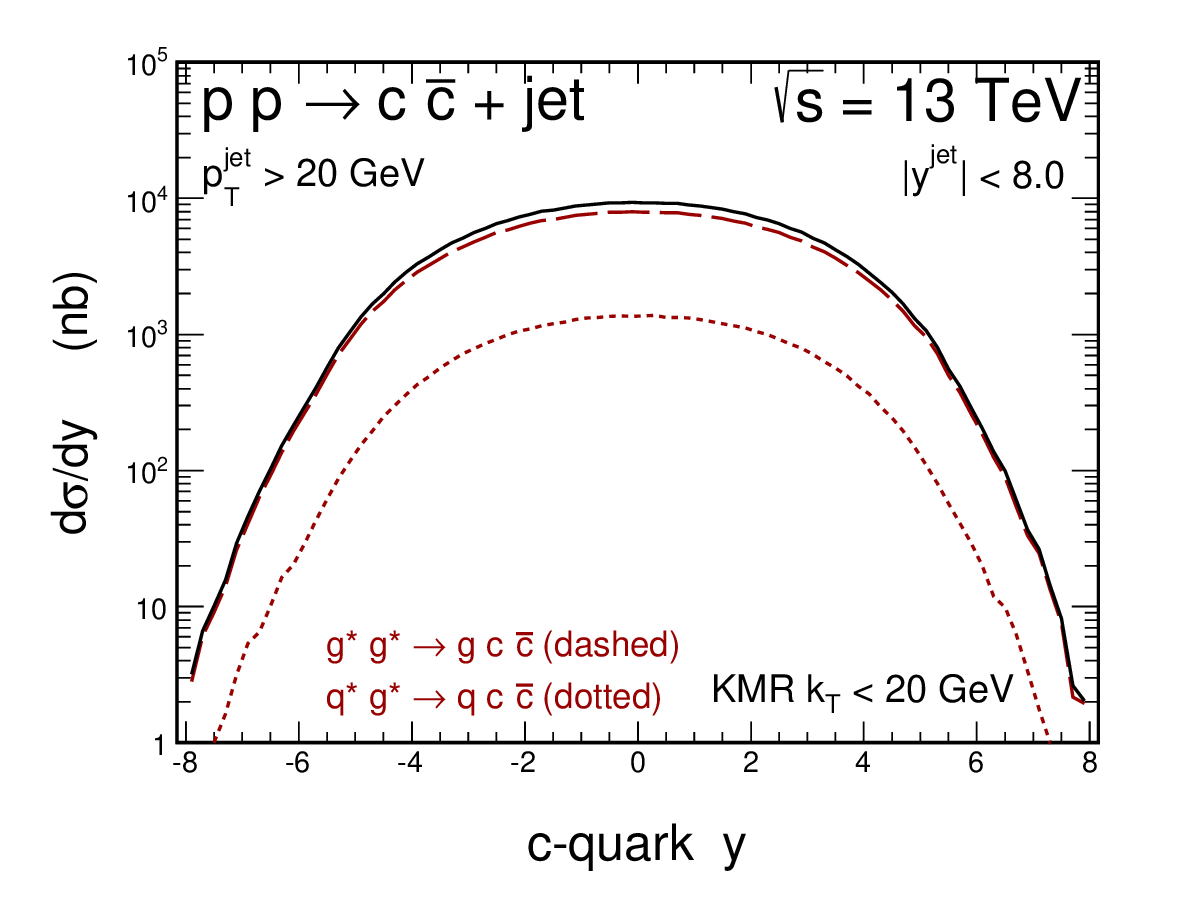}}
\end{minipage}
\hspace{0.5cm}
\begin{minipage}{0.47\textwidth}
 \centerline{\includegraphics[width=1.0\textwidth]{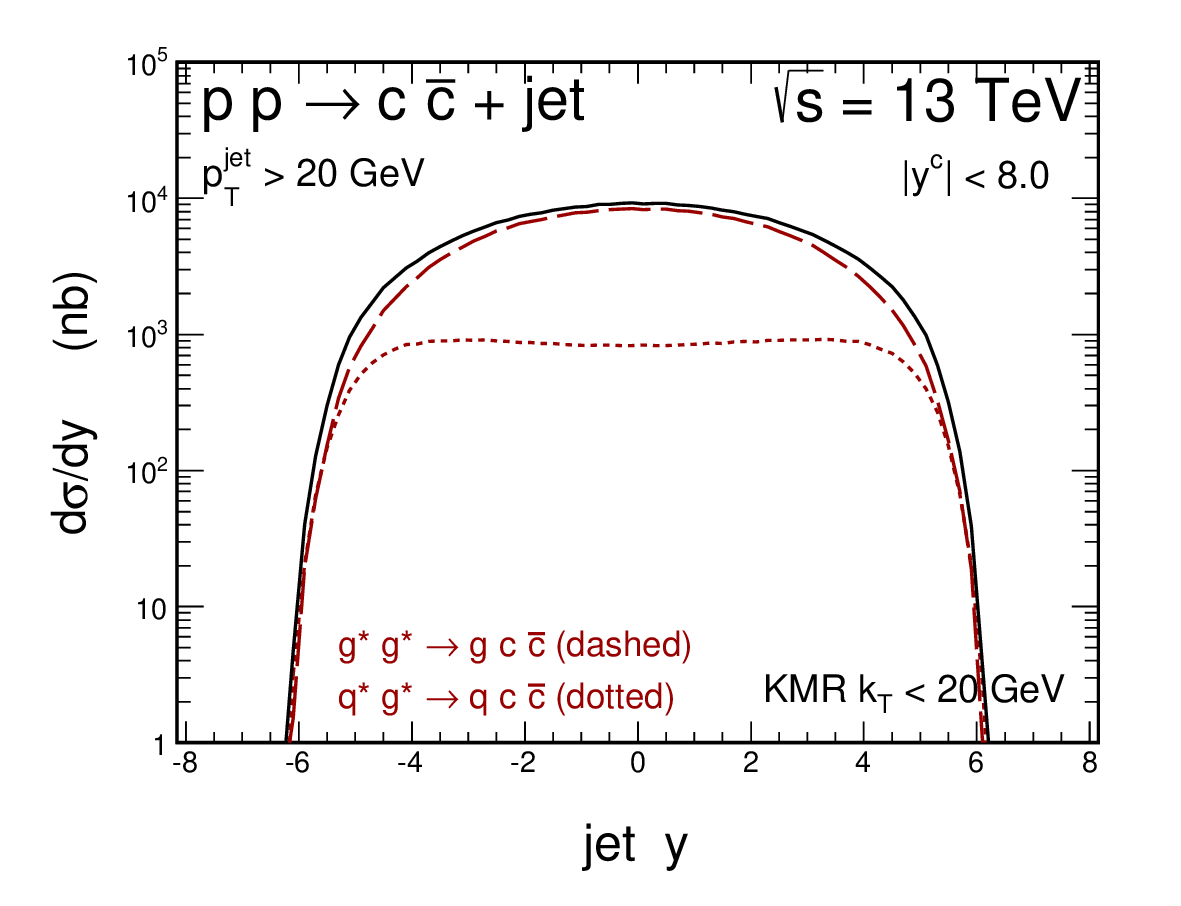}}
\end{minipage}
   \caption{
\small  Rapidity distribution of c-quark (left panel) and
associated jet (right panel) in the $k_T$-factorization approach
with the KMR UGDF and the extra cut on initial gluon transverse momenta.
 }
 \label{fig:ktKMR_ccbar_y}
\end{figure}

\begin{figure}[!h]
\begin{minipage}{0.47\textwidth}
 \centerline{\includegraphics[width=1.0\textwidth]{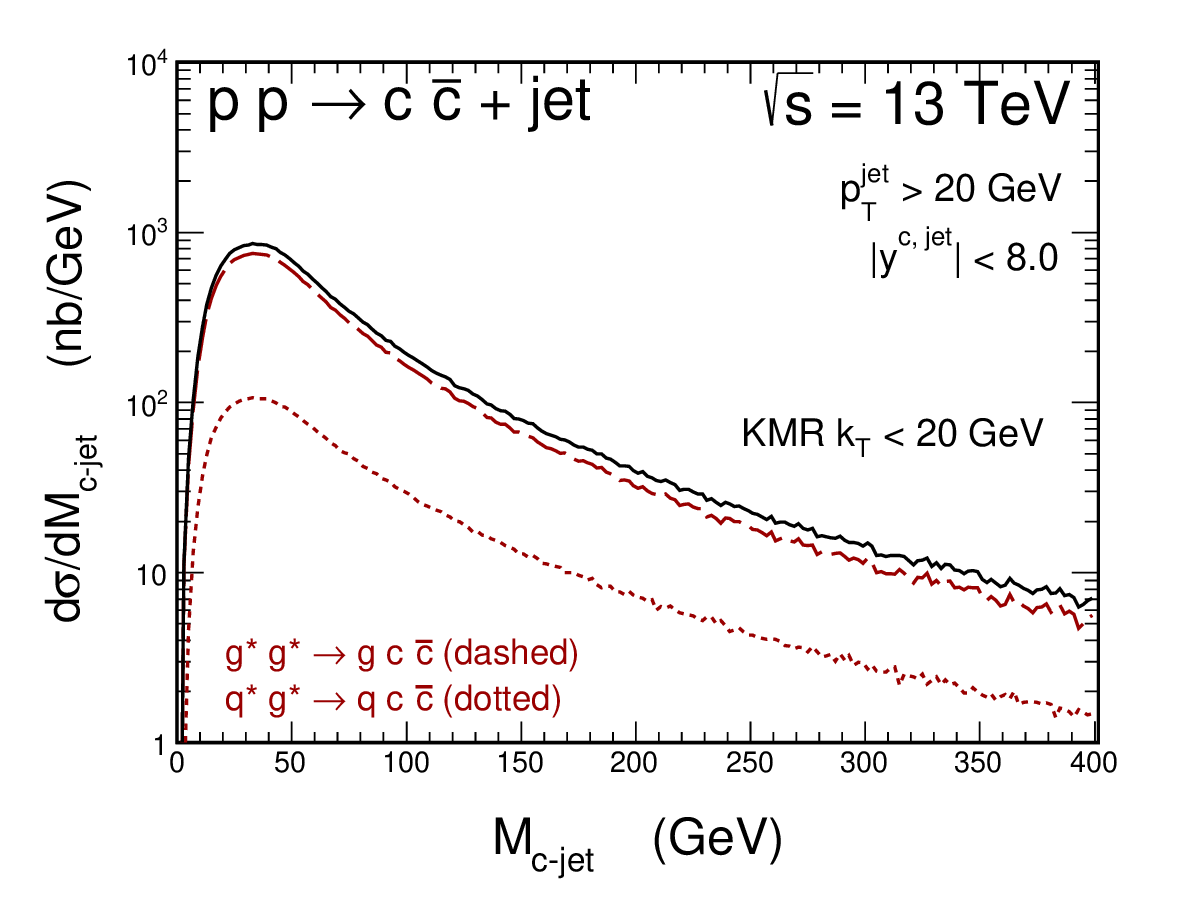}}
\end{minipage}
\hspace{0.5cm}
\begin{minipage}{0.47\textwidth}
 \centerline{\includegraphics[width=1.0\textwidth]{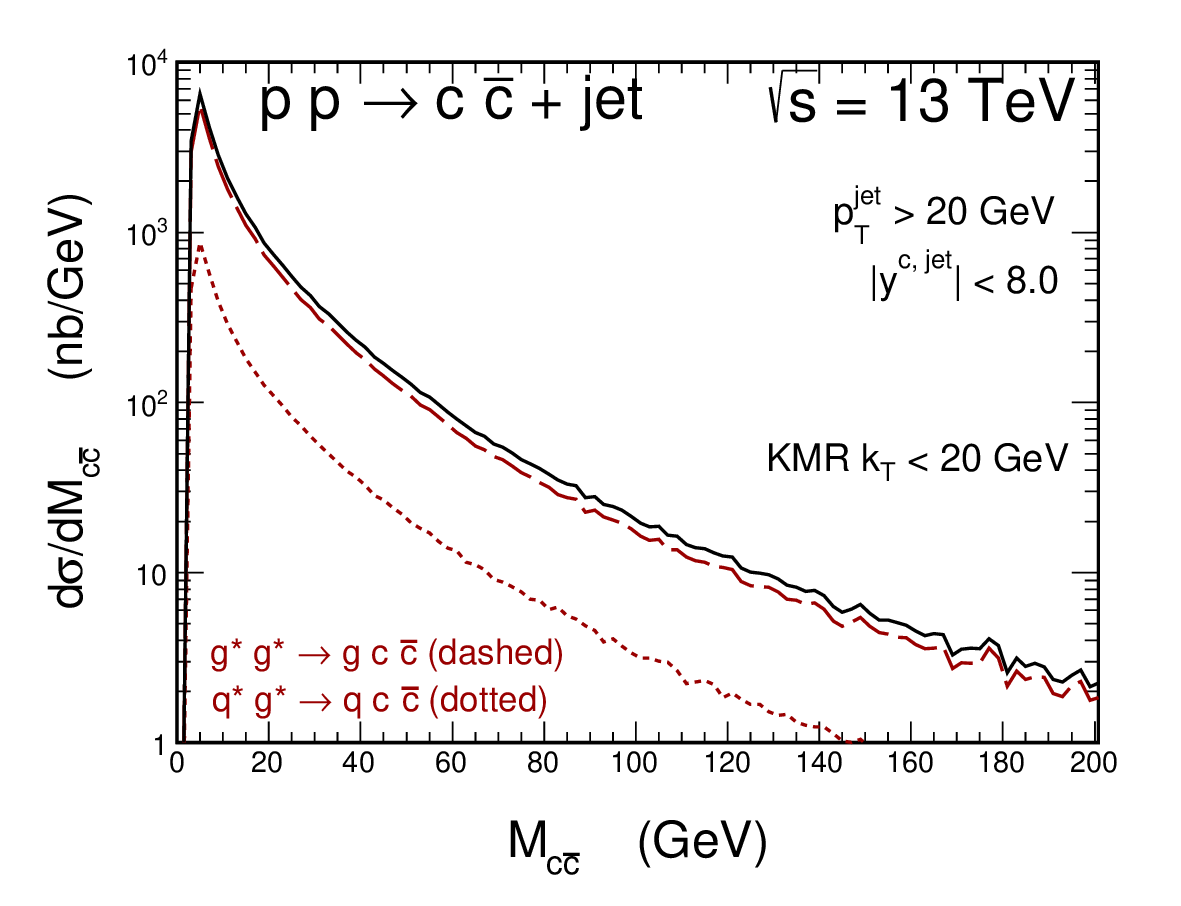}}
\end{minipage}
   \caption{
\small  Distribution in invariant mass of the c-quark-jet (left panel)
       and $c \bar c$ system (right panel) in the $k_T$-factorization
       approach with the KMR uGDF and the extra cut on initial gluon transverse momenta.
 }
 \label{fig:ktKMR_ccbar_M}
\end{figure}

\begin{figure}[!h]
\begin{minipage}{0.47\textwidth}
 \centerline{\includegraphics[width=1.0\textwidth]{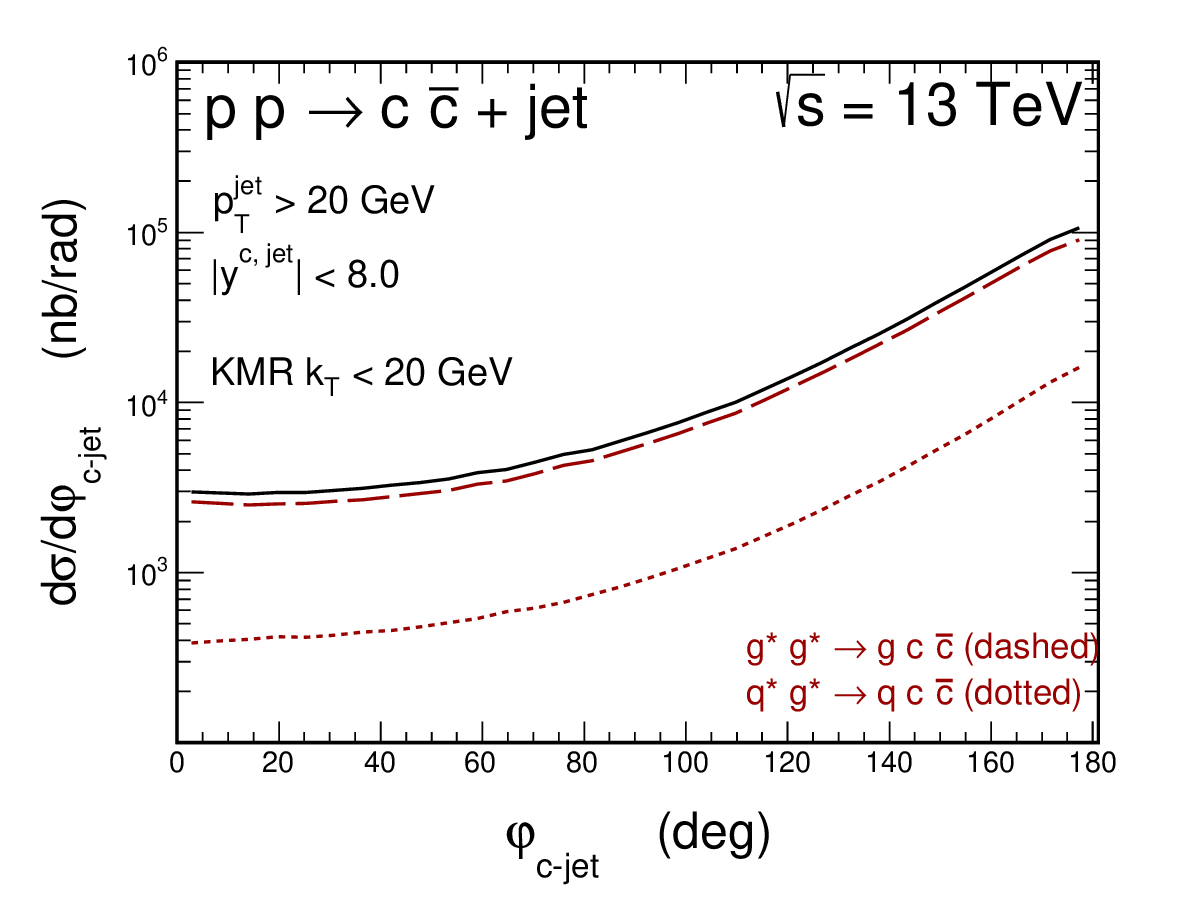}}
\end{minipage}
\hspace{0.5cm}
\begin{minipage}{0.47\textwidth}
 \centerline{\includegraphics[width=1.0\textwidth]{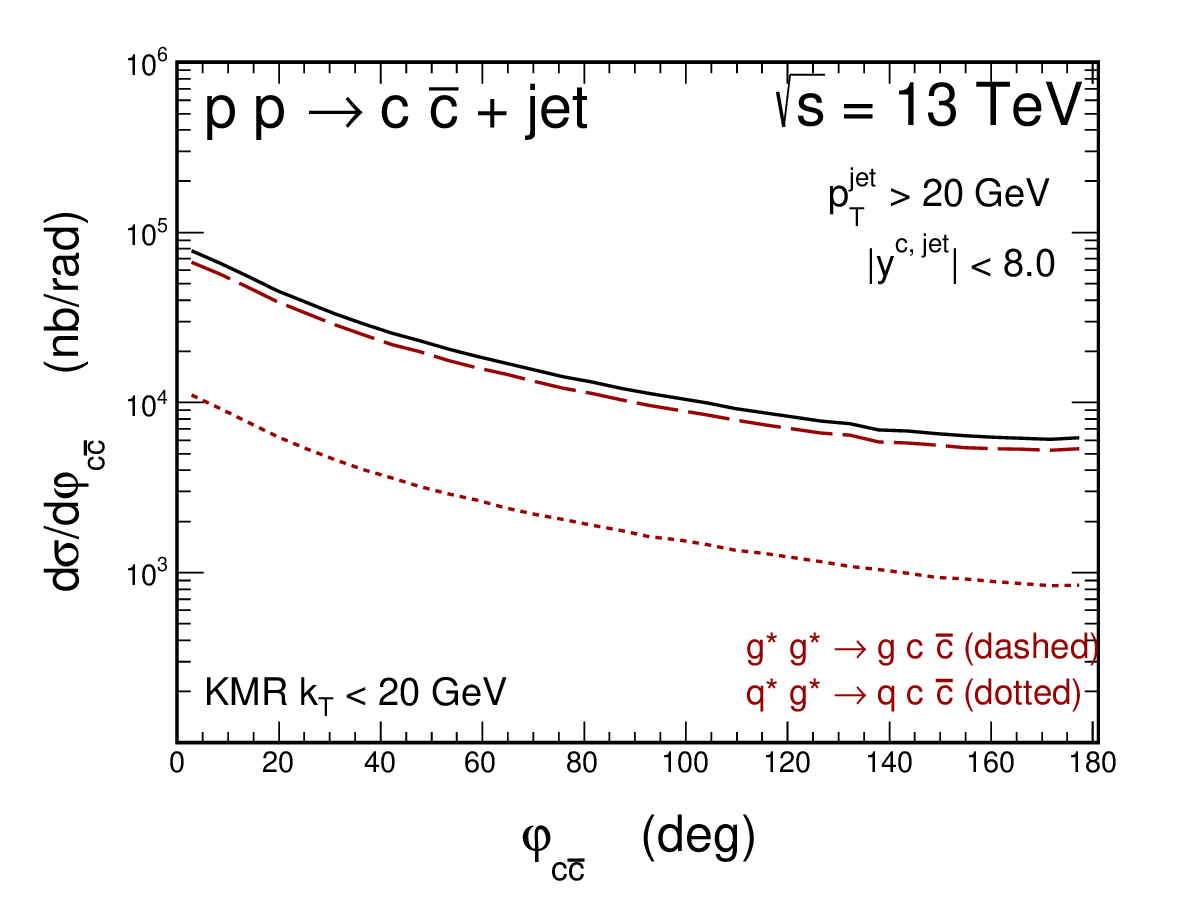}}
\end{minipage}
   \caption{
\small Distribution in azimuthal angle between c-quark and jet (left
panel) and between $c$-quark and ${\bar c}$-antiquark (right panel)
in the $k_T$-factorization approach with the KMR uGDF
and the extra cut on initial gluon transverse momenta.
 }
 \label{fig:ktKMR_ccbar_phi}
\end{figure}

The KMR method allows to construct not only uGDFs but also
unintegrated quark/antiquark distributions. Therefore we can compare
the contributions of the different subprocesses as it was done
in the previous subsection for the collinear case. 
In Figs.~\ref{fig:ktKMR_ccbar_pt}, \ref{fig:ktKMR_ccbar_y}, 
\ref{fig:ktKMR_ccbar_M} and \ref{fig:ktKMR_ccbar_phi}
we again show distributions in $c$/$\bar c$ transverse momentum, rapidity,
diparton invariant masses as well as in relative azimuthal angle
between different outgoing partons.
The results obtained here for the $k_T$-factorization approach are 
very similar to those obtained in the collinear approach.

%-----------------------------------------------------------------------------
%------------------------------------------------------------------------------

%-----------------------------------------------------------------------------
%------------------------------------------------------------------------------

%-----------------------------------------------------------------------------
%------------------------------------------------------------------------------

%-----------------------------------------------------------------------------
%------------------------------------------------------------------------------

%-------------------------------------------------------------
\subsection{Comparison of the two approaches}
%-------------------------------------------------------------

\begin{figure}[!h]
\begin{minipage}{0.47\textwidth}
 \centerline{\includegraphics[width=1.0\textwidth]{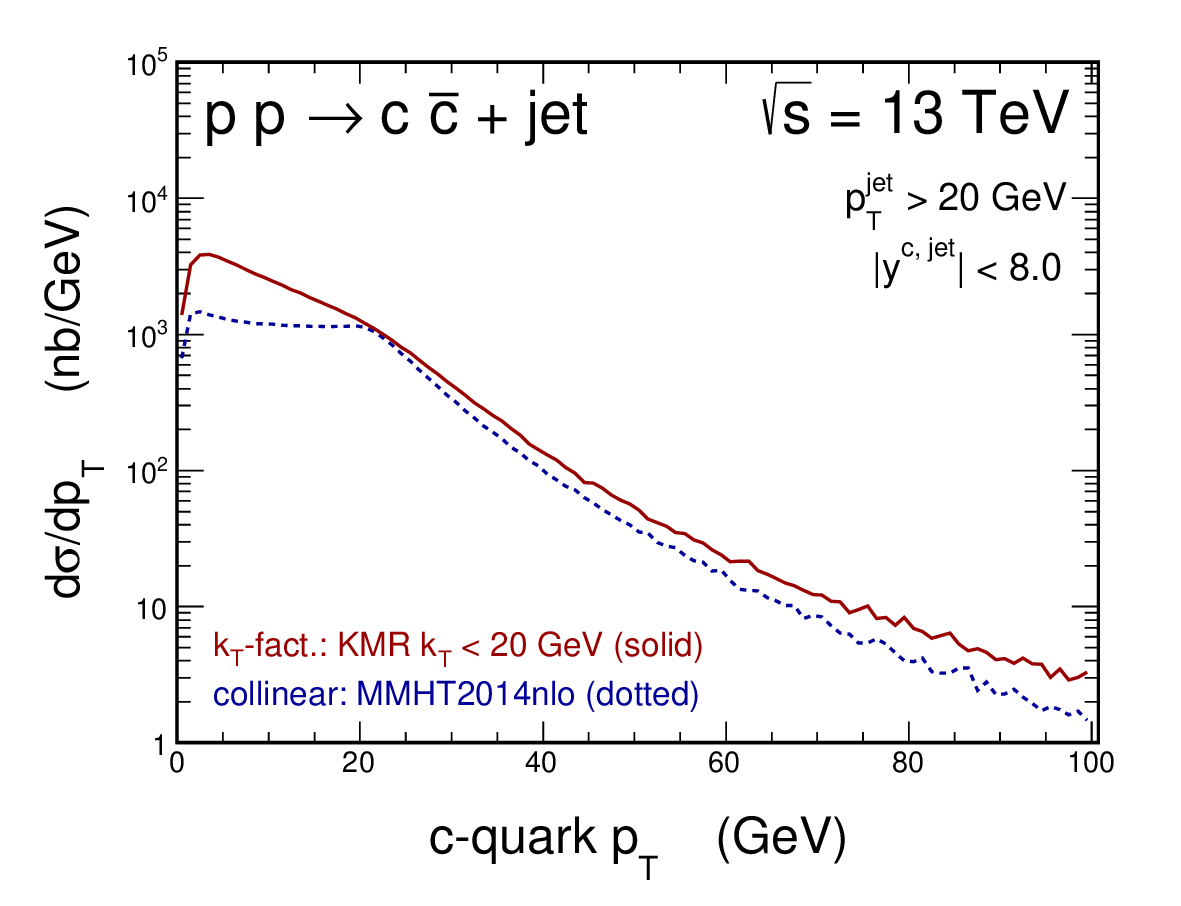}}
\end{minipage}
\hspace{0.5cm}
\begin{minipage}{0.47\textwidth}
 \centerline{\includegraphics[width=1.0\textwidth]{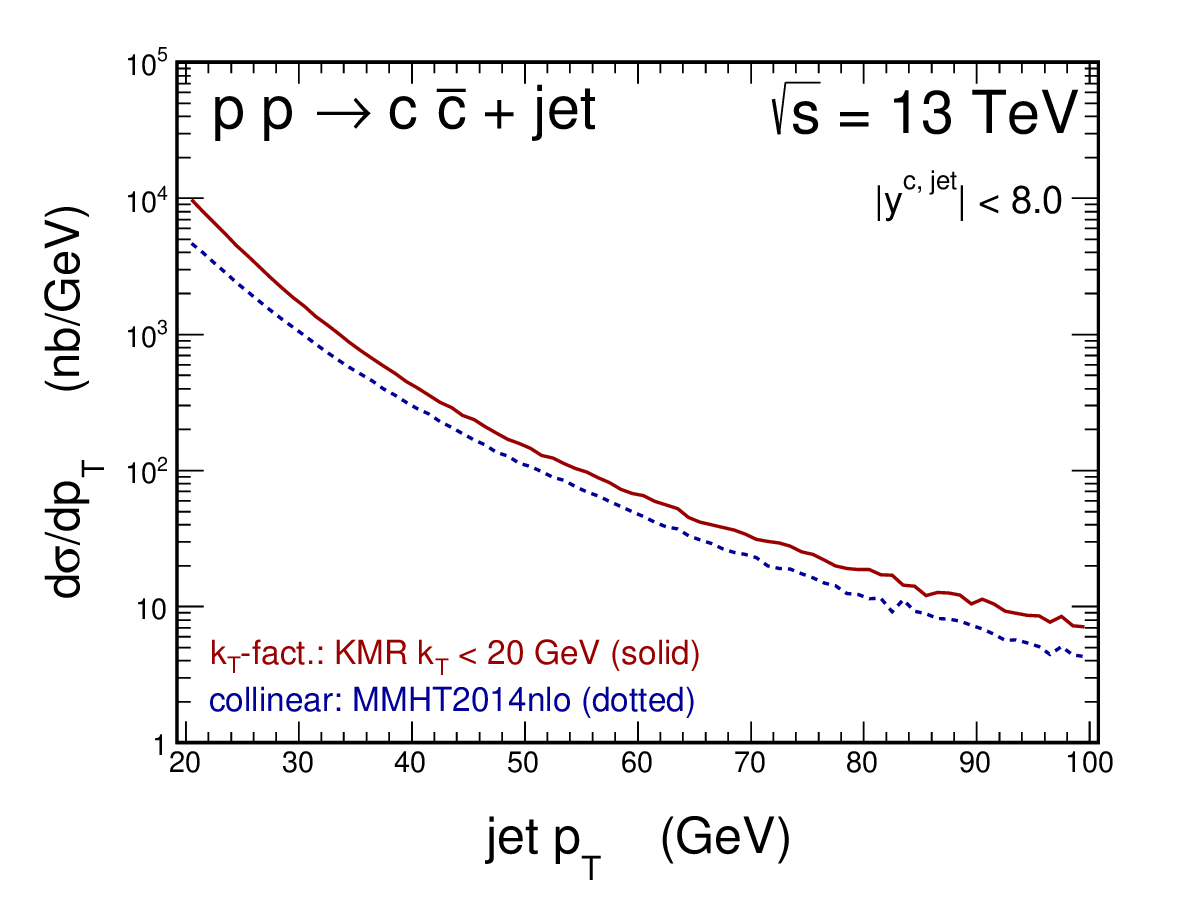}}
\end{minipage}
   \caption{ 
\small Comparison of transverse momentum distributions of c-quark (let panel) and
associated jet (right panel) for the collinear approach (dotted) and the
$k_T$-factorization approach
with the KMR uGDF and the extra cut on initial gluon transverse momenta (solid). 
 }
 \label{fig:compo_ccbar_pt}
\end{figure}

\begin{figure}[!h]
\begin{minipage}{0.47\textwidth}
 \centerline{\includegraphics[width=1.0\textwidth]{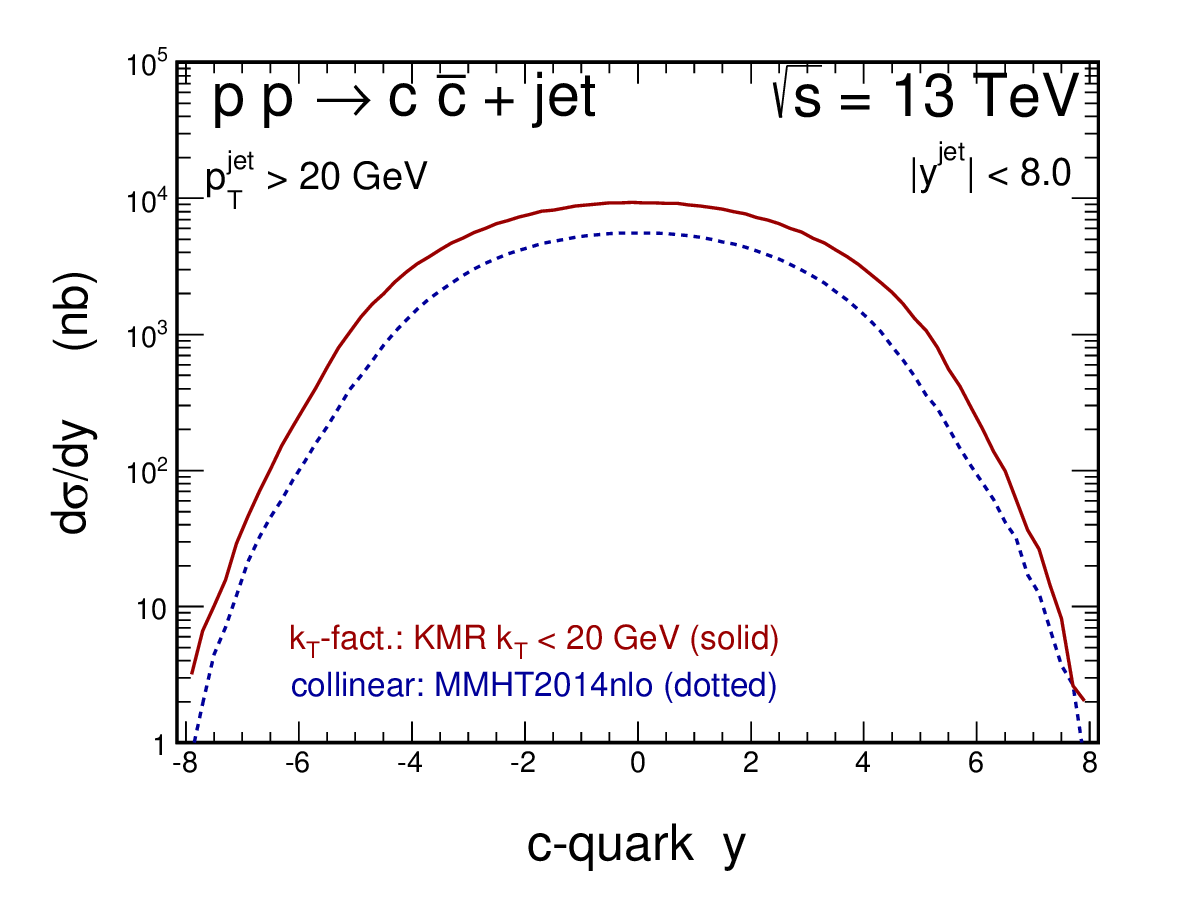}}
\end{minipage}
\hspace{0.5cm}
\begin{minipage}{0.47\textwidth}
 \centerline{\includegraphics[width=1.0\textwidth]{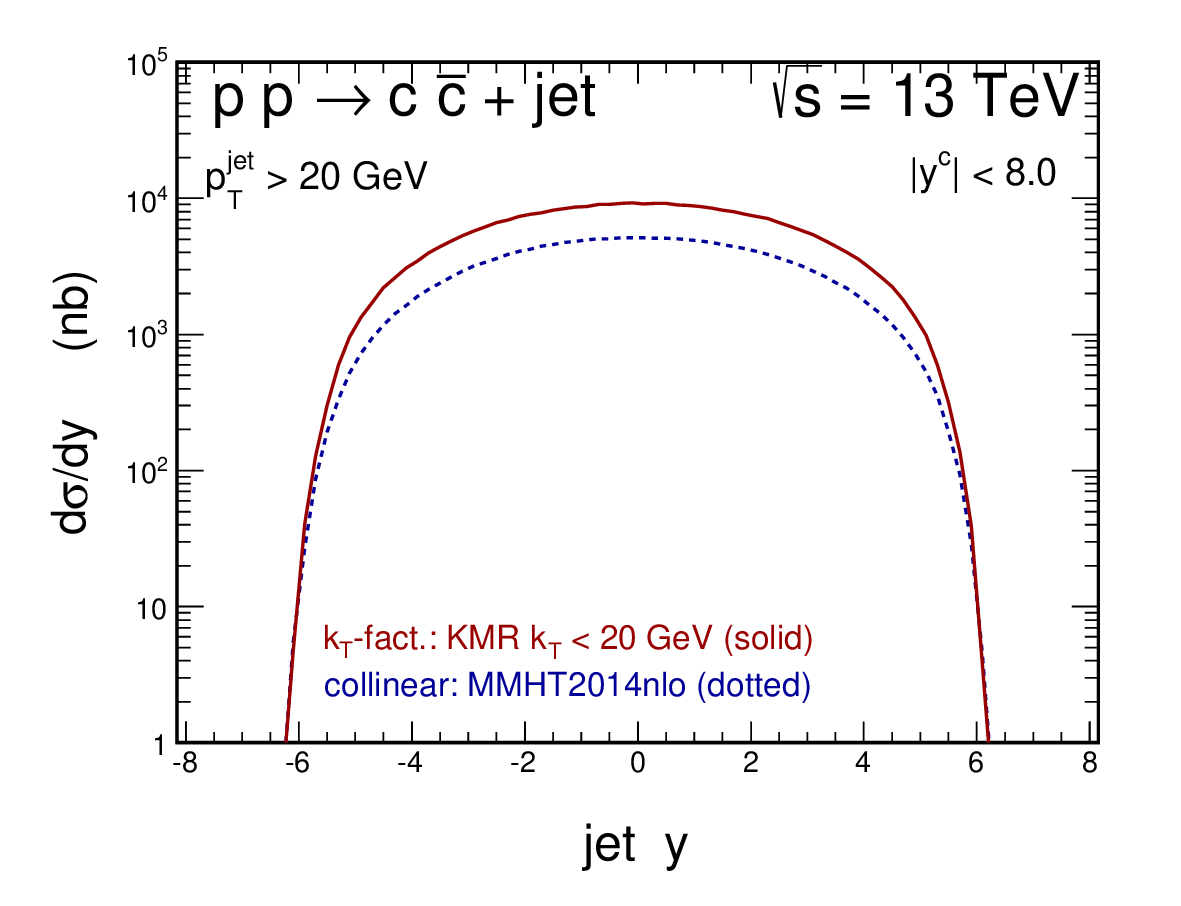}}
\end{minipage}
   \caption{
\small  Comparison of rapidity distribution of c-quark (left panel) and
associated jet (right panel) in the collinear approach (dotted) and the
$k_T$-factorization approach
with the KMR uGDF and the extra cut on initial gluon transverse momenta (solid).}
 \label{fig:compo_ccbar_y}
\end{figure}

\begin{figure}[!h]
\begin{minipage}{0.47\textwidth}
 \centerline{\includegraphics[width=1.0\textwidth]{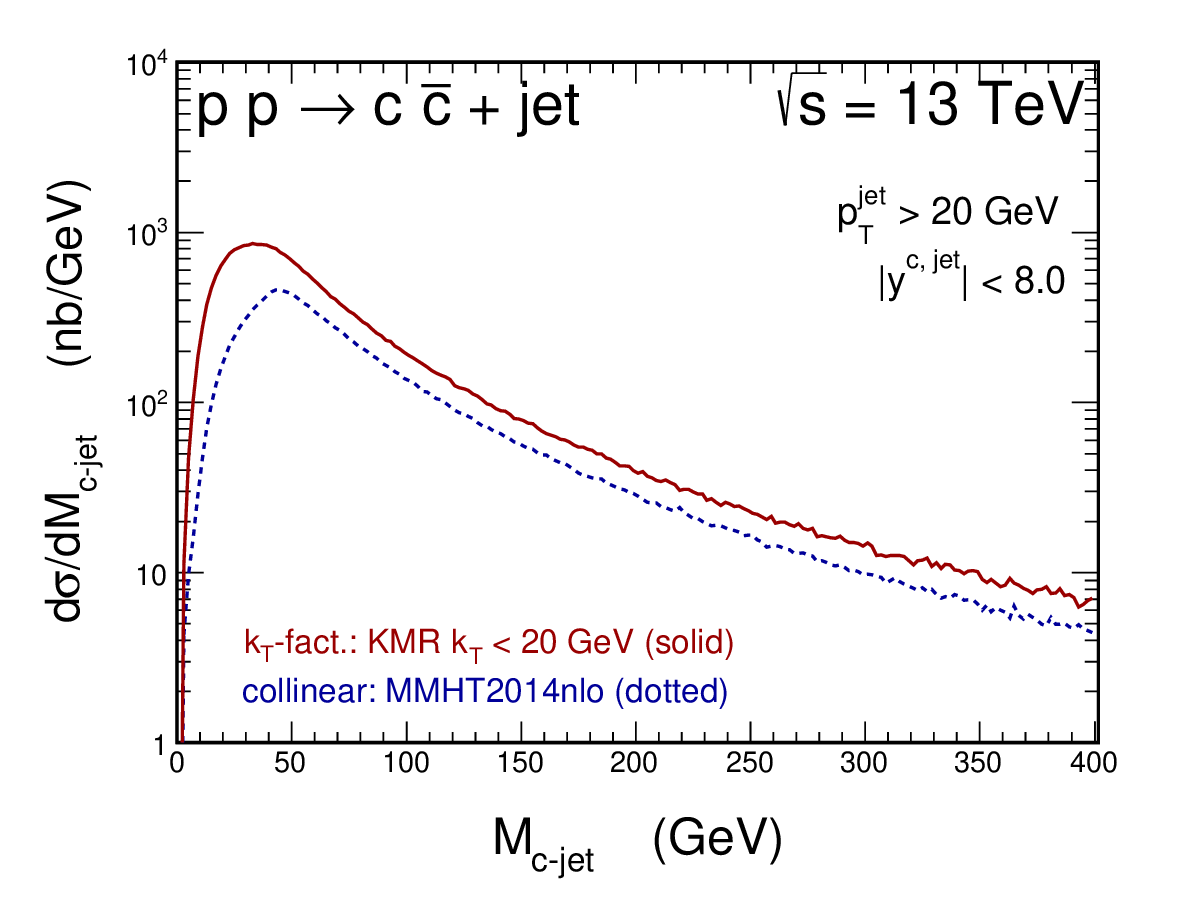}}
\end{minipage}
\hspace{0.5cm}
\begin{minipage}{0.47\textwidth}
 \centerline{\includegraphics[width=1.0\textwidth]{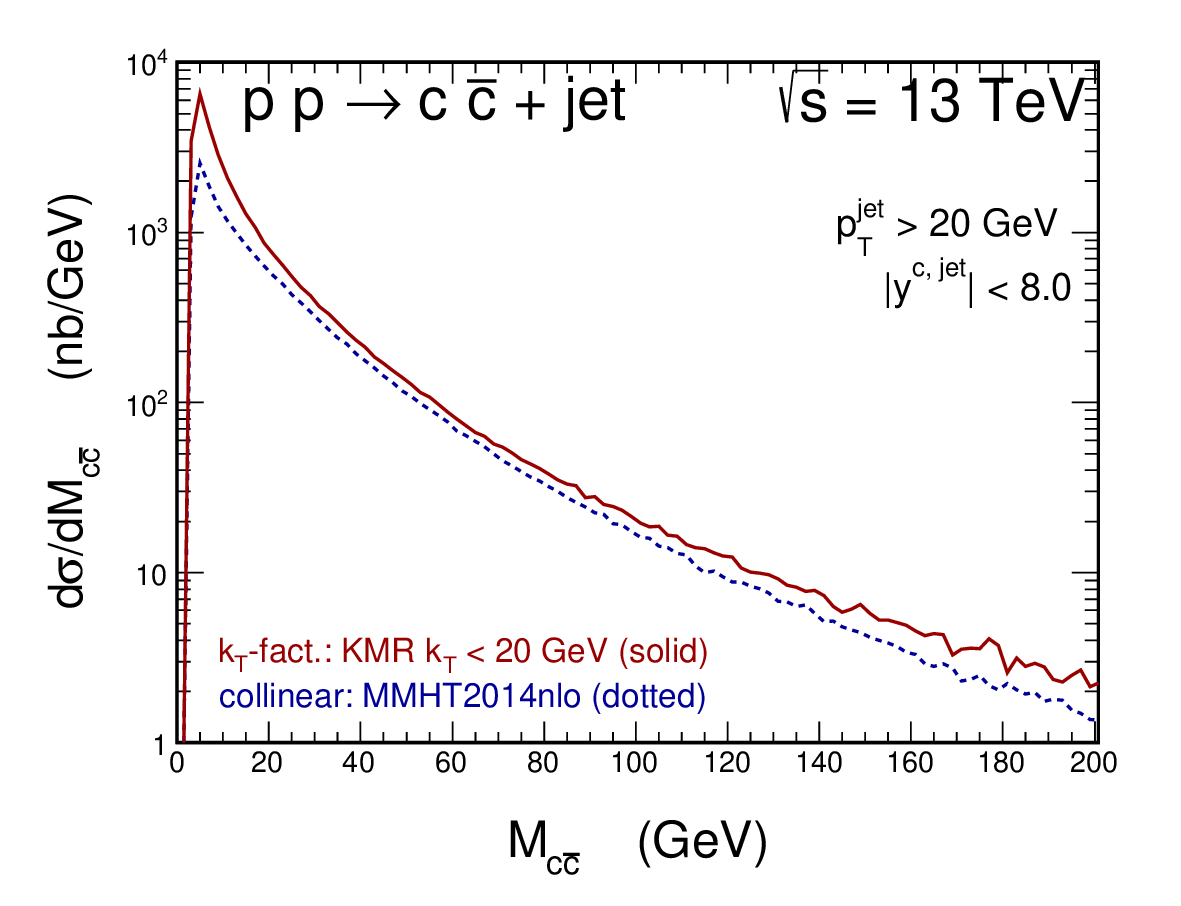}}
\end{minipage}
   \caption{ 
\small Comparison of distribution in invariant mass of the c-quark-jet (left panel)
       and $c \bar c$ system (right panel) in the collinear approach
       (dotted) and in the $k_T$-factorization
       approach with the KMR uGDF and the extra cut on initial gluon
       transverse momenta (solid).}
 \label{fig:compo_ccbar_M}
\end{figure}

\begin{figure}[!h]
\begin{minipage}{0.47\textwidth}
 \centerline{\includegraphics[width=1.0\textwidth]{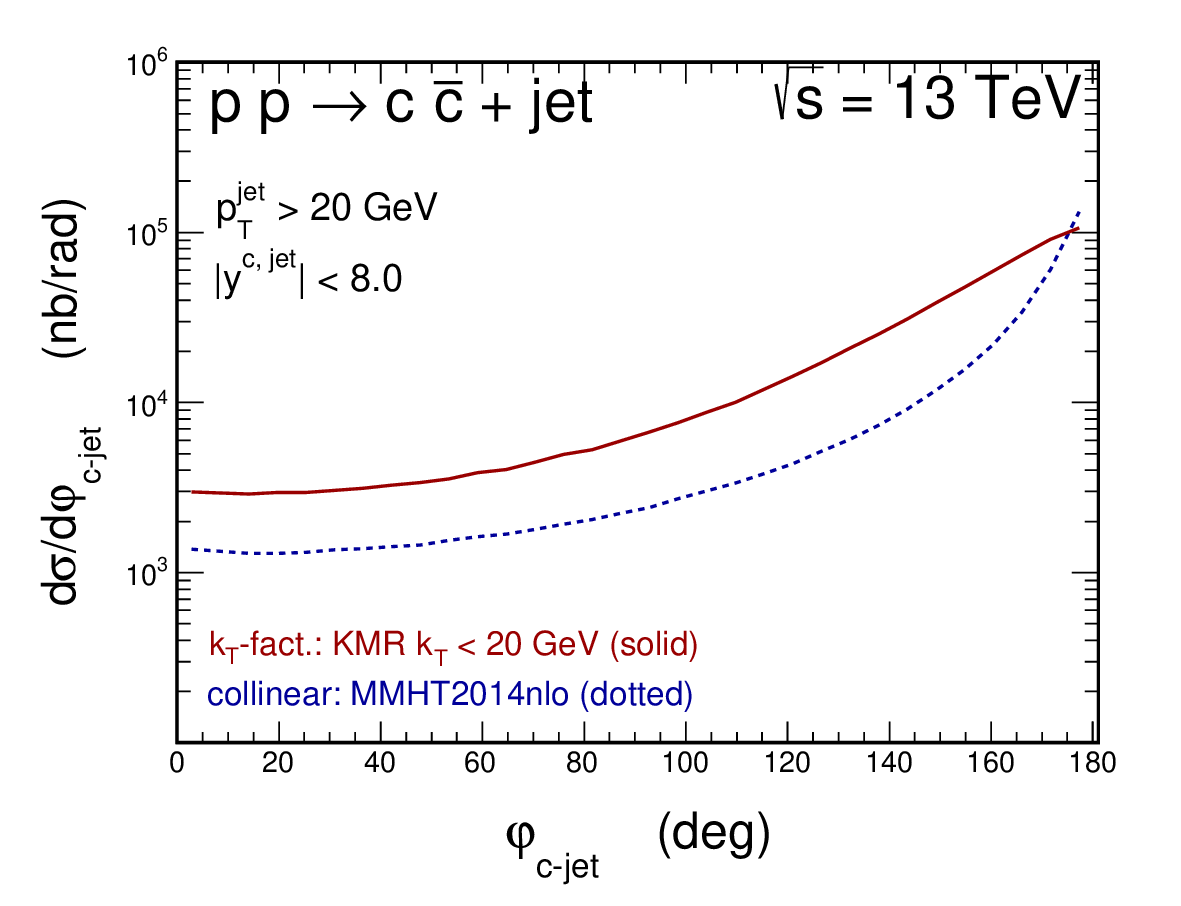}}
\end{minipage}
\hspace{0.5cm}
\begin{minipage}{0.47\textwidth}
 \centerline{\includegraphics[width=1.0\textwidth]{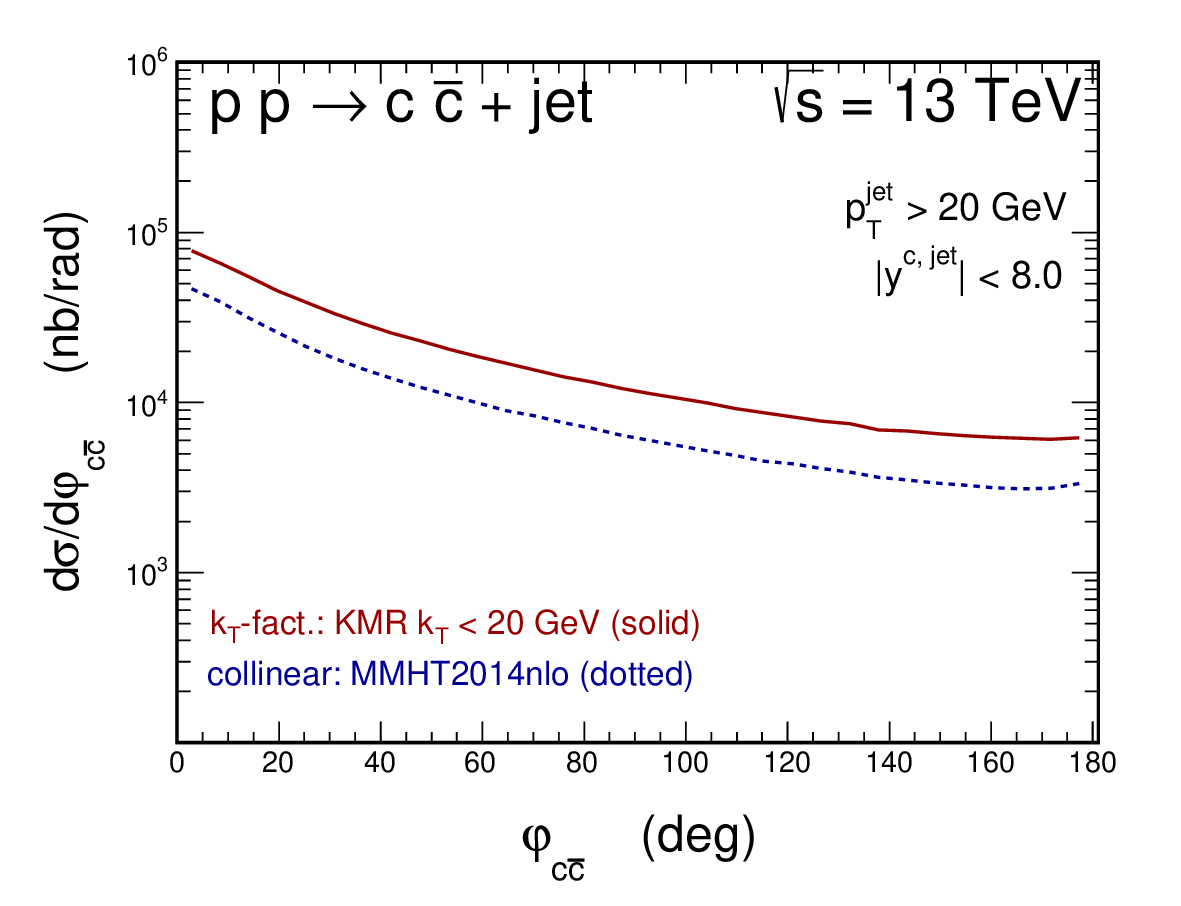}}
\end{minipage}
   \caption{  
\small Comparison of distribution in azimuthal angle between c-quark and jet (left
panel) and between $c$-quark and ${\bar c}$-antiquark (right panel)
in the collinear approach (dotted) and in the $k_T$-factorization approach with the KMR uGDF
and the extra cut on initial gluon transverse momenta (solid).}
 \label{fig:compo_ccbar_phi}
\end{figure}

In Figs.~\ref{fig:compo_ccbar_pt}, \ref{fig:compo_ccbar_y}, \ref{fig:compo_ccbar_M} and \ref{fig:compo_ccbar_phi}
we again show distributions in $c$/$\bar c$ transverse momentum, rapidity,
diparton invariant masses as well as in relative azimuthal angle
between different outgoing partons simultaneously for the collinear and $k_T$-factorization
approaches with the KMR uGDF and extra cut to effectively eliminate cases 
of more than one jet in the final state. 
In general, the results are rather similar.
The $k_T$-factorization approach gives slightly larger cross section.
However, the shapes of differential distributions are rather similar.
Dividing differential distributions for the $k_T$-factorization by corresponding ones
for the collinear approach one could get phase-space-point dependent $K$-factor.
Rough inspection of the figures shows that the $K$-factor is only weakly
dependent on kinematical variables. Approximately it is about $K \sim$ 1.5,
which is a typical value for NLO pQCD calculations for processes 
with initial gluons.

%-----------------------------------------------------------------------------
%------------------------------------------------------------------------------

%-----------------------------------------------------------------------------
%------------------------------------------------------------------------------

%-----------------------------------------------------------------------------
%------------------------------------------------------------------------------

%-----------------------------------------------------------------------------
%------------------------------------------------------------------------------

\begin{figure}[!h]
\begin{minipage}{0.33\textwidth}
 \centerline{\includegraphics[width=1.0\textwidth]{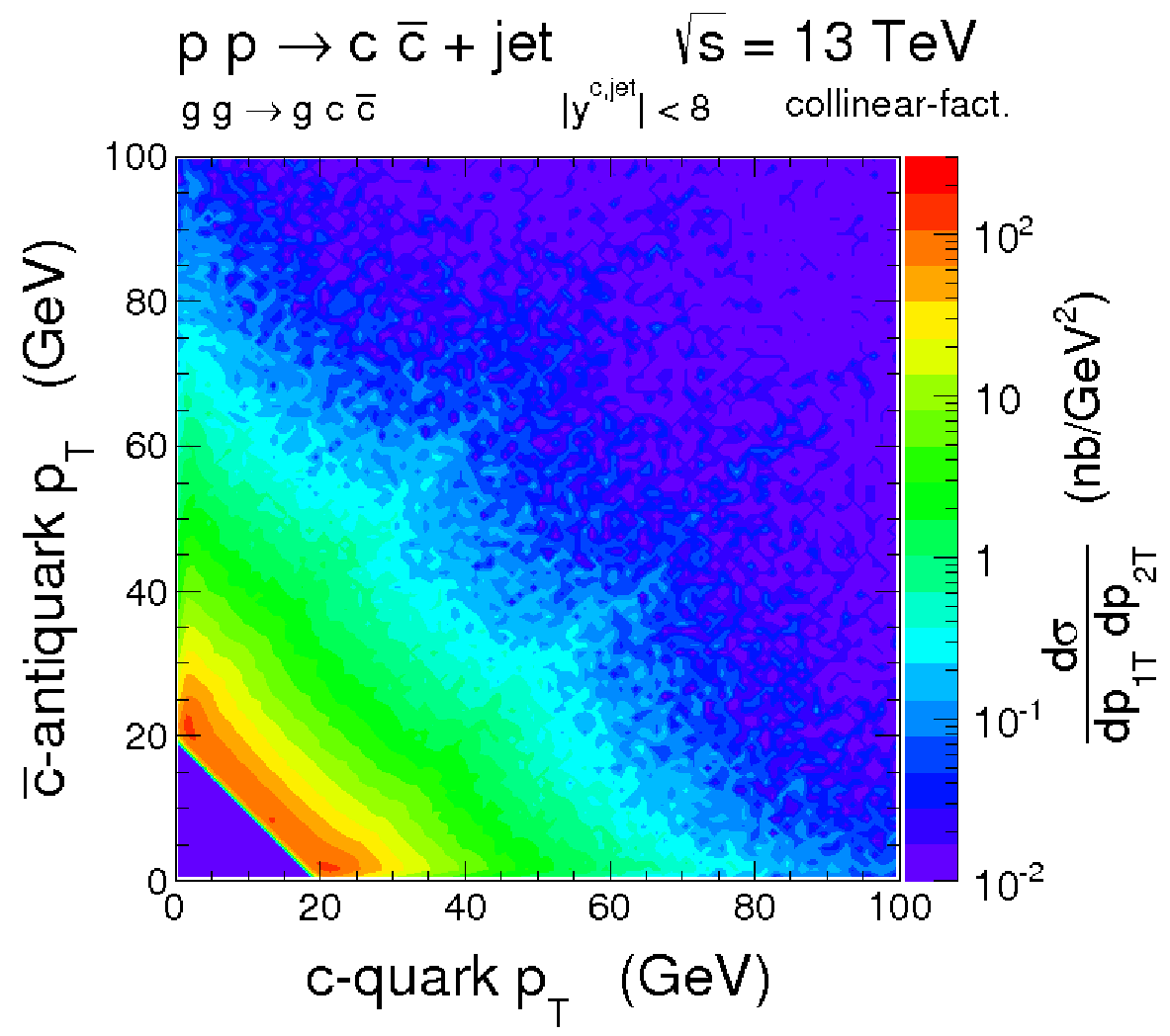}}
\end{minipage}
\hspace{0.5cm}
\begin{minipage}{0.33\textwidth}
 \centerline{\includegraphics[width=1.0\textwidth]{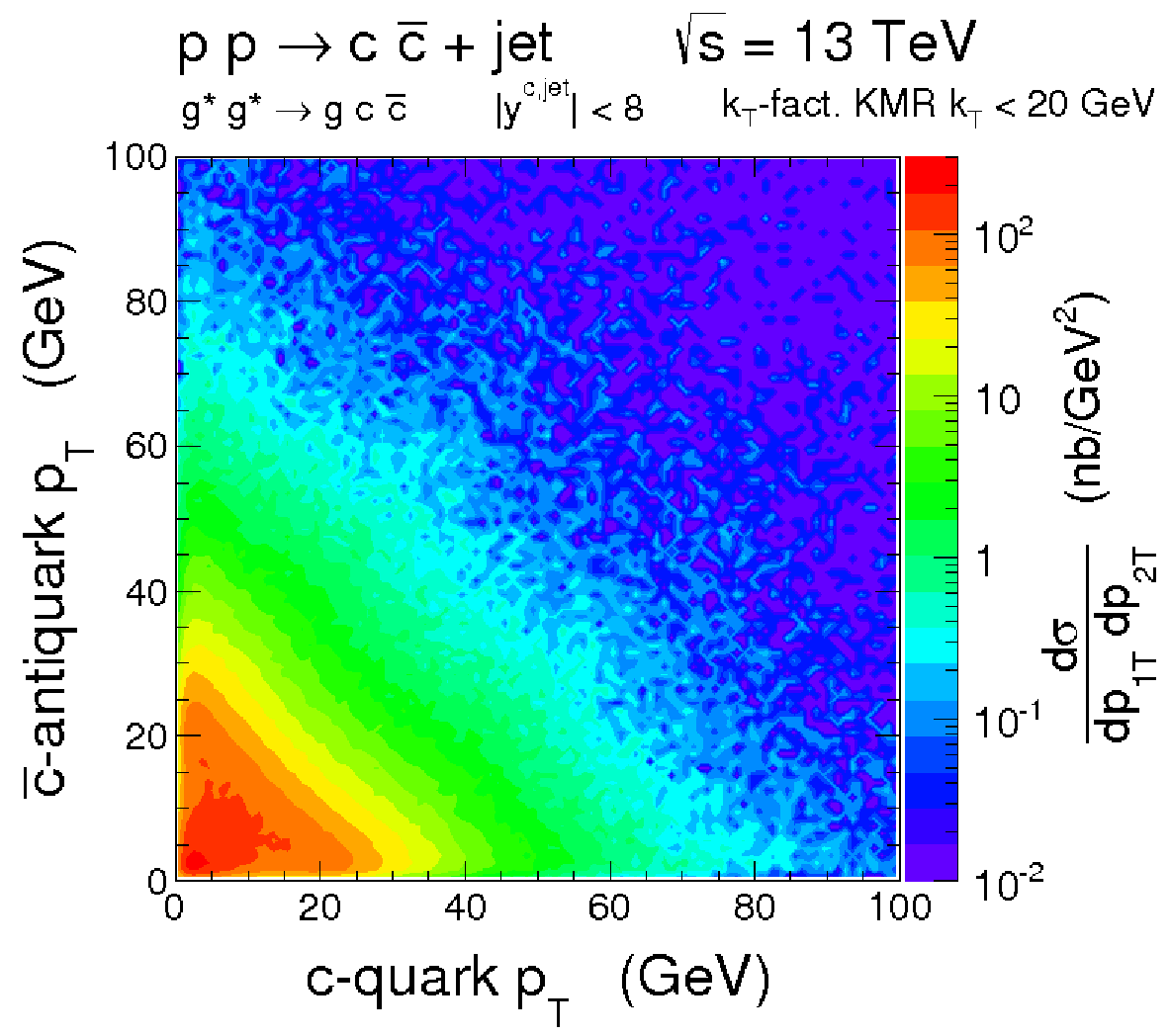}}
\end{minipage}\\
\begin{minipage}{0.33\textwidth}
 \centerline{\includegraphics[width=1.0\textwidth]{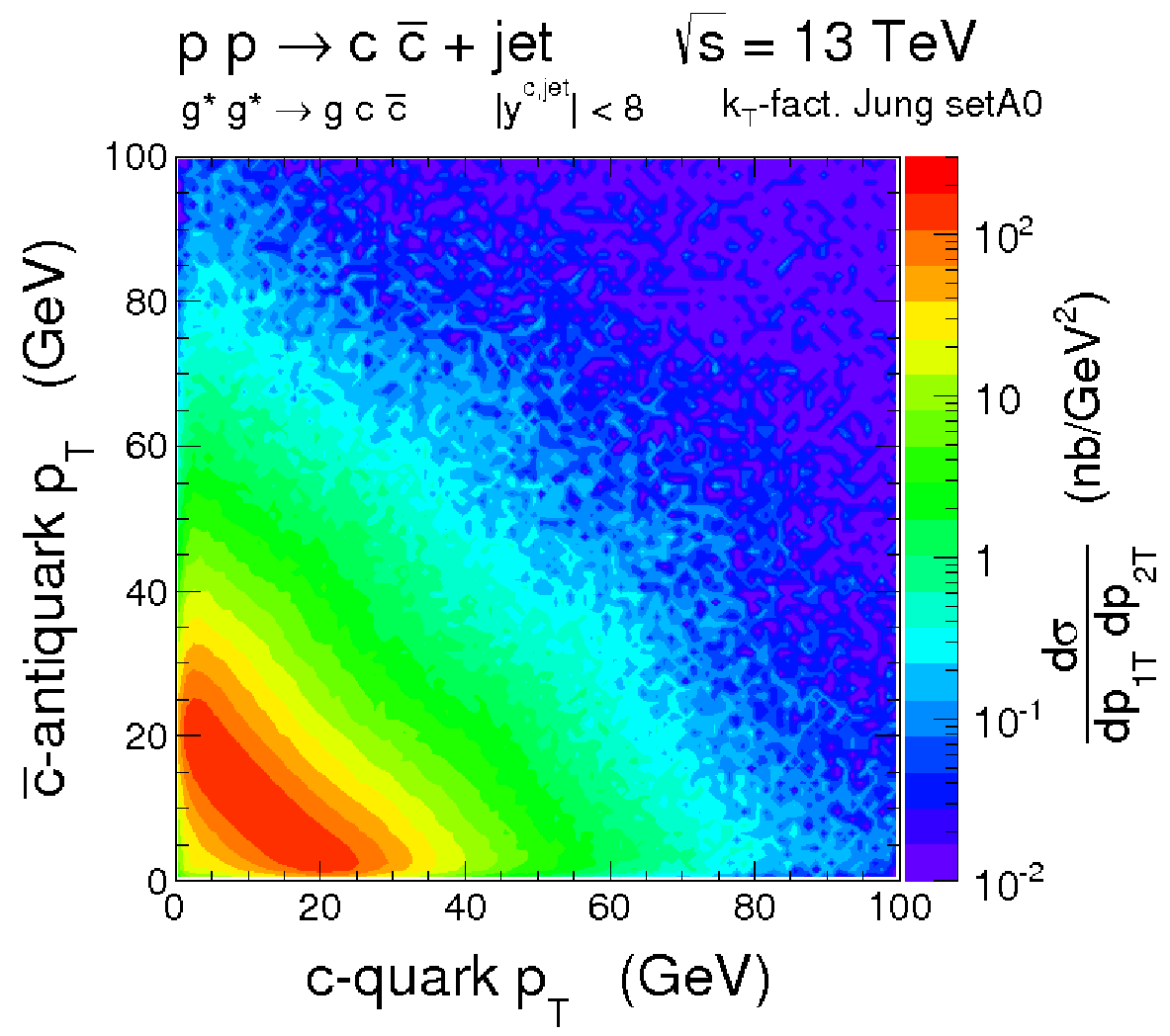}}
\end{minipage}
\hspace{0.5cm}
\begin{minipage}{0.33\textwidth}
 \centerline{\includegraphics[width=1.0\textwidth]{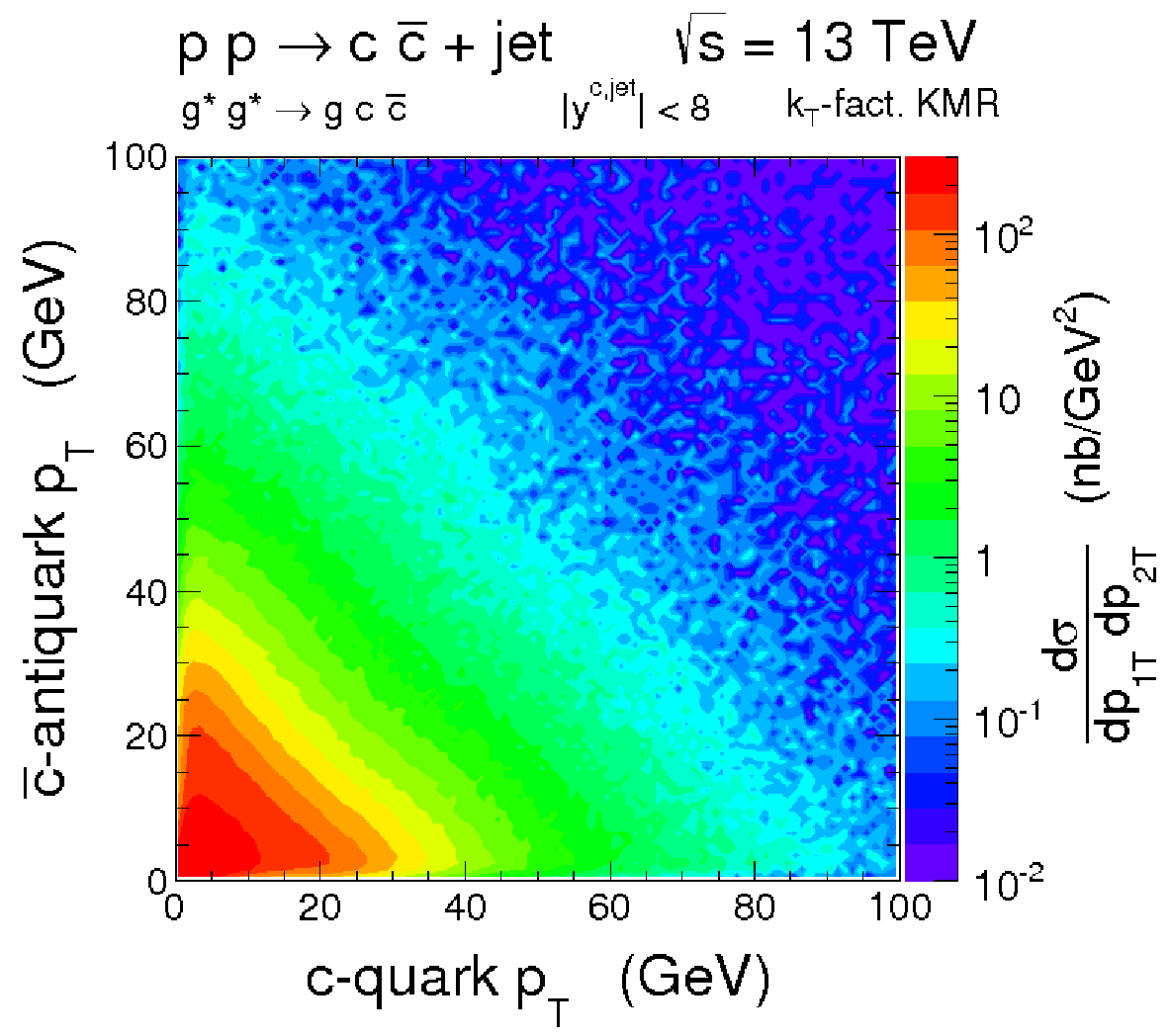}}
\end{minipage}

   \caption{
\small Two-dimensional distribution in transverse momenta of the
outgoing $c$-quark and $\bar c$-antiquark for the collinear and
$k_T$-factorization approaches. The details are specified in the figure captions.}
 \label{fig:compo_p1tp2t}
\end{figure}

\begin{figure}[!h]
\begin{minipage}{0.33\textwidth}
 \centerline{\includegraphics[width=1.0\textwidth]{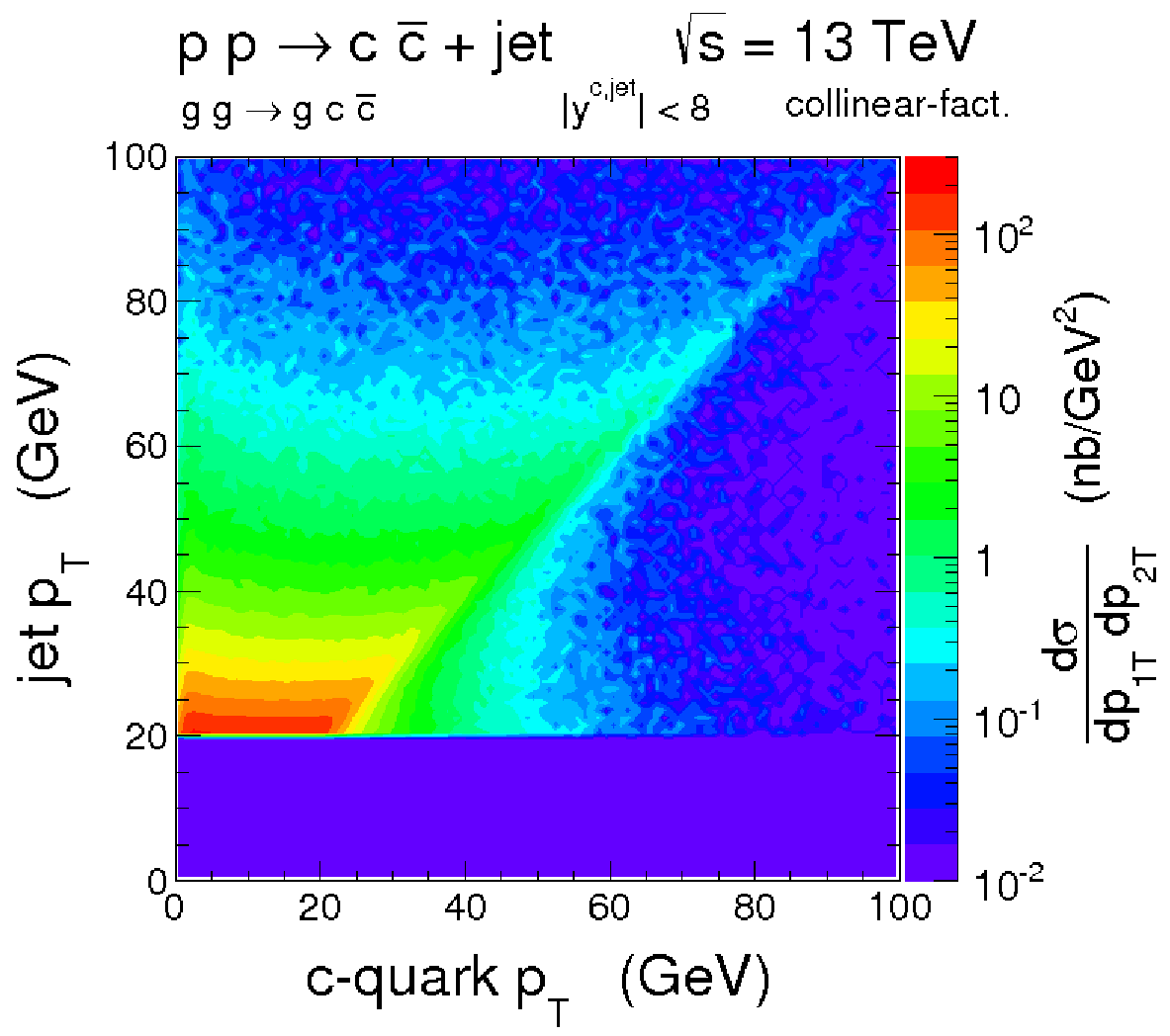}}
\end{minipage}
\hspace{0.5cm}
\begin{minipage}{0.33\textwidth}
 \centerline{\includegraphics[width=1.0\textwidth]{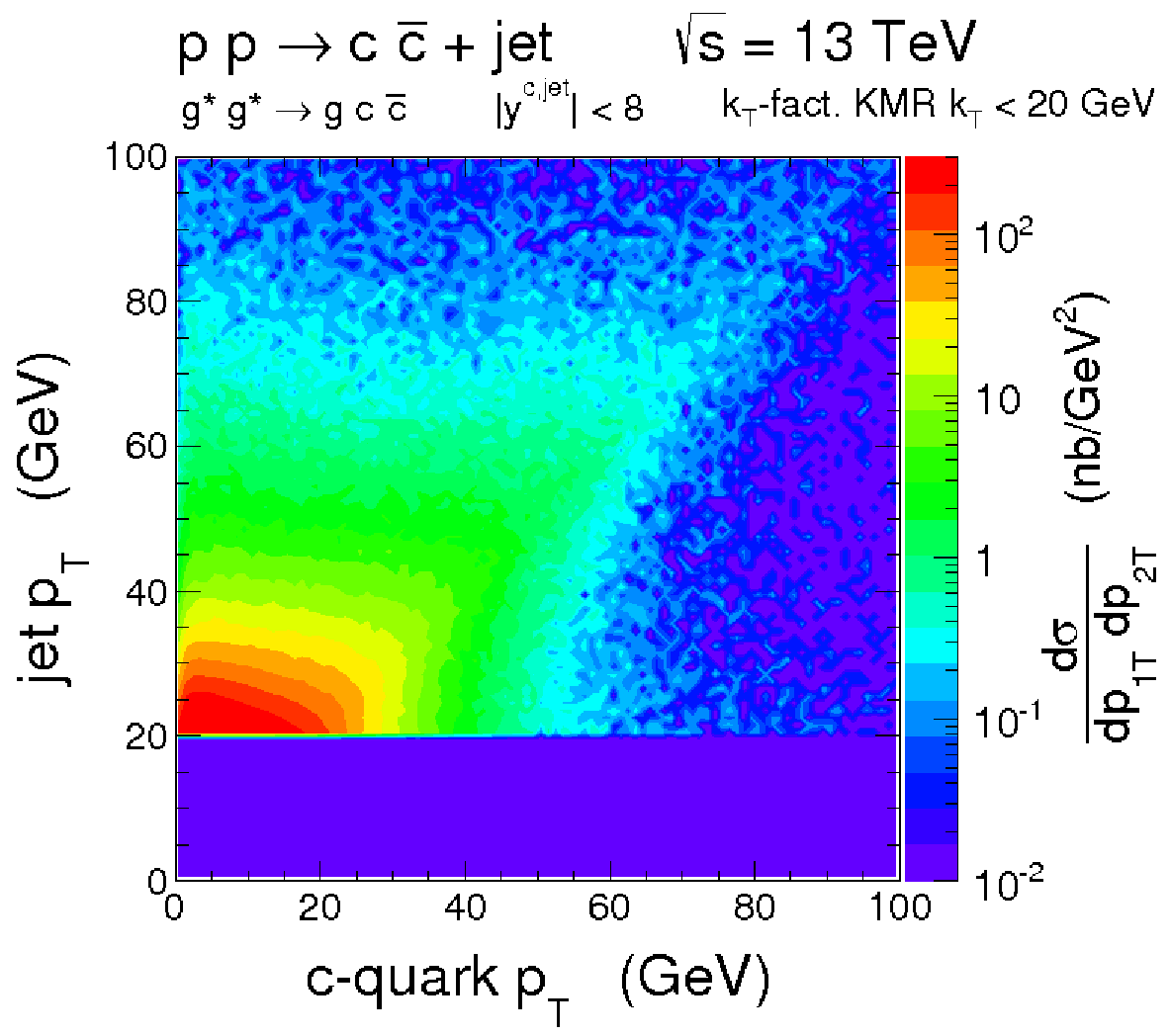}}
\end{minipage}\\
\begin{minipage}{0.33\textwidth}
 \centerline{\includegraphics[width=1.0\textwidth]{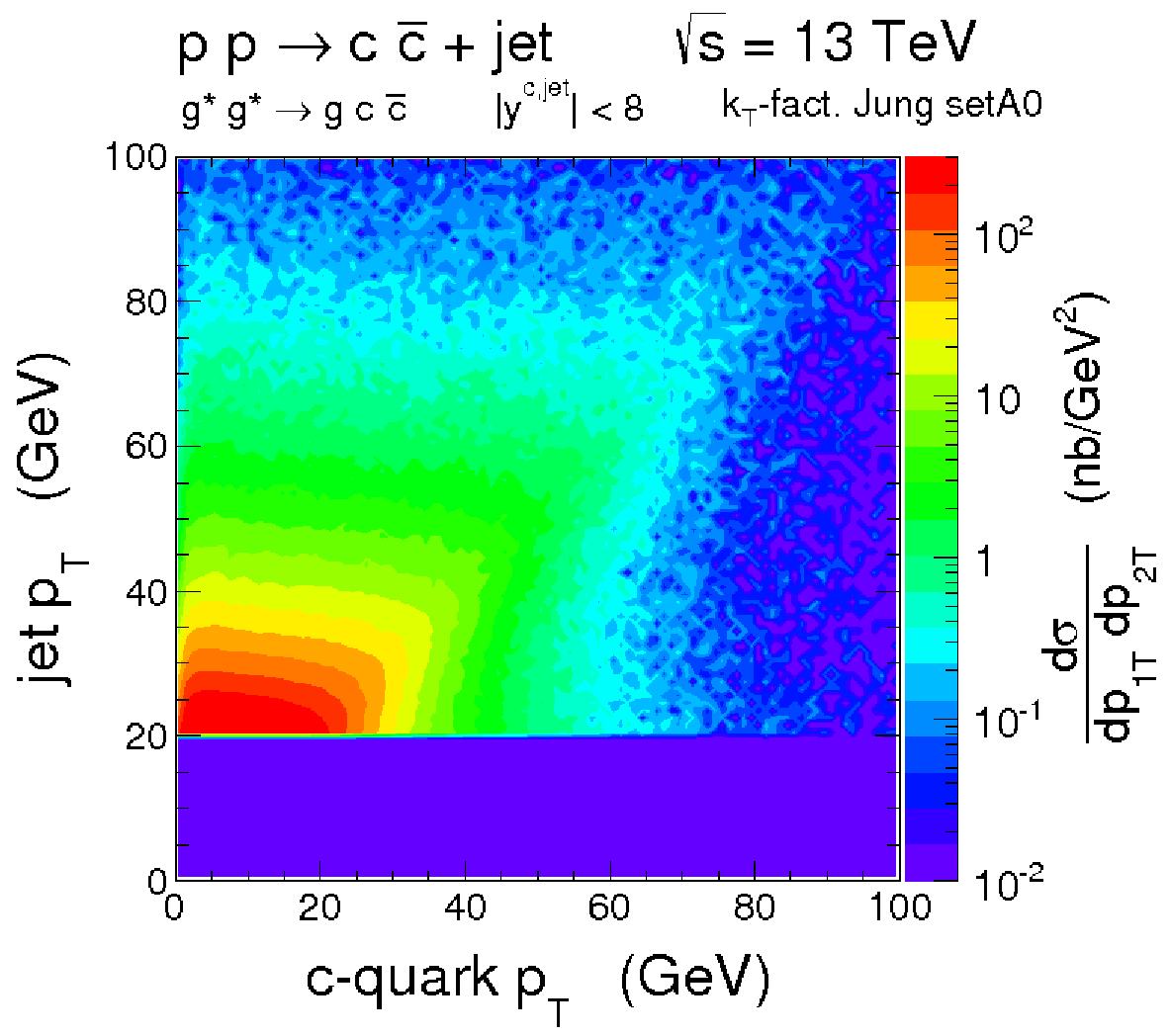}}
\end{minipage}
\hspace{0.5cm}
\begin{minipage}{0.33\textwidth}
 \centerline{\includegraphics[width=1.0\textwidth]{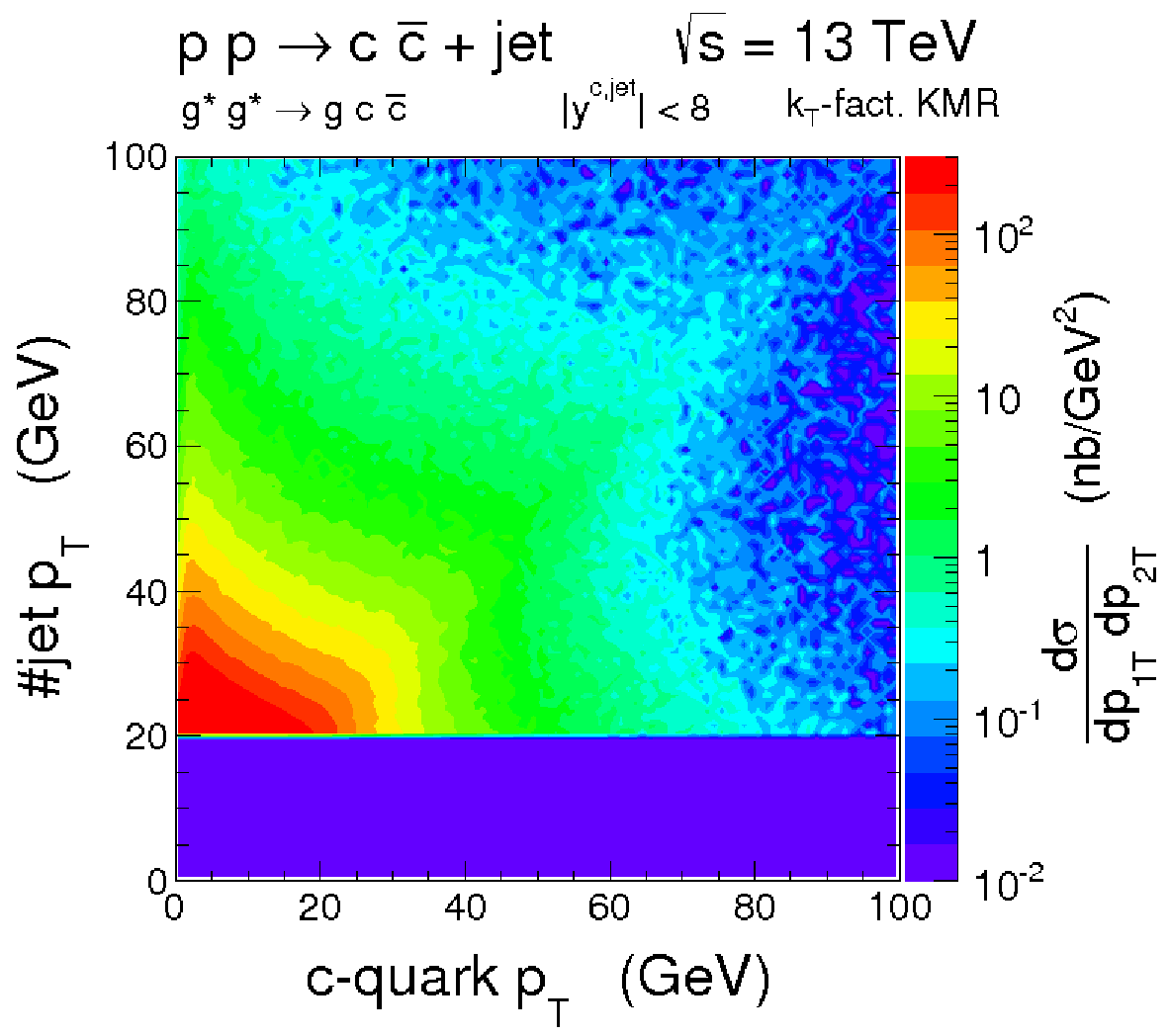}}
\end{minipage}

   \caption{
\small Two-dimensional distribution in transverse momenta of the
outgoing $c$-quark and jet for the collinear and $k_T$-factorization approaches.
The details are specified in the figure captions.
 }
 \label{fig:compo_pctpjt}
\end{figure}

\begin{figure}[!h]
\begin{minipage}{0.33\textwidth}
 \centerline{\includegraphics[width=1.0\textwidth]{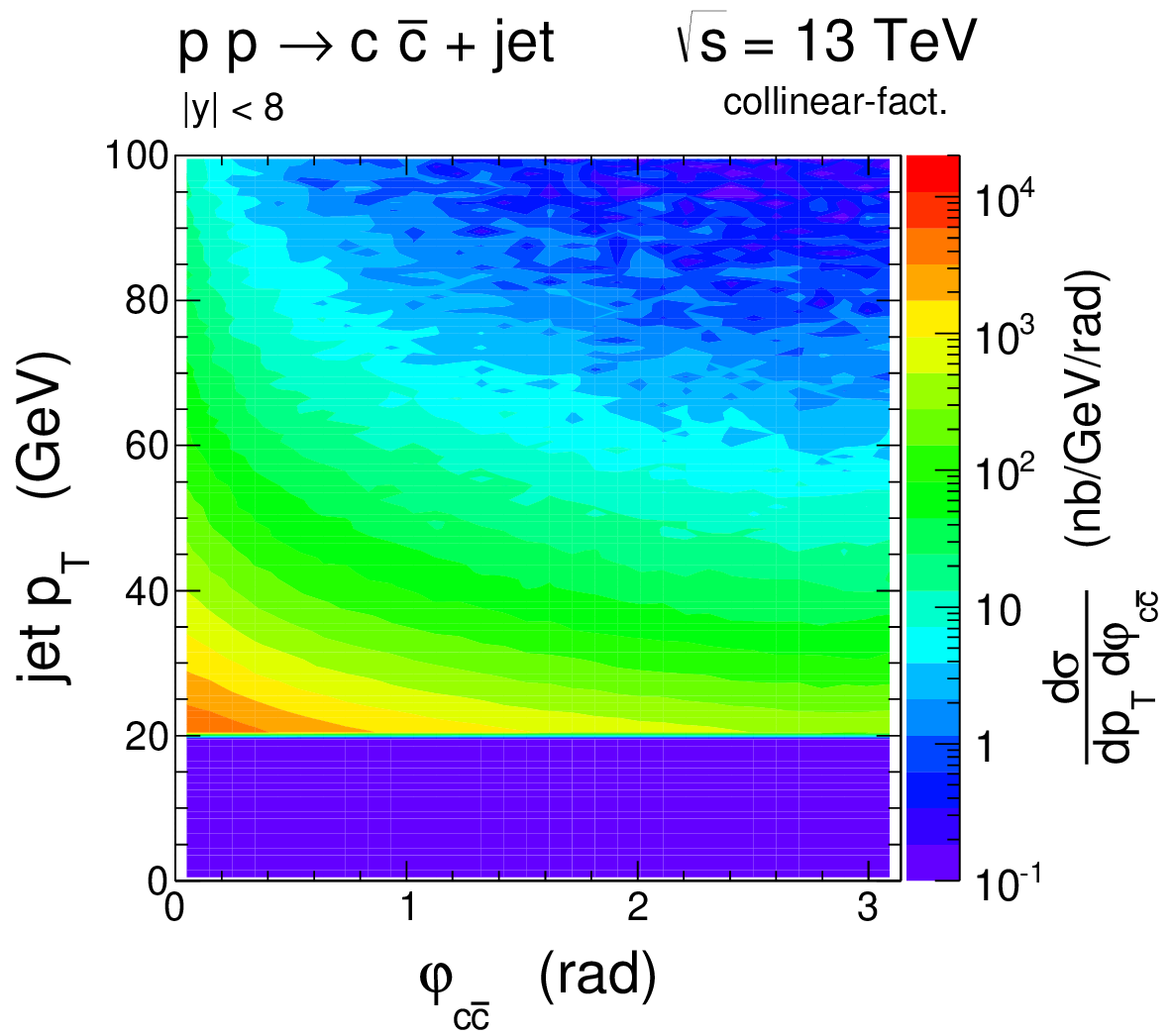}}
\end{minipage}
\hspace{0.5cm}
\begin{minipage}{0.33\textwidth}
 \centerline{\includegraphics[width=1.0\textwidth]{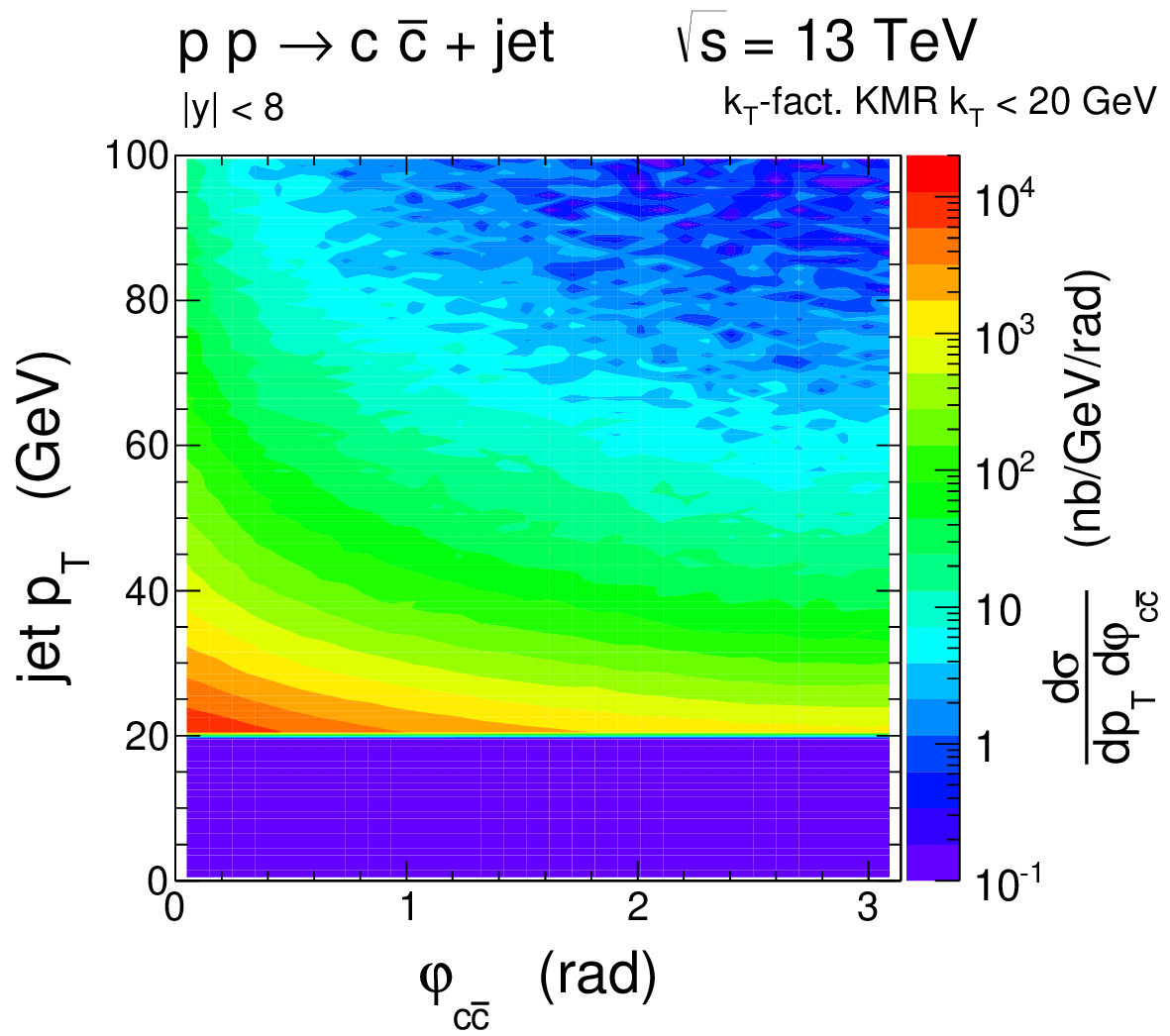}}
\end{minipage}\\
\begin{minipage}{0.33\textwidth}
 \centerline{\includegraphics[width=1.0\textwidth]{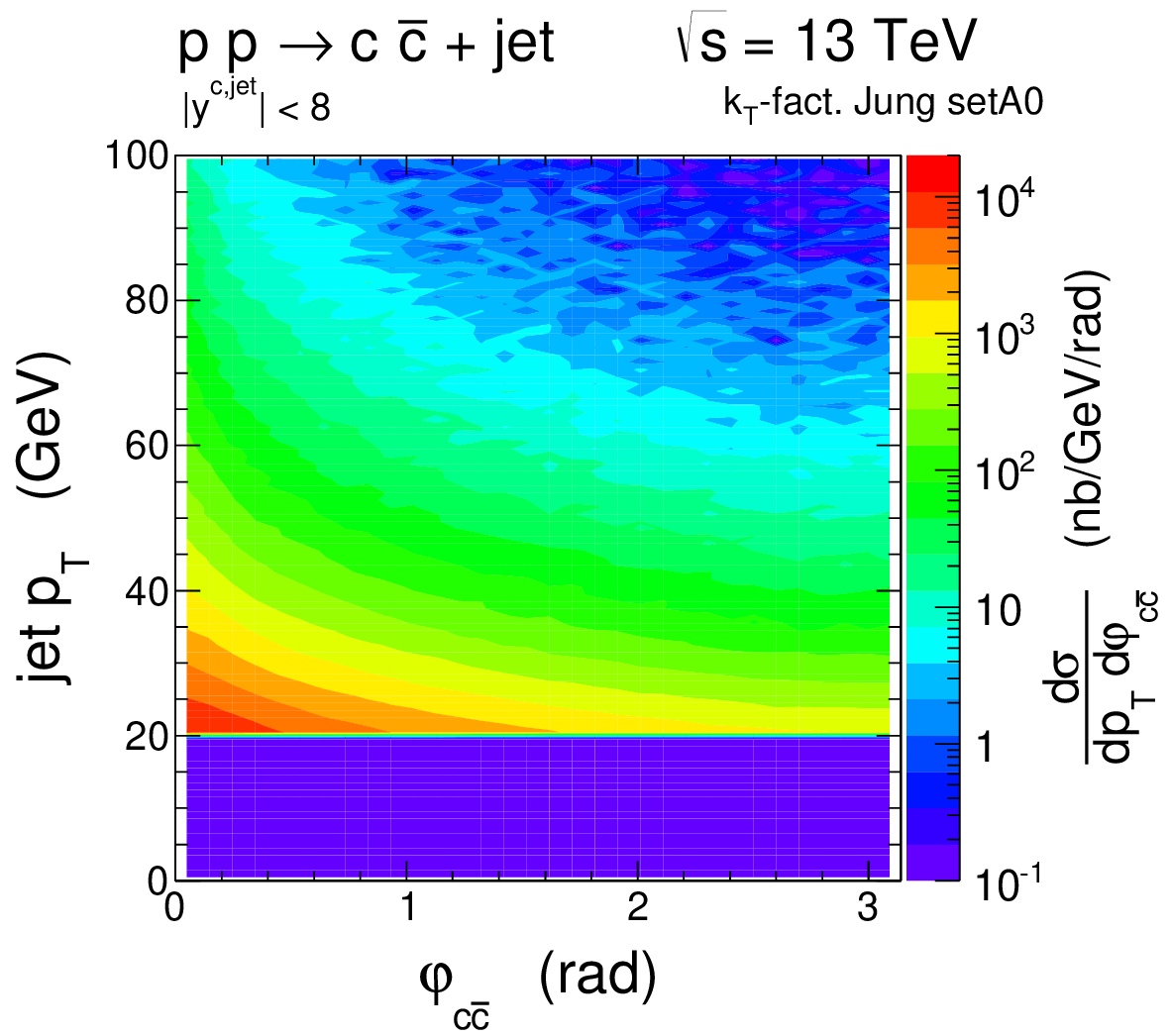}}
\end{minipage}
\begin{minipage}{0.33\textwidth}
 \centerline{\includegraphics[width=1.0\textwidth]{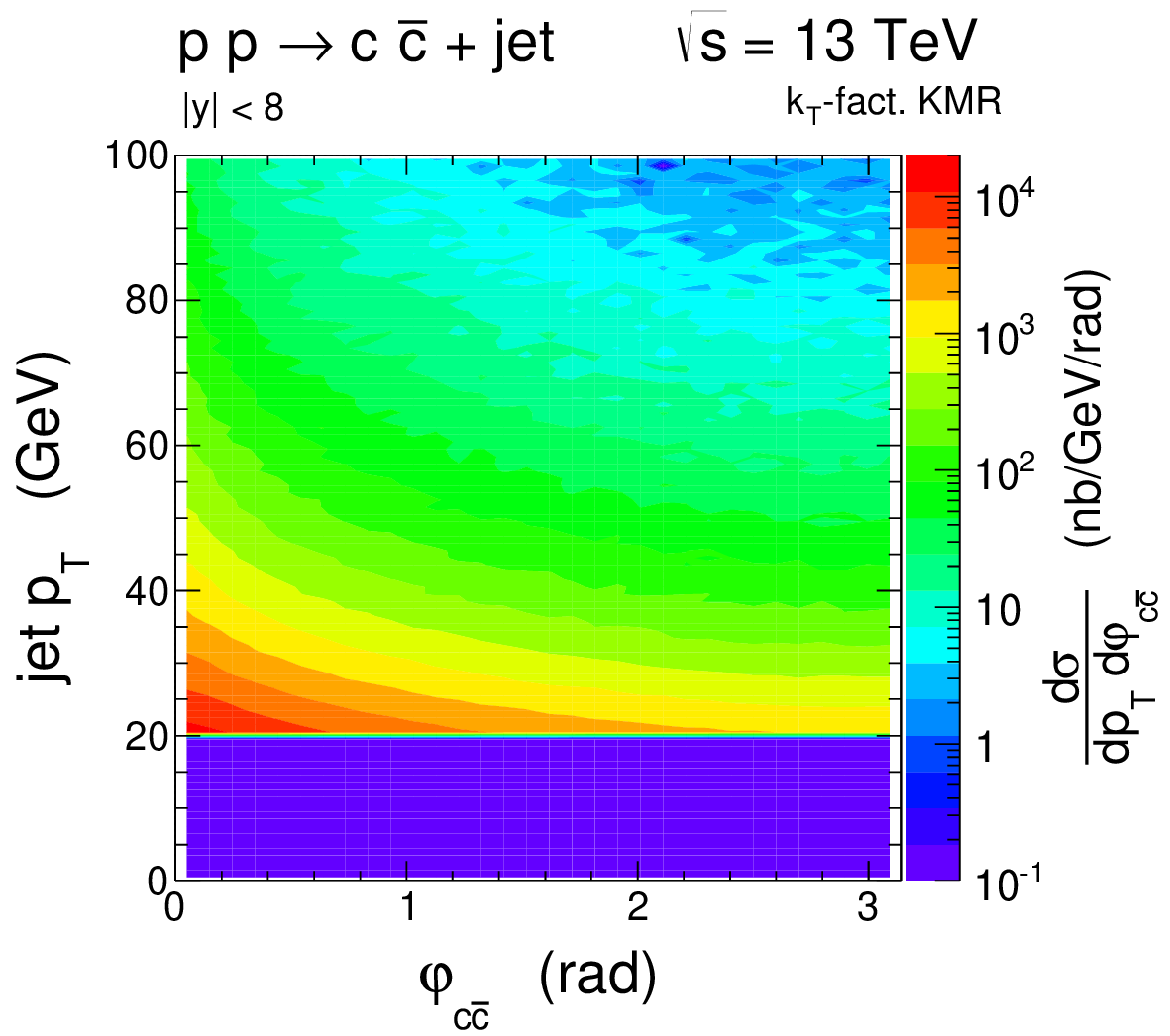}}
\end{minipage}

   \caption{
\small Two-dimensional distribution in relative azimuthal angle between
$c$-quark and $\bar c$-antiquark and jet transverse momentum
for the collinear and $k_T$-factorization approaches.
The details are specified in the figure captions.
 }
 \label{fig:compo:phiccbarpjt}
\end{figure}

The one-dimensional distributions in both approaches are rather similar.
In Fig.~\ref{fig:compo_p1tp2t}, \ref{fig:compo_pctpjt} 
and \ref{fig:compo:phiccbarpjt}
we compare also several two-dimensional distributions.
In all cases the result of the $k_T$-factorization approach with the 
KMR uGDF and the correction for exclusion of extra (more than one) jets 
gives rather similar results as those in the collinear approach.
Other uGDFs may give slightly different results.
In the LO collinear approach jet-$p_{T}$ is balanced by transverse momenta of $c$ and $\bar c$.
Therefore the sharp cut $p_{T}^{jet} > 20$ GeV generates an excluded region of the triangle shape at small $p_{T}^{c}$ and $p_{T}^{\bar c}$.
In contrast, there is no such excluded region in the $k_{T}$-factorization.
Clearly detailed studies of such two-dimensional distributions
would be an important test for uGDFs.

%-----------------------------------------------------------------------------
%------------------------------------------------------------------------------

%-----------------------------------------------------------------------------
%------------------------------------------------------------------------------

%-----------------------------------------------------------------------------
%------------------------------------------------------------------------------

%---------------------------------------------------------------------
\subsection{\boldmath{$p_{T}$} distributions of
\boldmath{$c/{\bar c}$} associated with jets and for the
inclusive case}
%---------------------------------------------------------------------

\begin{figure}[!h]
\begin{minipage}{0.47\textwidth}
 \centerline{\includegraphics[width=1.0\textwidth]{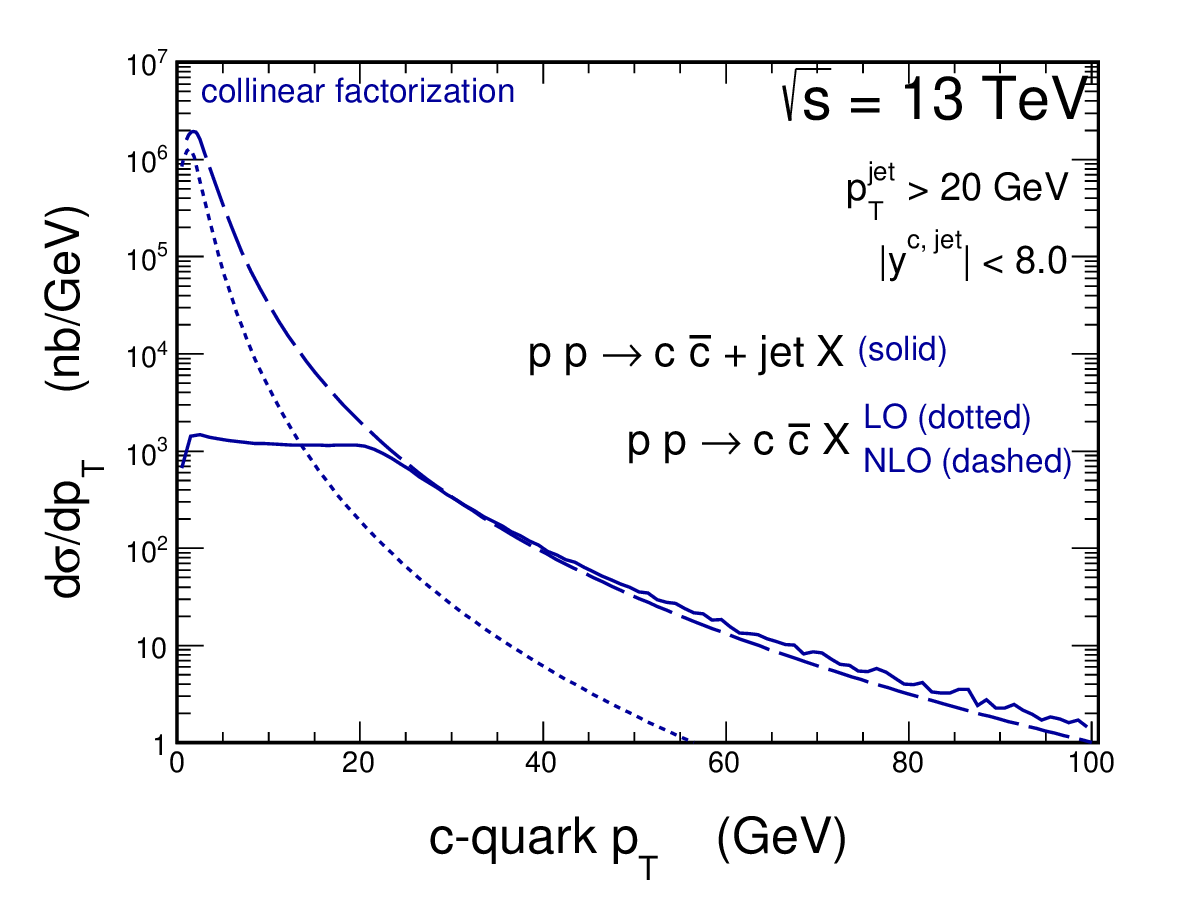}}
\end{minipage}
\hspace{0.5cm}
\begin{minipage}{0.47\textwidth}
 \centerline{\includegraphics[width=1.0\textwidth]{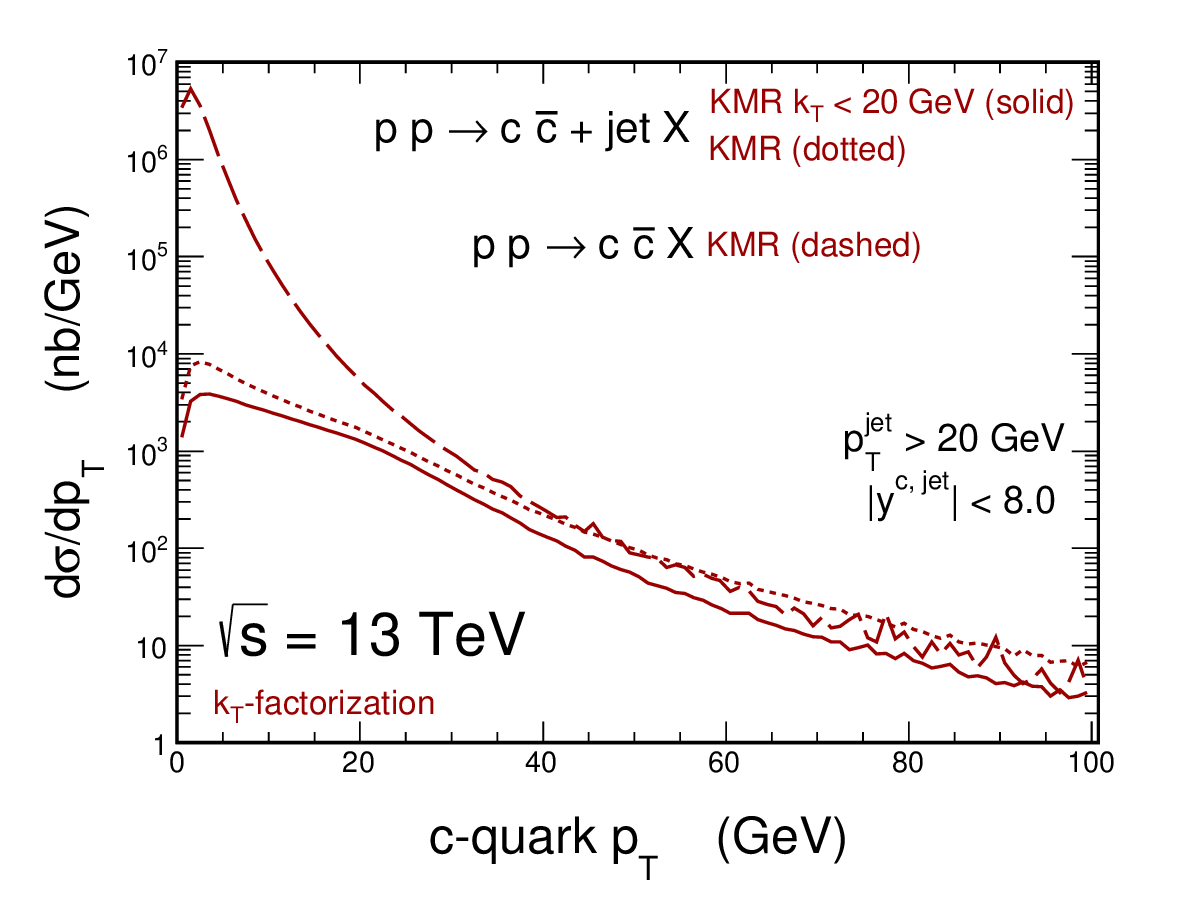}}
\end{minipage}
   \caption{  
\small Transverse momentum distribution of $c$-quarks in the collinear
(left panel) and $k_T$-factorization (right panel) approaches.
We compare results of $2 \to 2$ and $2 \to 3$ partonic calculations.
More detailed discussion is given in the main text.}
 \label{fig:ccbar_vs_ccbarjet}
\end{figure}

In this subsection we wish to discuss how the 
transverse momentum distributions of $c$ or ${\bar c}$ quarks/antiquarks
associated with jet compare to the so-called inclusive charm distributions
(see e.g. Ref.~\cite{Maciula:2013wg}).
In Fig.~\ref{fig:ccbar_vs_ccbarjet} we make such a comparison for the 
collinear-factorization (left panel) and for the $k_T$-factorization
(right panel).

%-----------------------------------------------------------------------------
%------------------------------------------------------------------------------

For the collinear case we show both LO (dotted line) and 
NLO \cite{Cacciari-Web} (dashed line) results. The NLO distribution is much
larger than the LO distribution especially for large transverse momenta
of $c$ or ${\bar c}$. The correponding $K$-factor is therefore
strongly dependent on $p_T^{c}$. For comparison we show the result for the
associated $c \bar c + jet$ production (solid line). We can see that the correponding
$c$ or $\bar c$ distribution almost coincides with that for the NLO
inclusive case for transverse momenta $p_{T}^{c} > p_{T,min}^{jet} = 20$ GeV.
This shows that the NLO distribution at large $p_T$ is practically
always associated with a (mini)jet. 

The same is true for the $k_T$-factorization with the KMR uPDFs. 
The $2 \to 2$ and $2 \to 3$ results coincide at large $p_{T}^{c}$.
This clearly shows that the KMR uPDFs already in the $2 \to 2$ case effectively
include higher-order corrections related to associated jet production ($2\to 3$ and even $2 \to 4$ cases). 
We get slightly lower cross section for $2 \to 3$ case when imposing the extra cut 
$k_{1,T},k_{2,T} < p_{T,min}^{jet}$ that in our opinion restricts the calculations to the case of production of the single-jet coming from hard-interaction and
does not allow for additional jets hidden in the KMR uPDFs (and not controlled in the calculation).

The $k_T$-factorization {\bf$2 \to 3$} result with the KMR uPDFs and the extra cut on
$k_{1,T}$ and $k_{2,T}$ almost coincides with the collinear NLO result at large $p^{c}_{T}$'s
as can be inferred by comparing the left and right panels.
However, there is a difference between the two approaches for $c \bar c + jet$ predictions at small
transverse momenta $p_{T}^{c} < p_{T,min}^{cut} = 20$ GeV.
Certainly a measurement of $D^0$ (${\bar D}^0$) mesons in association
with jets would be a valuable option at the LHC.

%------------------------------------------------------------------------------------
\subsection{Predictions for \boldmath{$D^{0} + jet$} production at the LHC}
%------------------------------------------------------------------------------------

\begin{figure}[!h]
\begin{minipage}{0.47\textwidth}
 \centerline{\includegraphics[width=1.0\textwidth]{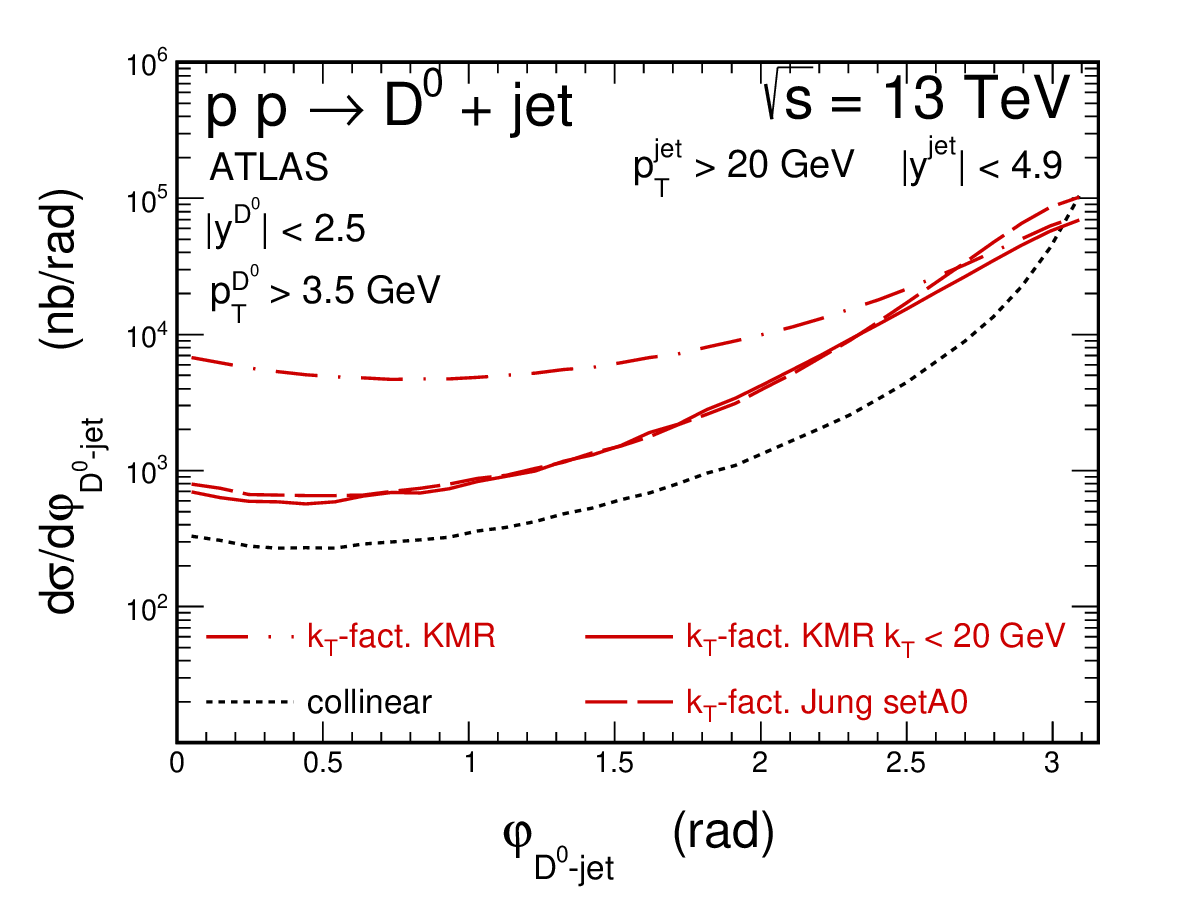}}
\end{minipage}
   \caption{
\small Azimuthal angle correlation between $D^0$ meson and jet for the
collinear and $k_T$-factorization approaches.
The cuts are specified in the figure caption.
 }
 \label{fig:phi_D0jet}
\end{figure}

Finally we present first measureable predictions for the LHC for the $D^0$ + jet
(including ${\bar D}^0$ + jet) production. As an example we consider
the case of the ATLAS apparatus. We assume that $D^0$ or ${\bar D}^0$ are registered
by the main ATLAS tracker and the jets are produced in the interval
$|y_{jet}| <$ 4.9. Similar experimental conditions for the considered reaction can be also achieved by the CMS experiment. The distribution in azimuthal angle between
$D^0$ and the jet is shown in Fig.~\ref{fig:phi_D0jet}.
This is a new possibility to test uGDFs.
We show result for different uGDFs as well as for the collinear result.

%-----------------------------------------------------------------------------
%------------------------------------------------------------------------------

Another interesting option is to look at $D^0$-${\bar D}^0$ 
correlations. This is a bit more complicated and will be discussed
elsewhere. ALICE and LHCb has smaller capability (smaller range of
pseudorapidities) to measure jets and therefore we leave corresponding calculations for a dedicated studies.

In Table I we present the integrated cross section for the ATLAS acceptance
(for $D^0$ mesons and jets) for different uGDFs and for the collinear
case. Clearly measurable cross sections of the order of a few to tens of $\mu$b (depending on the jet-$p_{T}$ cuts) 
are obtained.

%------------------------------------------------------------------------------------------------------------------------------
\begin{table}[tb]%
\caption{The calculated cross sections in microbarns for inclusive
  $D^{0}+\mathrm{jet}$ (plus $\bar{D^{0}}+\mathrm{jet}$) production in $pp$-scattering at $\sqrt{s} =$ 13 TeV for different cuts on transverse momentum of the associated jet. Here, the $D^{0}$ meson is required to have $|y^{D^{0}}| < 2.5$ and $p_{T}^{D^{0}} > 3.5$ GeV and the rapidity of the associated jet is $|y^{jet}| < 4.9$, that corresponds to the ATLAS detector acceptance.}

\label{tab:cross sections}
\centering %
%\newcolumntype{Z}{>{\centering\arraybackslash}X}
%\newcommand{\tn}{\tabularnewline}
\resizebox{\textwidth}{!}{%
\begin{tabularx}{0.9\linewidth}{c c c c c}
\\[-4.ex] 
\toprule[0.1em] %
\\[-4.ex] 
%\\[1.0ex]

%\multirow{1}{4.cm}{Jet $p_{T}$ cut} & \multirow{1}{2.6cm}{collinear} & \multirow{1}{1.5cm}{$\;\;\;$KMR} & \multirow{1}{3.2cm}{KMR {\small $k_{T} < 20$ GeV}} & \multirow{1}{2.cm}%{$\;$Jung setA0}  \\[+0.4ex]
%\bottomrule[0.1em]

\multirow{2}{3.cm}{$p_{T,min}^{jet}$ cuts}     & \multirow{1}{3.5cm}{collinear} & \multicolumn{3}{c}{$k_{T}$-factorization approach}   \\ [-0.2ex]
                      & \multirow{1}{3.5cm}{{\small MMHT2014nlo}}    &     \multirow{1}{1.5cm}{KMR}         &   \multirow{1}{3.5cm}{KMR {\small $k_{T} < p_{T,min}^{jet}$}}     &   \multirow{1}{3.0cm}{Jung setA0}  \\ [-0.2ex]
\bottomrule[0.1em]

\multirow{1}{3.cm}{$p_{T}^{jet} > 20$ GeV}  & \multirow{1}{3.5cm}{$\qquad 22.36$} & \multirow{1}{1.5cm}{49.20} & \multirow{1}{3.5cm}{$\qquad 33.12$} & \multirow{1}{3.0cm}{$\quad 43.45$} \\  [-0.2ex]
\multirow{1}{3.cm}{$p_{T}^{jet} > 35$ GeV}  & \multirow{1}{3.5cm}{$\qquad 3.70$} & \multirow{1}{1.5cm}{9.60} & \multirow{1}{3.5cm}{$\qquad 6.76$} & \multirow{1}{3.0cm}{$\quad 6.79$} \\  [-0.2ex]
\multirow{1}{3.cm}{$p_{T}^{jet} > 50$ GeV}  & \multirow{1}{3.5cm}{$\qquad 1.14$} & \multirow{1}{1.5cm}{3.32} & \multirow{1}{3.5cm}{$\qquad 2.45$} & \multirow{1}{3.0cm}{$\quad 1.94$} \\  [-0.2ex]

%\multirow{1}{4.cm}{ATLAS:} &  \multirow{2}{2.6cm}{$\;\;\;\;\;\;\;\;21.96$}   & \multirow{2}{1.5cm}{$\;\;\;48.41$}   & \multirow{2}{3.2cm}{$\;\;\;\;\;\;\;\;\;\;\;32.51$} & %\multirow{2}{2.cm}{$\;\;\;\;\;42.95$}  \\ [-0.2ex]
%\multirow{1}{4.cm}{??}     &                                    &                                     &   &  \\ [+0.4ex]

\hline

\bottomrule[0.1em]

\end{tabularx}
}
\end{table}
%-------------------------------------------------------------------------------------------------------------------------------

%--------------------------
\section{Conclusions}
%--------------------------

In the present paper we have presented first theoretical study 
related to associated production of charm and single-jet.
The related calculations were performed both in the collinear and
$k_T$-factorization approaches. The related matrix elements were obtained
with the help of the AVHLIB events generator.

We have performed first phenomenological study for $\sqrt{s} =$ 13 TeV.
In most cases we have limited to parton-level and full
phase space. 
Different uGDFs have been used in the $k_T$-factorization approach calculations.
We have compared results of the collinear approach with those within
the $k_T$-factorization. The results, one-dimesional and two-dimensional
distributions, with the KMR uGDF and a practical correction to exclude
production of more than one jet are very similar as those obtained
within the collinear approach.

We have discussed in addition how the transverse momentum distributions 
of $c/\bar{c}$ associated with jets relate to the inclusive charm case both 
for collinear LO/NLO calculation and for the $k_T$-factorization 
with the KMR uPDFs. We have shown that the production 
of $c$ (or $\bar c$) at large transverse momenta is
unavoidably related to an emission of extra parton (jet).
We have shown that the KMR uPDFs in the case of inclusive charm distributions effectively include
higher-order corrections related to associated jet production.
The distributions of $c/\bar c$ associated with jets obtained within the $k_T$-factorization
with the KMR uPDFs, when done without extra conditions, contain effectively the
situations with more than one jet (two or even three).
This can be approximately eliminated by imposing extra cut(s) on
transverse momenta of the initial partons. Then such a result
almost coincides with the NLO collinear result for inclusive charm production at large $c/\bar c$ transverse momenta.

We have also performed first feasibility studies of $D^{0}$ + jet production for ATLAS (and/or CMS)
cuts. We have obtained rather large cross sections.
We hope our first phenomenological studies will be an inspiration
for experimental groups to perform corresponding or similar analysis.

Associated production of charm (charmed mesons) and single-jet
is only a first step. In principle, production of charm and two jets
is equally interesting and can be done within considered here framework.
The program of activity sketched here may be very important in detailed
testing of pQCD dynamics for more complicated processes.
The large luminosity at the LHC now available opens such a possibility.
So far most of the efforts at the LHC concentrated on inclusive charm production. However, a study of correlation observables 
(some examples have been discussed here) may be very interesting 
and important in this context.

\vspace{1cm}

{\bf Acknowledgments}

We are particularly indebted to Andreas van Hameren for teaching us how
to use his code.
This study was partially
supported by the Polish National Science Center grant
DEC-2014/15/B/ST2/02528 and by the Center for Innovation and
Transfer of Natural Sciences and Engineering Knowledge in
Rzesz{\'o}w.

%-------------------------------------------------------------------------------------

\end{document}